\documentclass[trackchanges, twocolumn]{aastex701}

\usepackage{xspace}

\newcommand{\enzo}{\texttt{Enzo}\xspace}
\newcommand{\renaissance} {\texttt{Renaissance}\xspace}
\newcommand{\LCDM}{$\Lambda$CDM\xspace}

\newcommand{\yt} {\texttt{yt}\xspace}
\newcommand{\ytree} {\texttt{ytree}\xspace}
\newcommand{\fsps} {\texttt{FSPS}\xspace}
\newcommand{\powderday} {\texttt{Powderday}\xspace}
\newcommand{\hyperion} {\texttt{Hyperion}\xspace}
\newcommand{\cloudy} {\texttt{Cloudy}\xspace}

\newcommand{\sphinx} {\texttt{SPHINX}\xspace}
\newcommand{\thesan} {\texttt{THESAN}\xspace}

\newcommand{\illustristng} {\texttt{IllustrisTNG}\xspace}
\newcommand{\astrid} {\texttt{ASTRID}\xspace}
\newcommand{\megatron} {\texttt{MEGATRON}\xspace}
\newcommand{\thesanzoom} {\texttt{THESAN-ZOOM}\xspace}
\newcommand{\astropy} {\texttt{astropy}\xspace}

\begin{document}

\title{JWST Predictions for $z > 10$ Galaxies from the Renaissance Simulations - I: Photometry and Sizes}

\author[orcid=0009-0009-8242-0570,gname=Samantha, sname=Hardin]{Samantha E. Hardin}
\affiliation{Center for Relativistic Astrophysics, Georgia Institute of Technology, 837 State Street, Atlanta, GA 30332, USA}
\email[show]{shardin31@gatech.edu}  

\author[orcid=0000-0003-1173-8847,gname=John, sname=Wise]{John H. Wise} 
\affiliation{Center for Relativistic Astrophysics, Georgia Institute of Technology, 837 State Street, Atlanta, GA 30332, USA}
\email{jwise@physics.gatech.edu}

\author[orcid=0009-0004-3017-0673, gname=Emily]{Emily K. Troutman}
\affiliation{Center for Relativistic Astrophysics, Georgia Institute of Technology, 837 State Street, Atlanta, GA 30332, USA}
\email{etroutman6@gatech.edu}

\begin{abstract}

JWST has enabled new high redshift observations with 14 spectroscopically confirmed galaxies at $z > 10$ to date, leading to a need for high redshift, high resolution simulations to interpret these observations. We present the physical properties and mock observations of the galaxies in the \renaissance Simulations to add to the growing database of high redshift simulation data to guide and interpret observations. We find that they provide an accurate representation of the formation history of JWST's $z > 10$ discoveries and follow the trends observed in JWST galaxies but extended to lower masses. The stellar masses of the \renaissance galaxies range from $\approx 10^{3}$ to $10^8 \ \textup{M}_\odot$ and overlap well with the $z > 10$ JWST galaxies with a stellar mass range of about $10^{7}$ to $10^9 \ \textup{M}_\odot$. The simulated SFRs increase from $10^{-4}$ to $10^1 \ \textup{M}_\odot \textup{yr}^{-1}$, overlapping the JWST galaxies' lower SFRs in the range $1 - 20 \ \textup{M}_\odot \textup{yr}^{-1}$. These compact galaxies show minimal morphology change as their stellar masses increase with the majority of the half light radii between $1$ and $10$ pc and the majority of the half stellar mass radii around $0.1$ kpc; their Sersic indices vary between $0$ and $4$.  The \renaissance galaxies are bluer and generally transition well into the absolute UV magnitudes of the JWST galaxies in the main sequence of galaxies. Overall, our simulations agree well with JWST's discoveries to date, making them a valuable tool in the continued effort to understand the high redshift Universe.

\end{abstract}



\section{Introduction}
\label{intro}


JWST has obtained images and spectroscopy of galaxies at higher redshifts than ever before, granting us new and exciting insights into the early Universe. Thus far, JWST has found hundreds of galaxies with a redshift $z > 8$ and about 20 galaxies with $z > 10$ \citep{adamo2024billionyearsaccordingjwst}, with 14 of those spectroscopically confirmed. Currently, the highest redshift of a spectroscopically confirmed galaxy is $z = 14.44^{+0.02}_{-0.02}$ \citep{naidu2025cosmicmiracleremarkablyluminous}. This galaxy was found by the ``Mirage or Miracle'' (MoM) JWST NIRSpec survey focusing on galaxies previously observed with JWST's imaging instruments NIRCam + MIRI. Prior to JWST's launch, the highest redshift galaxy that had been discovered was GNz-11, found by the Hubble Space Telescope at $z = 11.09^{+0.08}_{-0.12}$ \citep{oesch_gnz11_hubble}. \par


The early JWST surveys, JWST Advanced Deep Extragalactic Survey \citep[JADES;][]{JADES_overview, eisenstein2023overview} and Cosmic Evolution Early Release Science \citep[CEERS;][]{CEERS_overview}, detected six spectroscopically confirmed galaxies with $z > 10$; we have used these surveys as references for our current and previous work \citep{mccaffrey2023tension}. JADES detected the galaxies GS-z10-0 ($z = 10.38$), GS-z11-0 ($z = 11.58$), GS-z12-0 ($z = 12.63$), GS-z13-0 ($z = 13.20$), and GN-z11 ($z = 10.60$) \citep{Bunker_2023, robertson_2022, curtislake2023spectroscopicconfirmationmetalpoorgalaxies}, while CEERS detected Maisie's galaxy ($z = 11.44$) \citep{arrabel_maisies_galaxy}. Both of these surveys make use of JWST's primary imager NIRCam and its spectrograph NIRSpec. JADES points at the Great Observatories Origins Deep Survey (GOODS) fields, while CEERS points at the Cosmic Assembly Near-infrared Deep Extragalactic
Legacy Survey (CANDELS) Extended Goth Strip \citep{grogin_candels, koekemoer_candels}, both of which were previously studied with the Hubble Space Telescope. \par

Through these galaxies and others, JWST has been used to make some exciting discoveries. Studies find prolonged production of UV photons in the early Universe and that very faint galaxies are the primary source of star formation in the early Universe because of the high number density. Through spatial and spectral modeling, stellar masses and star formation rates (SFRs) are inferred with strong rest-frame optical breaks as a reliable indicator for stellar ages \citep{Weibel_2024, Harvey_2024, adamo2024billionyearsaccordingjwst}. The star clusters in these galaxies at $z > 6$ are key players in reionization \citep[e.g.][]{stanway}. 
In this paper, we present results from our catalog of mock observations of high redshift galaxies from the \renaissance Simulations \citep{oshea2015probing} and their correlated physical properties that can be used to predict incredibly faint galaxies JWST may observe and those galaxies' formation histories. Currently there are no publicly available synthetic observational databases of galaxies in the high redshift ($z \geq 10$), low mass ($\textrm{M}_{*} \ \leq 10^7 \textrm{M}_{\odot}$) regime that is the \renaissance Simulations target. \par 

Early results found that \renaissance's UV luminosity function aligned well with observational results for brighter luminosities ($\textup{M}_{\textup{UV}} \lesssim -18$) and the luminosity function at $z >  8$. The bulk of the simulated galaxies are currently beyond the sensitivity limit of JWST that provide a glimpse of the progenitors of the detected galaxies. Additionally, the \renaissance Simulations have shown that small galaxies are a key factor in the reionization of the Universe \citep{Xu_2016}. Firstly, at higher redshifts, these small galaxies are experiencing strong but bursty star formation as a result of massive Population III (Pop III) stars, which are metal-free stars, in the shallow potential wells in halos with masses $\geq 10^7 \ M_{\odot}$ \citep{hazlett_renaissance}. Furthermore, at the beginning of reionization (z $\sim 10 - 15$), the majority of UV radiation comes from fainter, small galaxies because of their high abundances, while more massive galaxies produce the majority of UV radiation towards the end of reionization as the number of massive galaxies increases. They found that these galaxies' contribution to reionization and their characteristics are unrelated to their environments, meaning isolated galaxy formation during the beginnings of reionization is independent of redshift \citep{Xu_2016}. Overall, \renaissance's results have aligned well with other high redshift simulations, such as \sphinx \citep{sphinx_intro, sphinx} and \thesan \citep{thesan_intro, thesan, thesan_morphology, thesan_zoom, thesan_zoom_starbursts}; although, \renaissance, being a zoom-in simulation, is only able to integrate for shorter times but at a higher resolution than other large-box simulations. \par 


It is also worth noting that JWST's discoveries have led to some conversations about the validity of the $\Lambda$CDM  model in the high redshift, high mass regime. Specifically, after the initial JWST release, these high redshift galaxies had higher than expected stellar masses, and there were concerns that the abundance of baryonic matter available in the \LCDM model was not enough to produce these high stellar mass galaxies \citep{fake_news, boylan}. We found that the stellar masses of the galaxies in the \renaissance Simulations are consistent with those found by JWST \citep{mccaffrey2023tension} that compliment a similar study \citep{Keller_2023}, alleviating concerns that the \renaissance model, and theoretical and simulation models in general, do not align with JWST results. \par 


The goal of this paper is to fill a need for observational predictions of the first galaxies, that could uncover their complete star formation history, coinciding with the tidal wave of JWST observations. While other large box simulations such as \illustristng \citep{illustris_tng}, \astrid \citep{astrid, astrid_galaxies, astrid_synthetic}, \sphinx \citep{sphinx_intro, sphinx}, \thesan \citep{thesan_intro, thesan, thesan_morphology, thesan_zoom, thesan_zoom_starbursts}, and \megatron \citep{megatron_reproducingdiversityhighredshift, megatron_impactstarformationfeedback} have created mock observations \citep{illustris_synthetic_img_torrey, vogelsberger,  sphinx, thesan, astrid_synthetic, astrid_lrd} from which we have taken inspiration, the \renaissance Simulations will add high-redshift and high-resolution simulation results to the conversation. The synthetic observations presented here from \renaissance Simulations shows the evolution of these early galaxies elucidating on the origins of observed galaxies, which is timely in this age of telescopes that can observe nearly the beginning of the era of galaxy formation like JWST, 30-m class telescopes, and the Roman Space Telescope. The ultimate goal is for the \renaissance Simulations to be a guide for future observational campaigns, presenting the highest redshift mock observations to date from simulations with extremely high resolutions, and create a database of simulation data to which observations can be compared. \par

In this paper, we present correlations between mock observables and physical properties from the \renaissance Simulations. The paper is laid out as follows: Section \ref{method} describes how we calculated our catalog and created mock observations; Section \ref{results_global} presents the results for the global properties of the \renaissance observations; Section \ref{morph} overviews our results on the morphology of the \renaissance Simulations; Section \ref{photometry} gives our results on the photometry of the \renaissance Simulations; Section \ref{discussion} contextualizes the \renaissance Simulations in regards to other high redshift simulations and presents our outlook and caveats of our research; finally, Section \ref{conclusion} summarizes our findings. All length scales throughout the paper are in proper units unless otherwise specified.


\section{Methodology}
\label{method}

\subsection{The Renaissance Simulations}

We present mock observations of the \renaissance Simulations and their correlated physical properties. The \renaissance Simulations \citep[e.g.][]{oshea2015probing} are high resolution, high redshift zoom-in simulations that focus on the first stars and galaxies and were run using the open-source adaptive mesh refinement code \enzo \citep{bryan_enzo_2014, Brummel-Smith2019}. \renaissance simulates almost 2000 galaxies and 10,000 metal-free Pop III stars in a (40 Mpc)$^{3}$ comoving volume following the transformation from metal-free to metal-enriched star formation in three zoom-in regions named the Rarepeak, Normal, and Void regions that were evolved to redshifts $z = 15, 12.5,$ and $8$, respectively, with a maximum spatial resolution of 19 comoving parsecs. The simulations were halted at these respective redshifts because of the prohibitive computational expense required for radiation transport. The simulations have comoving volumes of $133.6$, $220.5$, and $220.5 \textup{Mpc}^3$ for each region, respectively. The Rarepeak's zoom-in region is centered on two $3 \times 10^{10} \textup{M}_\odot$ halos at a redshift of $z = 6$. As the simulation progresses, the Rarepeak decreases its zoom-in volume in order to only contain the the highest resolution dark matter particles, while the Normal and Void regions do not adjust their volumes. We analyze the simulation data with the visualization analysis toolkit \yt \citep{yt}. \par

The \renaissance Simulations use the Moray adaptive ray tracer to transport UV ionizing radiation created by metal-free (Pop III) and metal-enriched (Pop II) stars \citep{wiseandabel}. The Lyman-Werner $\textup{H}_{2}$ dissociating radiation from the stellar sources is calculated in the optically thin limit. The UV photon luminosities and stellar lifetimes are taken from the model in \citet{schaerer_2002}. The Pop III stars in the simulation form in dense $\textup{H}_{2}$ rich, metal poor gas with metallicities below $10^{-4} Z_{\odot}$ and an $\textup{H}_{2}$ fraction above $10^{-4}$. The masses of these Pop III stars are drawn randomly from a top heavy IMF with a characteristic mass of $40 \ \textup{M}_{\odot}$ \citep{wise_2012_popIII} with their stellar endpoints taken from \citet{heger} resulting in a mix of supernovae and black holes. Pop II stars are modeled in Jeans resolved star-forming clouds assuming a Salpeter IMF \citep{wise_2012_popIII}. Given its high resolution, \renaissance is somewhat unique in modeling these early galaxies through its inclusion of Pop III star formation and feedback providing an accurate thermodynamic state of the protogalactic gas that collapses to form the first galaxies.


The \renaissance Simulations’ data products and analysis tools
are accessible through a website titled The Renaissance Simulation Vault.\footnote{\url{https://www.firstgalaxies.physics.gatech.edu/}} On this website, users can download the catalog of physical properties and mock observations from the Rarepeak, Normal, and Void regions. This page also includes an interactive table of each region’s individual halos, allowing users to customize search parameters by region, redshift, and Pop II and III stellar mass and number.  From this table, users can download the data from each halo individually, rather than the data from the entire Rarepeak, Normal, or Void region. Downloading instructions, simulation visualizations, background information, and research results are also available through the \renaissance Simulation Vault. 



\subsection{Mock observations with Powderday}

We create the synthetic images and SEDs presented in this paper with \powderday \citep{Narayanan_2021}, a dust radiative transfer package that converts simulation data into spectral energy distributions (SEDs) and intrinsic images of the galaxy, using \hyperion for dust radiative transfer \citep{hyperion}, Flexible Stellar Population Synthesis (\fsps) for stellar population models \citep{conroy_2009, conroy_gunn_2009, conroy_gunn_2010}, and \cloudy to calculate nebula emission lines \citep{cloudy_2013, cloudy}. \par


\powderday computes the SEDs by first calling \fsps to obtain the intrinsic stellar spectra, given the masses, metallicities, and ages of the star particles.  Emission lines are included from unresolved ionized regions around the point sources with direct \cloudy calls, which will be the subject of a forthcoming paper, and then passes these radiation point sources to \hyperion that performs dust radiative transfer in the original \enzo adaptive mesh hierarchy. We consider a present-day dust model \citep{dust_draine_2001, dust_draine_2003, hyperion} in order to not introduce additional uncertainties, given our limited observational constraints on the dust content of the early universe. Photon packets are randomly sampled from the point sources, proportional to their luminosities, totaling $10^6 $ per source. They propagate through the computational domain, scattering through the dusty medium until they either escape the domain or are completely absorbed.  We create images at several wavelengths, corresponding to the NIRCam wide-band filters, while \powderday produces SEDs, covering wavelengths from the far UV to the radio.  We compute SEDs and images for 100 different orientations, regularly spaced in azimuthal and polar angles.

\subsection{Adding the point spread functions and noise to the mock observations}
\label{noise}

When creating our mock observations, we consider all wide-band NIRCam filters: F070W, F090W, F115W, F150W, F200W, F277W, F356W, and F444W. For each filter we generate a single image from a throughput-weighted average of monochromatic images from \powderday. We downsize these filters to only 10 wavelengths to decrease computation time and smooth each filter with \texttt{scipy.ndimage.gaussian\_filter} \citep{2020SciPy-NMeth}. We then convolve each image per filter with its respective point spread function (PSF) of the same filter \citep{STPSF_2012, STPSF_2014} using the \astropy \citep{astropy_1, astropy_2, astropy_3} routine \texttt{convolve\_fft}. Next, we coarsen this combined image to match the angular resolution of the short and long wavelength NIRCam filters, using 0.31 arcseconds per pixel for the short wavelength filters (F070W, F090W, F115W, F150W, and F200W) and 0.63 arcseconds per pixel for the long wavelength filters (F277W, F356W, and F444W). We have chosen to use the wide filters exclusively since they are more transmissive, allowing detection of dimmer galaxies. \par

We calculate noise in each filter using the sensitivities calculated from the JWST Exposure Time Calculator \citep[ETC;][]{Pickering_2016}. We compute the sensitivities assuming a signal-to-noise (SNR) ratio of $10$ and an exposure time of $10^4$ seconds. We assume a randomly distributed Gaussian noise \citep[e.g.][]{mock_galaxy_hst_jwst_illustris_tng} with $\sigma = \textup{signal}/\textup{SNR}$ and scale the exposure time up to $10^7$ seconds to mimic ultra-deep surveys. \par 

For the SEDs, we use the limiting point source sensitivities for the NIRSpec instrument with a low resolution of $\textup{R} = 100$, a $\textup{SNR} = 10$, and an exposure time of $10^4$ seconds. The $\sigma$ for each wavelength is found in the same way as they are found for the image, and they are interpolated to cover all wavelengths in that range. Once again, a randomly distributed Gaussian noise is used, and the exposure time is scaled to $10^7$ seconds. \par



\subsection{Calculating the catalog contents}
\label{calculations}

A major goal of this work is to create a publicly available database of physical properties and mock observations of the \renaissance Simulations. These properties and observations are stored in HDF5 files, covering all outputs; although, in this paper we only analyze data from the final outputs in each region. Next, we overview the catalog contents and how we calculate each quantity. \par

\subsubsection{Catalog contents calculated directly from the simulations}
\label{calculations_sim}

Directly computed from the \renaissance Simulations, we have included the following physical properties in our catalog: 

\begin{itemize}
    \item Masses: stellar, dark matter, cold gas (T $< 1000$ K), Pop III stellar, \ion{H}{1}, \ion{H}{2}, and total;
    \item Particle counts: Pop III stars, Pop II stars, and black hole particles;
    \item IDs: halo, halo descendants, subhalo, subhalo descendants, star, and black holes;
    \item Individual star particle information: age, position, velocity, angular momentum, mass, Pop II or III, and metallicity fraction;
    \item Halo: position, angular momentum, gas mass fraction, virial radius, last major merger redshift, specific stellar angular momentum, SFR over $10$ Myr and $100$ Myr, sSFR, half stellar mass radius, and star formation efficiency = total stellar mass/total mass.
\end{itemize}

Some of these values were calculated for the \renaissance simulations for prior work \citep{Xu_2016} with Rockstar \citep{behroozi_2012_rockstar} and are held in a \ytree \citep{ytree} merger tree file. We calculate the remaining quantities with a a \ytree analysis pipeline, specifically the total stellar mass, total gas mass, total Pop III stellar mass, total cold gas mass, total \ion{H}{1} mass, total \ion{H}{2} mass, total stellar mass over the last 10 Myr and 100 Myr, and the total stellar angular momentum. From these values, we calculate the total mass, the sSFRs, SFRs, star formation efficiency, half stellar mass radius, and specific stellar angular momentum. \par

\subsubsection{Catalog contents calculated from the mock observations}
\label{calculations_mock}

From the mock observations, we calculate the following physical properties:

\begin{itemize}
    \item Absolute UV magnitude 
    \item Apparent magnitude
    \item Half light radius
    \item Surface brightness
    \item NIRCam wideband colors
    \item UV continuum slope
    \item Sersic index
    \item SFR given a Salpeter IMF
\end{itemize}

We find the absolute UV magnitude from the flux values of the SED over all inclination angles for UV wavelengths bound by a top hat filter centered at $1500$ Å and a width of $100$ Å. We give the absolute UV magnitude in the AB system using the equation

\begin{equation}
    M_{\textup{UV}} = - 2.5 \log_{10} \left (\frac{\int f_\textup{v}(v)\frac{dv}{v}}{f_{\textup{AB}}} \right)
\end{equation}

\noindent where $f_v$ is the flux density per unit frequency given by the mock SED from \powderday and redshifted to the redshift of the galaxy, and $f_{AB} = 3630.78 \space \textup{Jy} $ is the zero-point for the AB magnitude system.

We calculate the apparent magnitude similarly using the AB magnitude system and the SED but in the filters F070W, F090W, F115W, F150W, F200W, F277W, F356W, and F444W, with 

\begin{equation}
    m_{\textup{AB}} = - 2.5 \log_{10} \left (\frac{\int f_\textup{v}(v)T(v)\frac{dv}{v}}{f_{\textup{AB}} \int T(v)\frac{dv}{v}} \right)
\end{equation}

\noindent where $T(v)$ is the throughput in each of the specified NIRCam filters and $f_v$ and $f_{AB}$ are the same as previously specified.

We calculate the half light radius by finding the radial profile of each galaxy from the synthetic images created by \powderday in each previously specified filter. The synthetic images used for the calculation of the half light radii, Sersic index, surface brightness and included in the catalog are created at a line of sight coordinate of $(\phi, \theta) = (0,0)$. We recenter our image, so the brightest point, which we consider the center of the galaxy, will be the center of the image. We calculate the new size after the recentering based on minimum length from the new center to the edge of the original image. That length in pixels becomes our maximum radius, which is the total number of radial bins used in each calculation. We take the cumulative sum of the radial profile then interpolate to find the radius at half of the total flux. We find the surface brightness similarly by taking the radial profile and dividing the flux at each radius by the area of the annuli.

We find the UV continuum slope $\beta$ \citep[e.g.][]{UV_continuum_slope} in the linear fit

\begin{equation}
    f_{\lambda} = \beta \log \lambda + q
\end{equation}

\noindent where $f_{\lambda}$ is the flux density per wavelength from the \powderday SEDs. We find the UV continuum slope within the 10 windows defined by \cite{UV_continuum_slope_windows} to avoid disruptions to the slope by the most prominent emission lines, fitting the equation with \texttt{scipy.optimize.curve\_fit}. 

We find the Sersic index in a similar fashion, by fitting a Sersic profile to the flux radial profile

\begin{equation}
    \ln [I(R)] = \ln (I_e) \times  \left\{ -b_\textup{n} * \left[ \frac{R}{R_\textup{e}}^{1/n} - 1 \right] \right\} 
\end{equation}

\noindent where $n$ is the Sersic index, $I_\textup{e}$ is the flux at the half light radius, $R_\textup{e}$ is the half light radius, and 

\begin{eqnarray}
    b_n = 2n - \frac{1}{3}  + \frac{4}{405n} + \frac{46}{25515n^2} + \frac{131}{1148175n^3} 
    \nonumber \\ + \frac{2194697}{30690717750n^4}    
\end{eqnarray}

\noindent for $n > 0.36$. 

To validate with the SFR in the past 100 Myr calculated from data in the \ytree analysis pipeline, we find the SFR $\Psi$ using the common empirical conversion

\begin{equation}
     \Psi \; (\textup{M}_{\odot} \; \textup{yr}^{-1}) = (1.4 \times 10^{-28}) \; L_{\nu} \; (\textup{erg} \; \textup{s}^{-1} \; \textup{Hz}^{-1})
\end{equation}

\noindent for a Salpeter IMF to convert between the UV luminosity of the \powderday images and the SFR rates of these galaxies \citep{kennicutt}. This relation applies in the UV range of $1500 - 2800$ Å, where the UV spectrum is relatively flat, allowing for this conversion between the luminosity and SFR; therefore, we take the average of this empirical SFR estimate in this range for each halo.

We calculate specific star formation rates (sSFRs) from the SFRs taking the stellar mass $M_*$ from the \ytree analysis pipeline using the relation $\Psi_{\textup{S}} = \Psi/M_{*}$.  \par

We find all of these values with the images and SEDs directly from \powderday without noise added and from the images and SEDs with noise added as described in Section \ref{noise}. In this paper, we only present results from data without the PSF and noise added.


%

\begin{longrotatetable}
\begin{deluxetable*}{cccccccc}
\digitalasset
\tablewidth{0pt}
\tablecaption{JWST galaxy properties with z $>$ 10 \label{tab:galaxy_z}}
\tablehead{
\colhead{Galaxy} & \colhead{z} & \colhead{log($\textup{M}_*$) [$M_{\odot}$]} & \colhead{SFR [$M_{\odot}/yr$]} & \colhead{$\textup{M}_{\textup{UV}}$} & \colhead{Sersic index} & \colhead{Half light radius [kpc]} &\colhead{Ref.}\\
}
\startdata
GS-z10-0 & $10.38^{+0.07}_{-0.06}$ & $7.76^{+0.12}_{-0.11}$ & $1.17^{+0.17}_{-0.15}$ & $-18.51^{+0.07}_{-0.07}$ & $< 2$ & $0.05 - 0.165$ & 1, 2, 3\\
GS-z11-0 & $11.58^{+0.05}_{-0.05}$ & $8.81^{+0.04}_{-0.08}$ & $2.03^{+0.25}_{-0.21}$ & $-19.33^{+0.04}_{-0.04}$ & $< 2$ & $0.05 - 0.165$ & 1, 2, 3\\
GS-z12-0 & $12.63^{+0.24}_{-0.08}$ & $8.10^{+0.27}_{-0.10}$ & $1.27^{+0.48}_{-0.37}$ & $-18.64^{+0.08}_{-0.08}$ & $< 2$ & $0.05 - 0.165$ & 1, 2, 3 \\
GS-z13-0 & $13.20^{+0.04}_{-0.07}$ & $7.89^{+0.19}_{-0.07}$ & $1.08^{+0.27}_{-0.20}$ & $-18.71^{+0.06}_{-0.06}$ & $< 2$ & $0.05 - 0.165$ & 1, 2, 3\\
GN-z11 & $10.60$ & $8.73^{+0.06}_{-0.06}$ & $18.78^{+0.81}_{-0.69}$ & $-21.70$ & $2.42^{+1.12}_{-1.12}$ & $0.15^{+0.025}_{-0.025}$ & 2, 4, 5, 6, 7\\
UNCOVER-z12 & $12.39^{+0.00}_{-0.00}$ & $8.63^{+0.20}_{-0.22}$ & $1.40^{+2.81}_{-1.16}$ & $-19.2^{+0.50}_{-0.50}$ & $2.77^{+1.04}_{-0.81}$ & $0.64^{+0.06}_{-0.06}$ & 8\\
UNCOVER-z13 & $13.08^{+0.01}_{-0.00}$ & $7.92^{+0.23}_{-0.16}$ & $1.95^{+0.41}_{-0.39}$  & $-19.40^{+1.8}_{-1.8}$ & $0.78^{+0.09}_{-0.19}$ & $0.56^{+0.19}_{-0.13}$ & 8 \\
Maisie's galaxy & $11.42^{+0.01}_{-0.01}$ & $8.15^{+0.33}_{-0.12}$ & $1.46^{+0.87}_{-0.10}$ & $-20.08^{+0.05}_{-0.04}$ & --- & $0.34^{+0.01}_{-0.01}$ & 3, 9, 10, 11, 12\\
CEERS2-588 & $11.04^{+0.00}_{-0.00}$ & $8.70^{+0.10}_{-0.10}$ & $12.70^{+9.70}_{-4.90}$  & $-20.01^{+0.07}_{-0.10}$ & --- & $0.48$ & 3, 12, 13, 14, 15\\
GLASS-z12 & $12.117^{+0.00}_{-0.00}$ & $8.23^{+0.32}_{-0.08}$ & $19.00^{+14.00}_{-10.00}$  & $-21.00^{+0.09}_{-0.09}$ & $1.00$ & $0.50$ & 16, 17 \\
GS-z14-0 & $14.32^{+0.08}_{-0.20}$ & $8.66^{+0.47}_{-0.15}$ & $22.26^{+4.24}_{-3.84}$  & $-21.08^{+0.02}_{-0.02}$ & $0.877^{+0.027}_{-0.027}$ & $0.26^{+0.002}_{-0.002}$ & 18, 19\\
GS-z14-1 & $13.90^{+0.17}_{-0.17}$ & $8.00^{+0.40}_{-0.30}$ & $1.20^{+0.70}_{-0.90}$  & $-19.09^{+0.16}_{-0.13}$ & $\sim 1$ & $< 0.05$ or $< 0.165$ & 18, 20, 21\\
MoM-z14 & $14.44^{+0.02}_{-0.02}$ & $8.10^{+0.30}_{-0.20}$ & $13.00^{+3.70}_{-3.50}$ & $-20.23^{+0.06}_{-0.06}$ & $1.00^{+0.02}_{-0.02}$ & $0.07^{+0.015}_{-0.012}$ & 22\\
PAN-z14-1 & $13.53^{+0.05}_{-0.06}$ & $8.23^{+1.14}_{-0.21}$ & $4.80^{+13.6}_{-4.8}$ & $-20.00^{+0.20}_{-0.20}$ & $\sim 1$ & $0.23^{+0.01}_{-0.01}$ & 23\\
\enddata
\tablerefs{[1] \cite{robertson_2022}, [2] \cite{curtislake2023spectroscopicconfirmationmetalpoorgalaxies}, [3] \cite{harikane_spectroscopic_constraints}, [4] \cite{Bunker_2023}, [5] \cite{tacchella_JADES_imaging}, [6] \cite{hainline_cosmos_infancy}, [7] \cite{Baldwin_GNz11_size_estimate}, [8] \cite{Wang_UNCOVER}, [9] \cite{finkelstein_maisies}, [10] \cite{harikane_maisies}, [11] \cite{Ono_maisies_morphology}, [12] \cite{arrabel_maisies_galaxy}, [13] \cite{donnan_corey}, [14] \cite{CEERS_overview}, [15] \cite{bouwens_UV_luminosity_CEERS}, [16] \cite{naidu_glass}, [17] \cite{bakx_glass}, [18] \cite{Carniani_2024}, [19] \cite{helton_photodetection_GSz14_1}, [20] \cite{Ono_morphology_GSz14}, [21] \cite{Wu_GSz14_1}, [22] \cite{naidu2025cosmicmiracleremarkablyluminous}, [23] \cite{donnan2026spectroscopicconfirmationlargeluminous}}
\end{deluxetable*}
\end{longrotatetable}
\label{table}

\subsection{JWST $z > 10$ galaxy properties}

In this paper, we compare the \renaissance galaxies to spectroscopically confirmed JWST galaxies with $z > 10$. We compile the physical properties of these JWST galaxies from several papers into Table \ref{table}. In Table \ref{tab:galaxy_z}, we list the weighted averages of these values that we plot on our figures and include the references of the original values. We note that UNCOVER-z12 and UNCOVER-z13 do not have available $\textup{M}_\textup{UV}$ values, so we choose the absolute UV magnitudes in the F200W filter since it is the closest we can find to $\textup{M}_\textup{UV}$. We also note that the Sersic index for GN-z11 given by \cite{Baldwin_GNz11_size_estimate} is an average value they calculated. \par

We do not include the apparent magnitudes that we calculated in each filter, but we calculate them with available fluxes using 

\begin{equation}
    M_{\textup{UV}} = - 2.5 \log_{10} \left (\frac{f_\textup{v}(v)}{f_{\textup{AB}}} \right)
\end{equation}

\noindent where $f_v$ is the flux given in each wideband filter derived from observations, and $f_{AB} = 3630.78 \space \textup{Jy} $ is the zero-point for the AB magnitude system. \par

\section{Global properties of the Renaissance galaxies} 
\label{results_global}

\subsection{Stellar masses}

\begin{figure}[t!]
\includegraphics[width = \columnwidth]{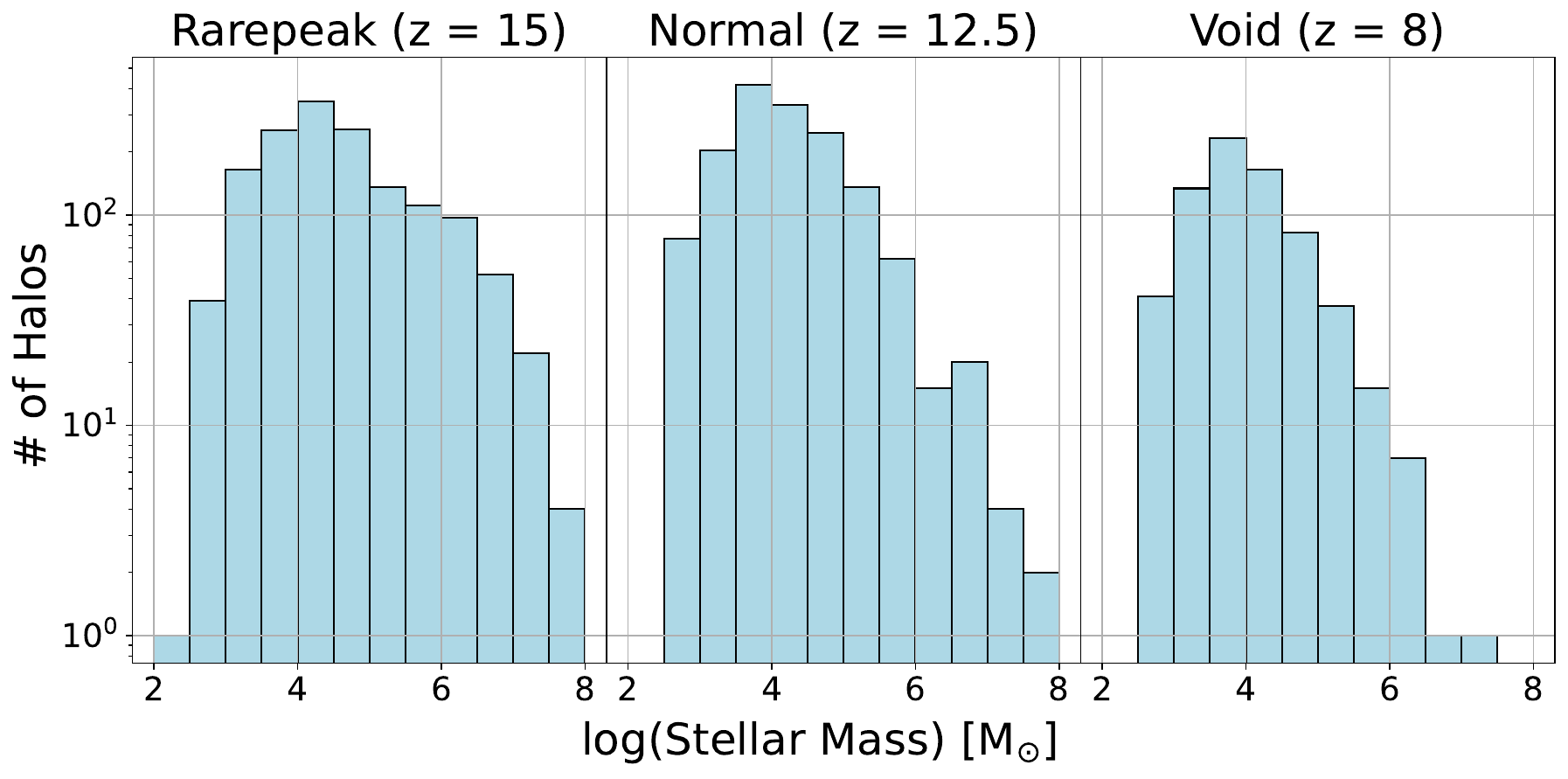}
\caption{The number of galaxies binned by stellar masses in the Rarepeak, Normal, and Void regions of \renaissance from left to right, respectively.
\label{fig:stellar_mass_hist}}
\end{figure}

Figure \ref{fig:stellar_mass_hist} shows the number of halos as a function of stellar mass for the \renaissance simulations for the Rarepeak, Normal, and Void region from left to right, respectively, at their final redshifts. We use this plot to contextualize the \renaissance Simulations' stellar mass coverage, which focuses on the lower-mass regime of galaxies as seen by each region in Figure \ref{fig:stellar_mass_hist}'s distribution and peak at around $10^4 \ \textup{M}_{\odot}$. We also use this figure to point out that the Void region has significantly less galaxies than the Normal or Rarepeak regions. Our simulated galaxies have clearly much lower stellar masses than most of the JWST galaxies we reference throughout this paper, which lie in the range of $10^{7.76} - 10^{8.81} \ \textup{M}_\odot$. As Figure \ref{fig:stellar_mass_hist} shows, there is very little overlap between the \renaissance galaxies' masses and the JWST galaxies' masses. Therefore, the \renaissance galaxies can be viewed as ancestors to the JWST galaxies, allowing us to infer the evolution of these galaxies throughout this paper.

\subsection{Star formation versus stellar mass}

\begin{figure*}[t!]
\includegraphics[width = \textwidth]{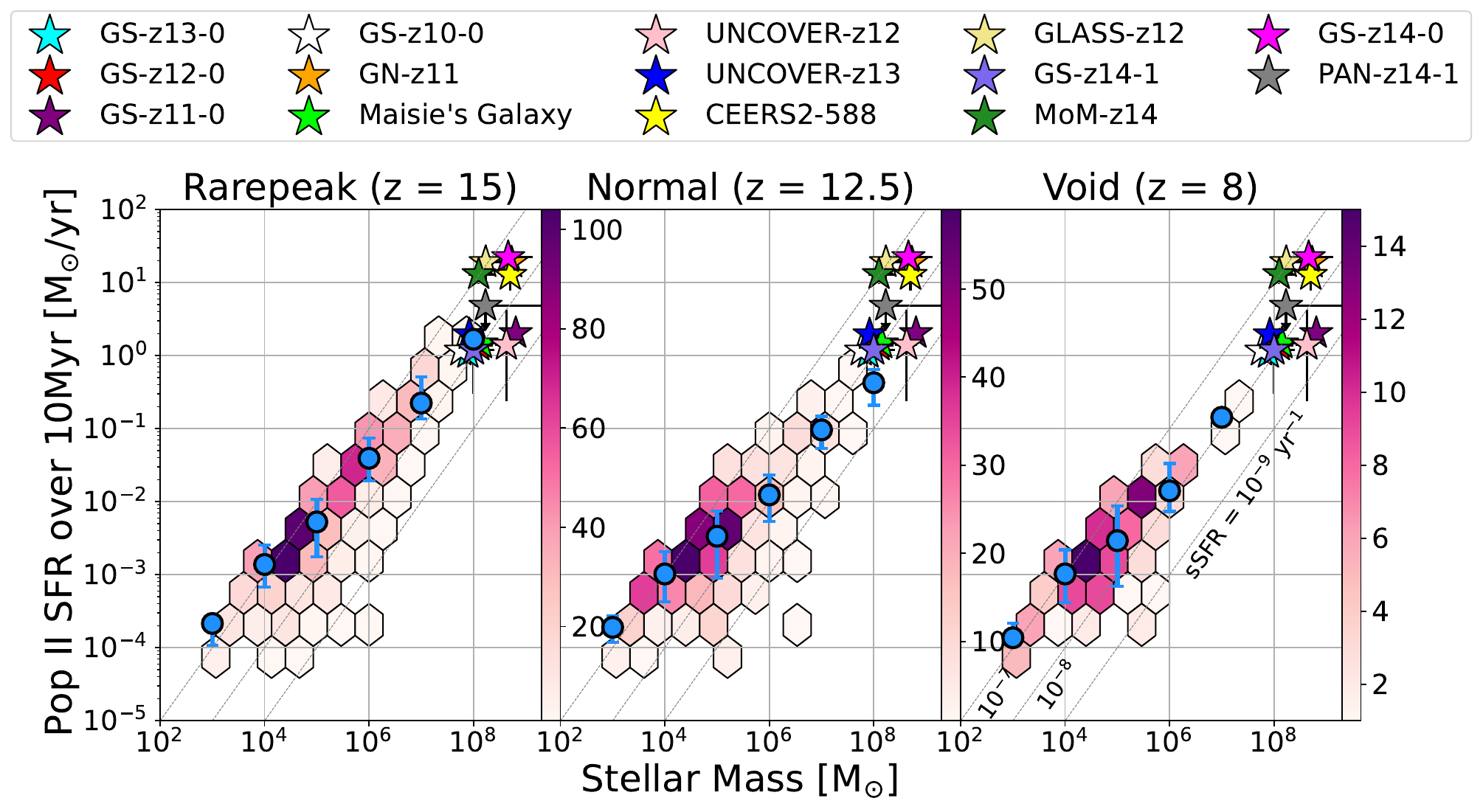}
\caption{The Pop II SFRs over $10 \ \textup{Myr}$ of the galaxies in the Rarepeak, Normal, and Void regions of \renaissance, from left to right respectively, as a function of their stellar masses. The red hexbins depict the number of galaxies. The stars represent the $z > 10$ spectroscopically confirmed galaxies observed by JWST. The dashed lines show sSFRs that overlap most with the data. The blue points represent the median of the SFR with error bars showing a range of $1 \sigma$.
\label{fig:stellar_mass_SFR}}
\end{figure*}

Figure \ref{fig:stellar_mass_SFR} shows the Pop II SFR in \renaissance in the last 10 Myr as a function of the stellar mass. The amount of galaxies in each region are represented as hexbins with the colorbar showing the number of galaxies; they are shown in the Rarepeak, Normal, and Void regions from left to right, respectively. Overplotted in each panel as dashed lines are sSFRs of $10^{-7}$, $10^{-8}$, and $10^{-9} \ \textup{yr}^{-1}$. Spectroscopically confirmed $z > 10$ JWST galaxies are shown in each region represented as stars. The blue points are the median of the SFRs with $1 \sigma$ error bars.  \par

The Rarepeak and Void region do not overlap any of the JWST galaxies in redshift as seen in Table \ref{table}, with Rarepeak being earlier and Void being later than the JWST galaxies selected. The Rarepeak and Normal regions have a higher sampling with more massive galaxies overlapping directly with JWST galaxies; although, all regions have more lower mass galaxies from the hierarchical nature of structure formation. This is particularly of note in the Rarepeak and Normal region with redshifts most similar to those of the $z > 10$ JWST galaxies, furthering the point that JWST is detecting galaxies at higher masses than we are modeling. The underdensity in the Void region leads to fewer massive galaxies. Studies have found that the scatter in absolute UV magnitude at high redshift decreases with increasing stellar mass and decreasing redshift \citep{stellar_pop_dwarf_scatter, legrand_scatter, fulanetto_bursty_scatter, hopkins_scatter_bursty, morphology_cosmic_dawn_scatter, kravtsov2024stochasticstarformationabundance, gelli_mass_stochasticity_scatter, stark_scatter}. We plot the SFR median with $1 \sigma$ error bars since $\textup{M}_{\textup{UV}}$ is a proxy to SFR of a galaxy, where a bursty SFR causes the scatter in the $\textup{M}_{\textup{UV}}$ \citep{sparre_scatter, fulanetto_bursty_scatter, sun_seen_unseen_scatter}. We find that the SFR standard deviation in each stellar masses bin are similar in all three regions at their different redshifts. The scatter is small at the lowest masses scales ($10^3$ M$_\odot$) explored by this work, resulting from a single star formation event in the last 10 Myr. The SFR scatter increases to a maximum of an order of magnitude at $M_\star = 10^5$ M$_\odot$ caused by variations in how quickly the halo can regain sufficient gas after its initial star formation that expels the majority of the gas through strong outflows \mbox{\citep{hazlett_renaissance}}.  At higher stellar masses, the SFR scatter in all three regions decrease with increasing $M_\star$, consistent with the expected trend found in previous work.


As mentioned in Section \ref{intro}, there has been some discussion that the stellar masses of the galaxies observed by JWST are too high, too early. However, our results, while at a lower masses than the JWST galaxies, smoothly transition at constant sSFRs in all regions to the JWST galaxies as also seen in Figure 4 in \cite{mccaffrey2023tension}. The majority of the \renaissance galaxies in all regions have a high specific star formation rate (sSFR) with most lying in between $\Psi = 10^{-7}$ and $10^{-8} \textup{yr}^{-1}$, while the JWST galaxies show slightly more variations in their sSFRs, suggesting that these lower mass galaxies are forming stars more intensely relative to their mass than the higher mass counterparts. This feature is particularly prominent in the Rarepeak region, where all of the galaxies have a higher redshift than the JWST galaxies, showing that as galaxies age and grow, their sSFRs might decrease, indicating that the initial star formation events have a larger relative impact on these nascent galaxies. This evolution aligns with our understanding of burstier star formation in the early Universe with galaxies becoming quiescent as they age. In general, our galaxies represent the earliest stages of galaxy formation that will eventually reach the high mass regime seen by JWST. \par

\section{Morphologies of the Renaissance galaxies}
\label{morph}

\subsection{Galaxy sizes}
\label{results_mass_radii}

\begin{figure}[t!]
\includegraphics[width = \columnwidth]{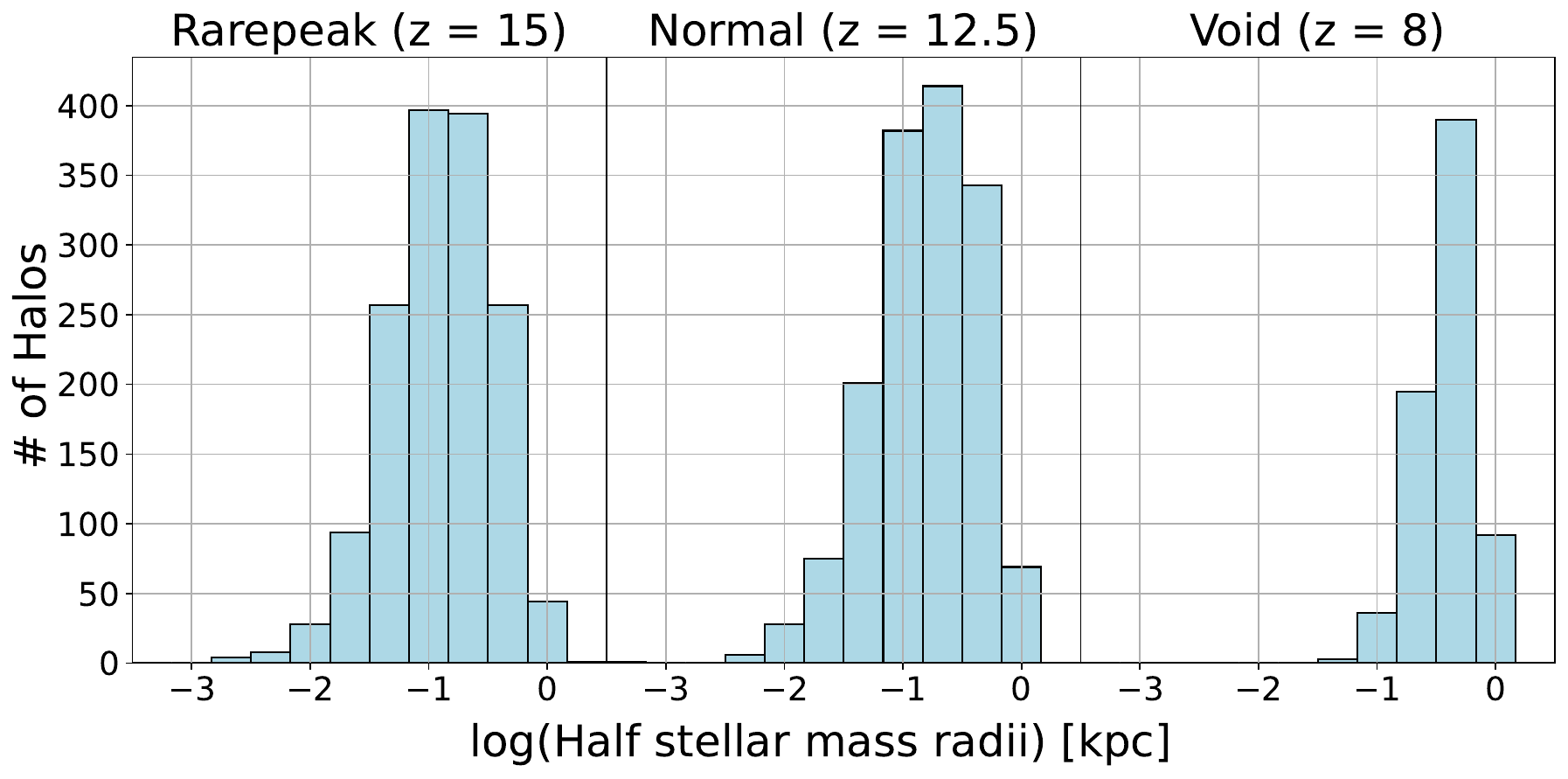}
\caption{The number of galaxies in the Rarepeak, Normal, and Void regions of \renaissance, from left to right respectively, as a function of half stellar mass radius. Nearly all galaxies have sizes below 1 kpc with a median around 100 pc.
\label{fig:bin_half_stellar_mass}}
\end{figure}

Figure \ref{fig:bin_half_stellar_mass} shows an overview of the half stellar mass radii of the galaxies in the \renaissance Simulations. While the half stellar mass radius is not a value that has been published yet for the spectroscopically confirmed $z > 10$ JWST galaxies, the half stellar mass radii give us an idea of the morphology and compactness of the ancestors of these JWST galaxies with nearly all of the \renaissance galaxies smaller than $1$ kpc.


\begin{figure}[t!]
\includegraphics[width = \columnwidth]{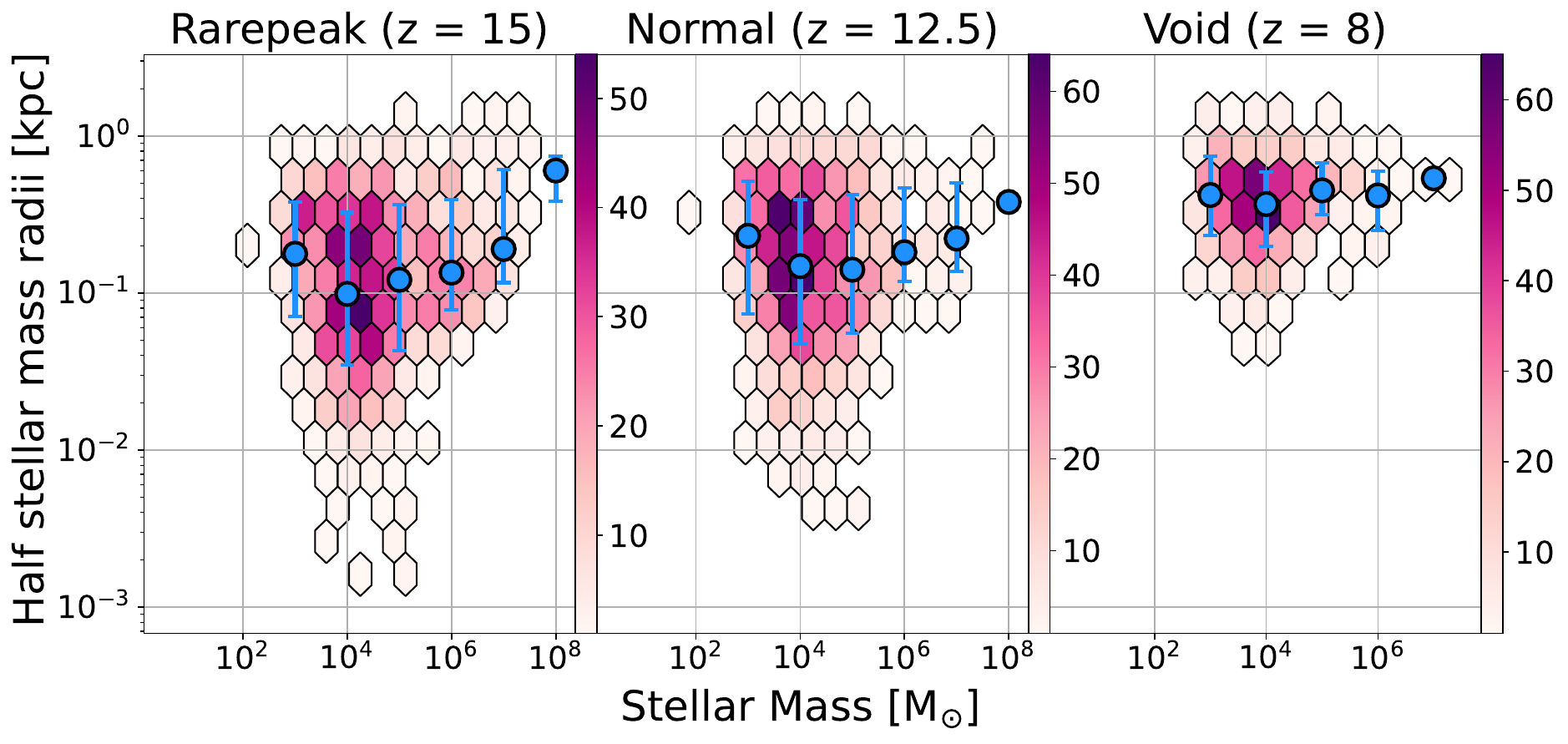}
\caption{The number of galaxies in the Rarepeak, Normal, and Void regions of \renaissance, from left to right respectively, as a function of stellar mass and half stellar mass radius. The blue points represent the medians of the half stellar mass radii with $1\sigma$ error bars.
\label{fig:stellar_mass_half_mass_radii}}
\end{figure}

Figure \ref{fig:stellar_mass_half_mass_radii} shows the half stellar mass radii of the \renaissance galaxies in the Rarepeak, Normal, and Void regions, shown from left to right, respectively, as a function of their stellar mass with the hexbin intensities representing the number of galaxies. 
We find much more scatter for lower mass galaxies than higher mass galaxies, as seen by the decreasing $1\sigma$ as the stellar mass increases, and the majority of these galaxies have half stellar mass radii between $0.1$ and $1$ kpc regardless of stellar mass. Below $10^6 \ \textup{M}_{\odot}$, we find a population of compact galaxies with radii below $30$ pc; however, as the stellar mass increases, their frequency decreases. Despite this decrease in extremely compact galaxies, there is no significant increase in half stellar mass radii beyond $1$ kpc as the stellar mass increases above $10^4 \ \textup{M}_{\odot}$, with the half stellar mass radii similar to the majority of the galaxies at low stellar masses. This lack of a trend indicates some level of compactness is maintained in the higher mass galaxies in \renaissance.
The majority of the $z>10$ JWST galaxies have approximately the same compactness less than 1~kpc, measured by their light, indicating that the sizes of galaxies do not grow substantially as they undergo their initial burst of growth at high redshifts. The majority of the galaxies in the Void region are at larger half stellar mass than the galaxies in Rarepeak and Normal, likely due to the fact it is at $z = 8$, so the galaxies have had more time to equilibrate. In this paper, we only discuss the stellar contribution to the luminosity, as \renaissance{} does not include any AGN formation or feedback.  It is still unclear the level of AGN feedback on the galaxy morphology and size in these small galaxies, such as Little Red Dots \citep[LRDs; e.g.][]{naidu2025_BHstar} and the frequency of this population that host a central massive black hole.  Nevertheless, our results point toward the collapse of protogalactic gas is intense, creating high sSFRs in compact, low-mass galaxies that do not grow much as they initially assemble.


\subsection{Distribution of the half light radii}

\begin{figure}[t!]
\includegraphics[width = \columnwidth]{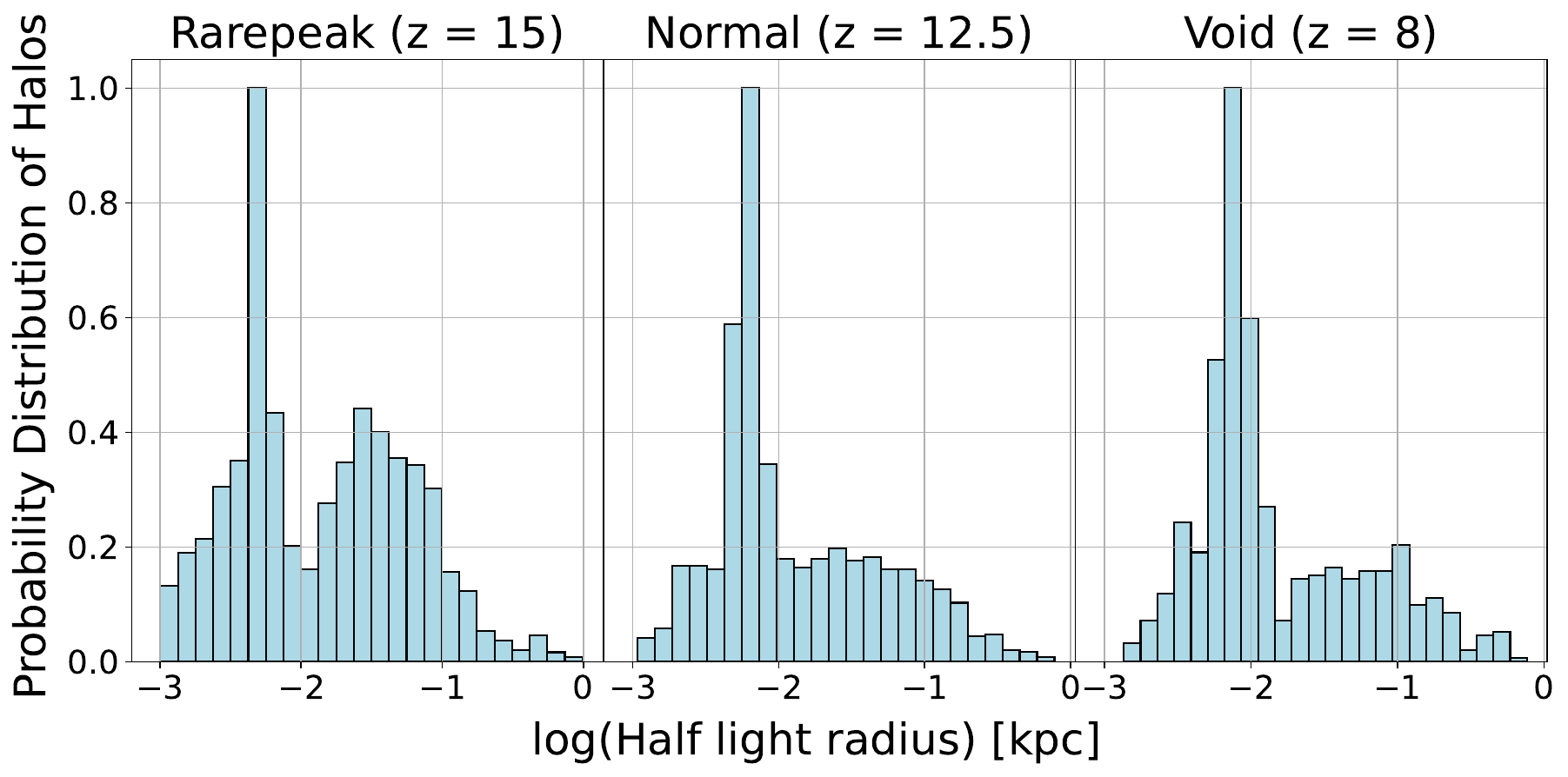}
\caption{The probability distribution of the half light radii of the galaxies in the Rarepeak, Normal, and Void regions of \renaissance, from left to right respectively for the F200W filter, chosen because the F200W is redward of the Ly$\alpha$ line. We exclude galaxies with half light radii that fall below the resolution limit of \renaissance. 
\label{fig:half_light_prob}}
\end{figure}

Figure \ref{fig:half_light_prob} shows the probability distribution of the half light radii of the galaxies in the \renaissance Simulations in the Rarepeak, Normal, and Void regions shown from left to right, respectively. In each region, the half light radii probability distribution peaks around $10$ pc, indicating most of the rest-frame UV radiation is highly centrally concentrated, as Figure \ref{fig:bin_half_stellar_mass} and Figure \ref{fig:sersic_index_prob} show as well. There is another, smaller peak around $30$ pc in the Rarepeak region arising from the more massive galaxies. We present this plot to further highlight that the \renaissance galaxies are low-mass galaxies that are ancestors of the JWST galaxies that are generally larger. We note that we exclude any galaxies with half light radii below the resolution limit of \renaissance of $1$ pc. \par 

\subsection{Half light radii versus stellar mass}

\begin{figure}[t!]
\includegraphics[width = \columnwidth]{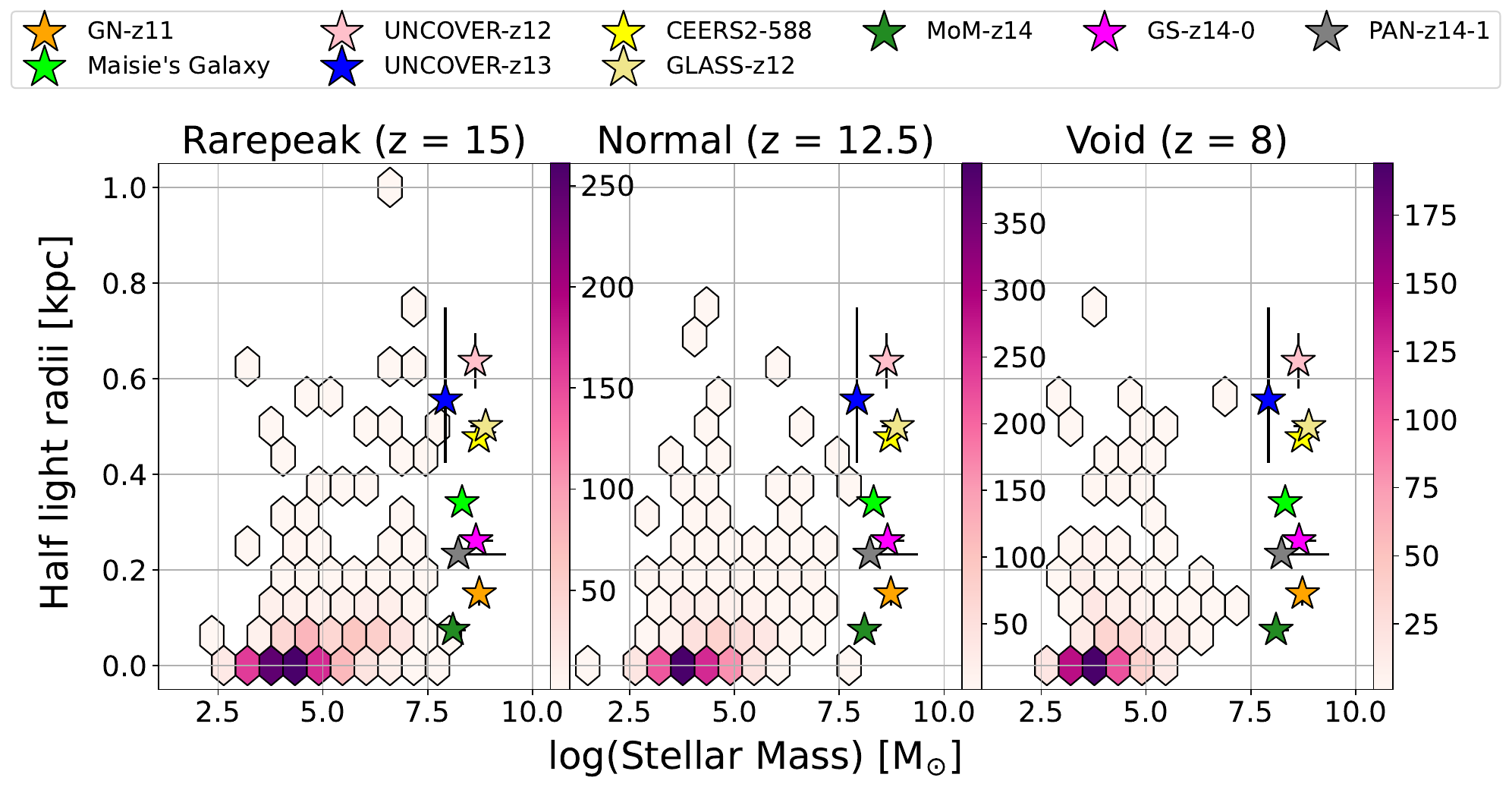}
\caption{The half light radii of the \renaissance Simulations as a function of stellar mass in the Rarepeak, Normal, and Void regions of \renaissance, from left to right respectively. The hexbins depeict the number of galaxies. The stars represent the $z > 10$ spectroscopically confirmed galaxies observed by JWST.
\label{fig:stellar_mass_half_light}}
\end{figure}

We now expand the half light radii distribution with respect to stellar mass in Figure \ref{fig:stellar_mass_half_light} showing the Rarepeak, Normal, and Void regions from left to right. As before, the stars represent $z > 10$ JWST galaxies. As also shown in Figure \ref{fig:half_light_prob}, the majority of the \renaissance galaxies are smaller than the JWST galaxies, representing a population that assembled to form galaxies similar to the observed galaxies. Figure \ref{fig:stellar_mass_half_light} shows no overlap in stellar mass with the JWST galaxies in the Void or Normal region, but there is some overlap with UNCOVER-z13 and MoM-z14 in the Rarepeak region. UNCOVER-z13 falls above the half light radii of the majority of the \renaissance galaxies making MoM-z14 the closest match to the \renaissance galaxies' half light radii and stellar masses. This outlier indicates that MoM-z14 has much more compact star formation than its fellow JWST galaxies at a similar stellar mass, hinting at a particular formation sequence that resulted in a high SFR in a compact region. However, all of the JWST galaxies with available half light radii fall in the same range the \renaissance galaxies; although, many fall in the upper end of the range. \par 

Figure \ref{fig:stellar_mass_half_light} further demonstrates that the \renaissance galaxies have very compact sizes with extremely centralized star formation, most likely due to their youth. However, some of the lower mass \renaissance galaxies are more diffuse than these extreme cases, indicating a variety of formation sequences in the initial build-up of stellar mass in these galaxies, similar to the spread in the JWST galaxies' half light radii. This spread continues with increasing stellar mass in the Rarepeak and Normal regions, much like the JWST galaxies. \par 

The \renaissance galaxies maintain these small half light radii even with increasing stellar masses, thus showing that most of the \renaissance galaxies are likely forming stars at a rapid rate without feedback disrupting their gas supply. Likewise, some of the JWST galaxies with lower half light radii suggest that some of these early galaxies may experience a fast formation sequence and remain as compact as seen in the \renaissance Simulations in their nascent stages; this rapid formation and compactness may be maintained in some cases as the \renaissance galaxies evolve into ones resembling JWST observations.  \par

\subsection{Half light radii and half mass radii comparison}

\begin{figure}[t!]
\includegraphics[width = \columnwidth]{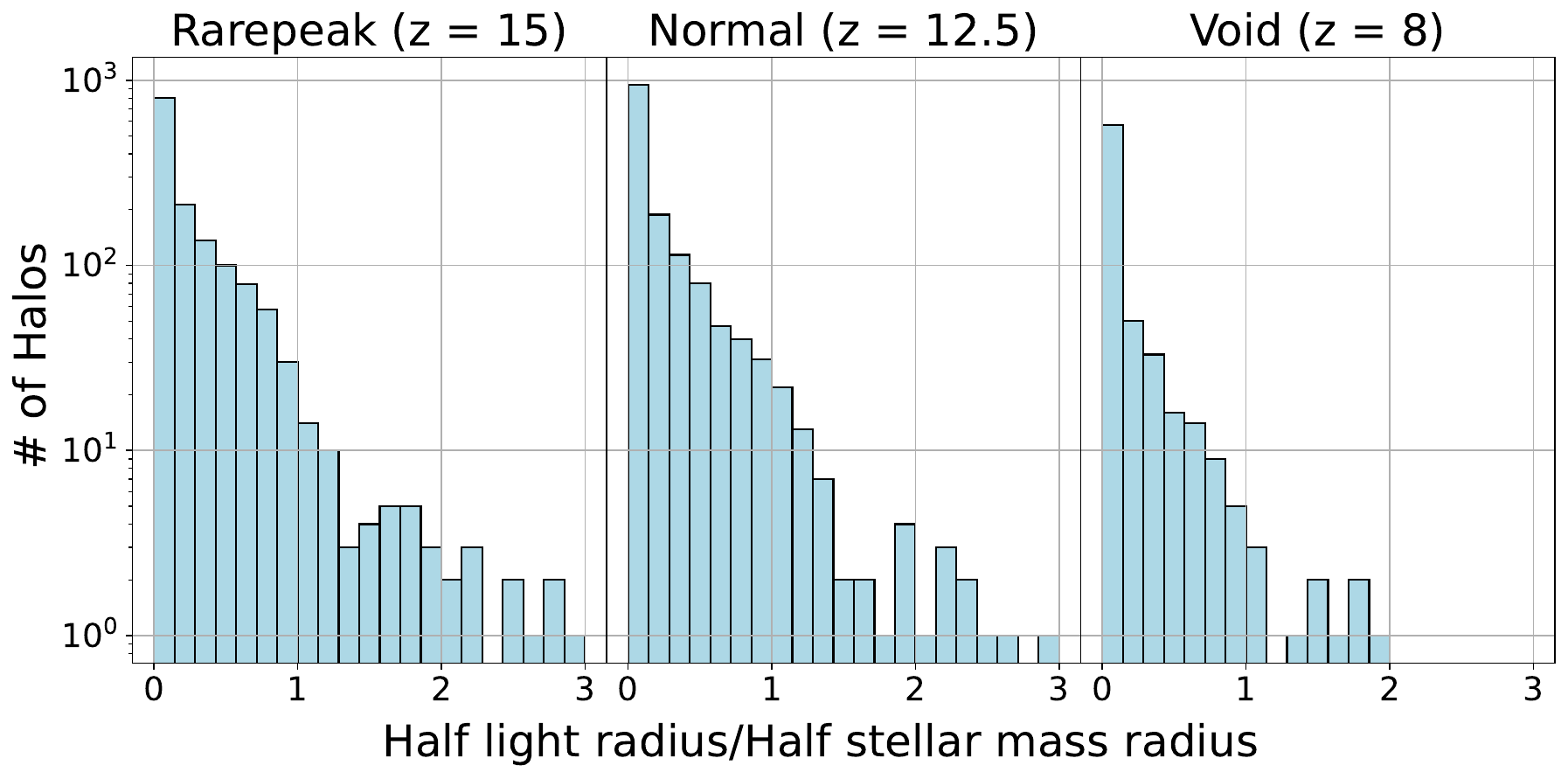}
\caption{The ratio of half light to half stellar mass radii represented as a histogram for the Rarepeak, Normal, and Void regions of \renaissance, from left to right. This distribution highlights that the emitted light is more concentrated than the stellar mass in most low-mass galaxies. 
\label{fig:bin_half_light_mass}}
\end{figure}

The fraction of half light radii to half stellar mass radii shown in Figure \ref{fig:bin_half_light_mass} gives insight into where the majority of the rest-frame UV radiation is emitted in the \renaissance galaxies. It informs us that the bright, young massive stars from recent star formation primarily reside in the center of the galaxy, as is common in young, high redshift galaxies. Unsurprisingly, since \renaissance contains very young galaxies, the majority of the galaxies in each region have a much smaller half light radii than their half stellar mass radii with the Rarepeak, Normal, and Void regions having $96\%$, $96\%$, and $99\%$ of their galaxies with ratios $< 1$; $81\%$, $85\%$, and $93\%$ of their galaxies with ratios $<0.5$; and $47\%$, $55\%$, and $76\%$ of their ratios $<0.1$. These low ratios could also be caused by lack of resolution in the mock observations used to calculate the half light radii, which is not present in the raw data used to calculate the half stellar mass radii.\par  

The frequencies in all three regions steadily and gradually decline with increasing ratios, with some of the galaxies having a larger half light radii than their stellar mass counterpart. The majority of the galaxies with a ratio $< 1$ have stellar masses around $10^4 \ \textup{M}_{\odot}$ with some of the higher masses still having this small ratio. These galaxies with large half light radii are much more diffuse and are relatively older galaxies without the intense central star formation and dense central gas contained in young galaxies; their centers contain older stars that have consumed much of the central gas reservoir in early star formation. Furthermore, galaxies with these large ratios may be undergoing mergers with values reaching up to $2-3$. The galaxies that have a half light radii/half stellar mass radii ratio of unity, have dispersed star formation throughout them without a dominant central star burst. Overall, the \renaissance Simulations galaxies clearly demonstrate that it is not uncommon for young galaxies to have intense central star formation with the galaxy light being more concentrated than the overall stellar radial distribution. \par

\subsection{Distribution of the Sersic indices}
\label{prob_dis_sersic}

\begin{figure}[t!]
\includegraphics[width = \columnwidth]{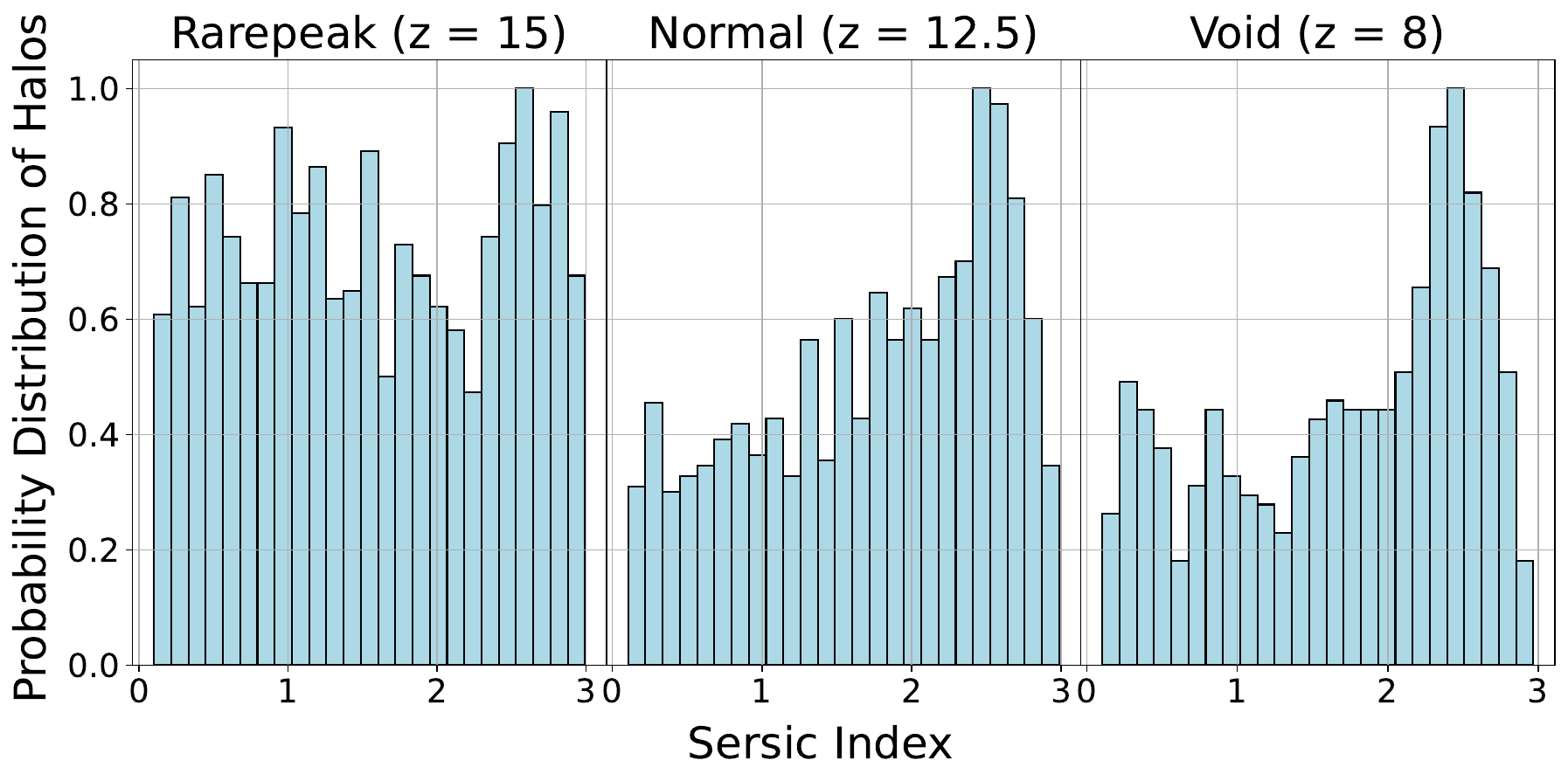}
\caption{The probability distribution of the Sersic index in the Rarepeak, Normal, and Void regions of \renaissance, from left to right respectively for the F200W filter. The data in this filter was chosen because its wavelength is redward of the Ly$\alpha$ line. We disregard Sersic indices higher than than 3 in this plot to focus on the bulk of the simulated galaxies.
\label{fig:sersic_index_prob}}
\end{figure}

Figure \ref{fig:sersic_index_prob} shows the probability distribution of the Sersic index of the galaxies in the \renaissance Simulations in the Rarepeak, Normal, and Void regions shown from left to right, respectively. Similar to the half light radii, we exclude the galaxies that fell below the resolution limit of \renaissance from this analysis, and we exclude galaxies with extremely high, outlier Sersic indices. In the Rarepeak region, the Sersic index probability distribution is uniformly distributed in the range $0 - 3$. For the Normal and Void regions, the Sersic index distribution peaks between $2$ and $3$, indicating the light profiles more likely to be closer to a power law. This large variation originates from the fact that these early galaxies are rigorously forming and are often very irregular in shape, so they do not necessarily resemble the $n = 1$ spiral range or $n = 4$ elliptical profile. The Rarepeak region has even more spread than the Normal and Void regions because it only evolves to $z \approx 15$, whereas the other regions evolve longer, giving them more time to equilibrate, while still having many irregular galaxies that have not recovered from an epoch of rapid assembly.  

\subsection{Sersic index versus stellar mass} 

\begin{figure*}[t!]
\includegraphics[width = \textwidth]{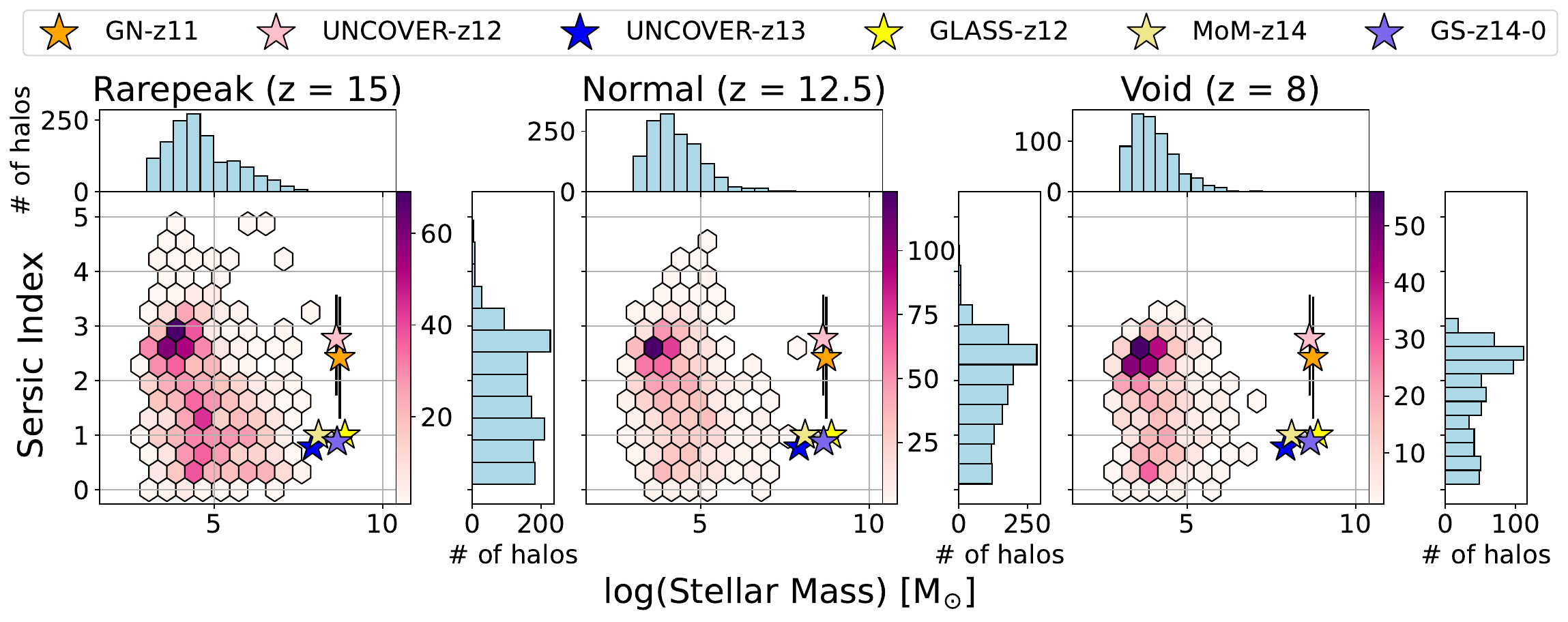}
\caption{The Sersic index as a function of stellar mass for the F200W filter represented as hexbins colored by the number of galaxies. There is a histogram of the number of halos with each Sersic index on the x-axis and a histogram with the number halos with each stellar mass on the y-axis for the Rarepeak, Normal, and Void regions of \renaissance, from left to right respectively. The stars represent spectroscopically confirmed JWST galaxies that have a $z > 10$ and an available Sersic index.
\label{fig:sersic_index_stellar_mass}}
\end{figure*}

Figure \ref{fig:sersic_index_stellar_mass} shows the Sersic index of the galaxies in \renaissance in the F200W filter as a function of their stellar masses for the Rarepeak, Normal, and Void regions from left to right, respectively, represented as hexbins with colorbars scaled to the number of halos. We compare against the $z > 10$ JWST galaxies with available Sersic indices, represented as stars of different colors. The histograms along the axes show the distribution with respect to stellar mass and Sersic index, on the $x$ and $y$-axis, respectively. The Sersic indices of GS-z10-0, GS-z11-0, GS-z12-0, and GS-z13-0 are not included in the figure, but they are all found to have $n < 2$ \citep{robertson_2022}, which is less than the majority of the values found in the \renaissance Simulations galaxies. PAN-z14-1 is also not included in the Figure because it has not been found to converge to a specific value in the filter F200W; although, it does have n $\simeq 1$ in  F277W, F356W and F444W filters. This is also less than the majority of the \renaissance galaxies but not beyond the \renaissance galaxies' range \citep{donnan2026spectroscopicconfirmationlargeluminous}. We again emphasize that, the stellar masses of the \renaissance galaxies are much lower than the $z > 10$ JWST galaxies we present here, and our results predict the properties of their progenitors. \par 

The Sersic indices in \renaissance have similar scatter in the range $n=0-3$ in all three regions with the Rarepeak and Normal regions having outliers at $n > 3$.  This is broadly consistent with the values found in the $z>10$ JWST galaxies that either have $n=1$ or $n=2-3$ with the latter set having larger uncertainties.  Our findings demonstrate that early galaxies can have a range of profile shapes across orders of magnitude in stellar mass, which is similar to the trends in half stellar mass and half light radii.


\section{Photometry of the Renaissance galaxies}
\label{photometry}

\subsection{Apparent magnitudes}

\begin{figure}[t!]
\includegraphics[width = \columnwidth]{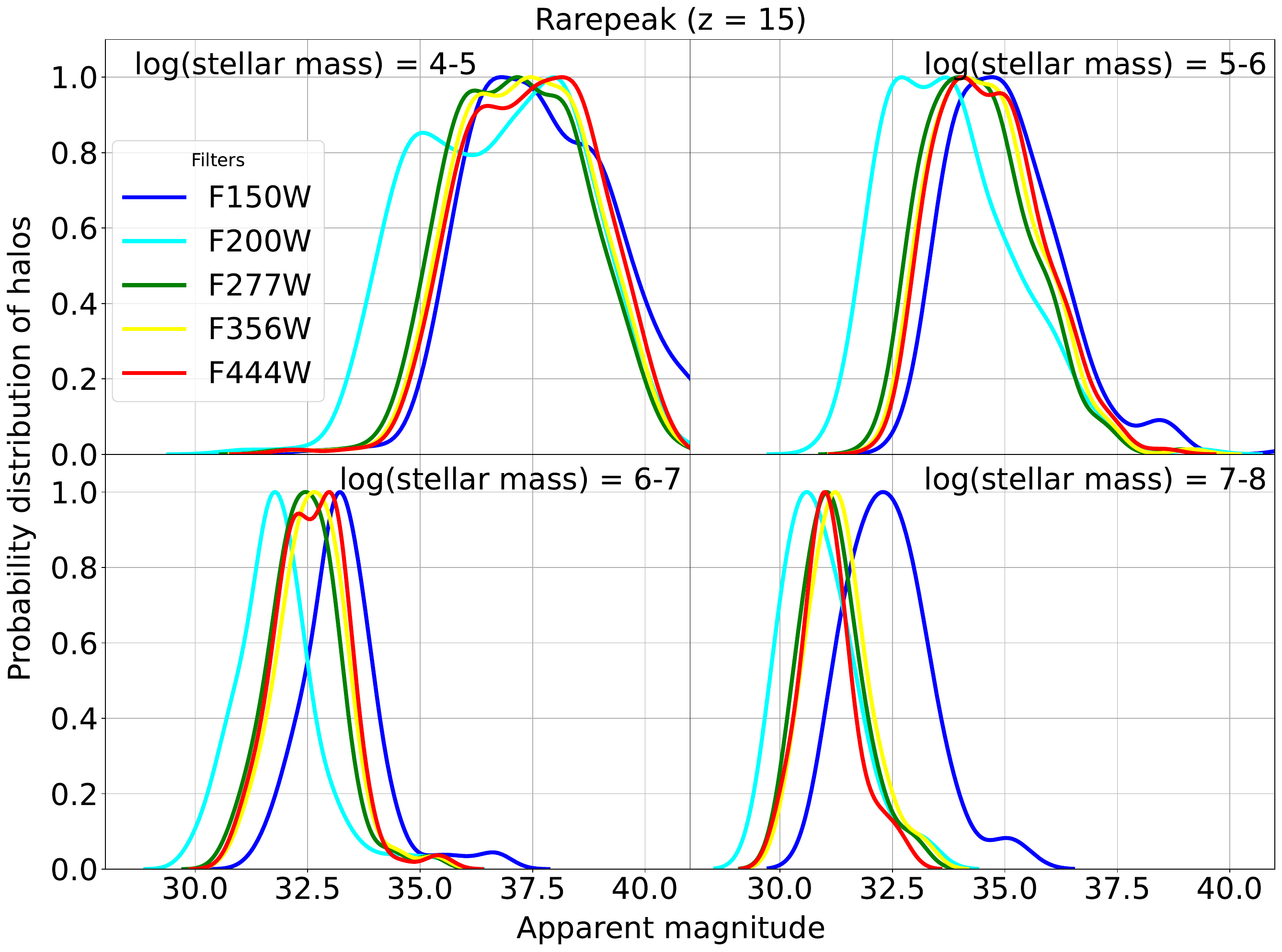}
\caption{The probability distribution of the apparent magnitude in JWST's wideband filters in the Rarepeak region of \renaissance in increasing stellar mass cuts in each panel. Each probability distribution line is colored based on the filters. 
\label{fig:app_mag_prob_RP}}
\end{figure}

\begin{figure}[t!]
\includegraphics[width = \columnwidth]{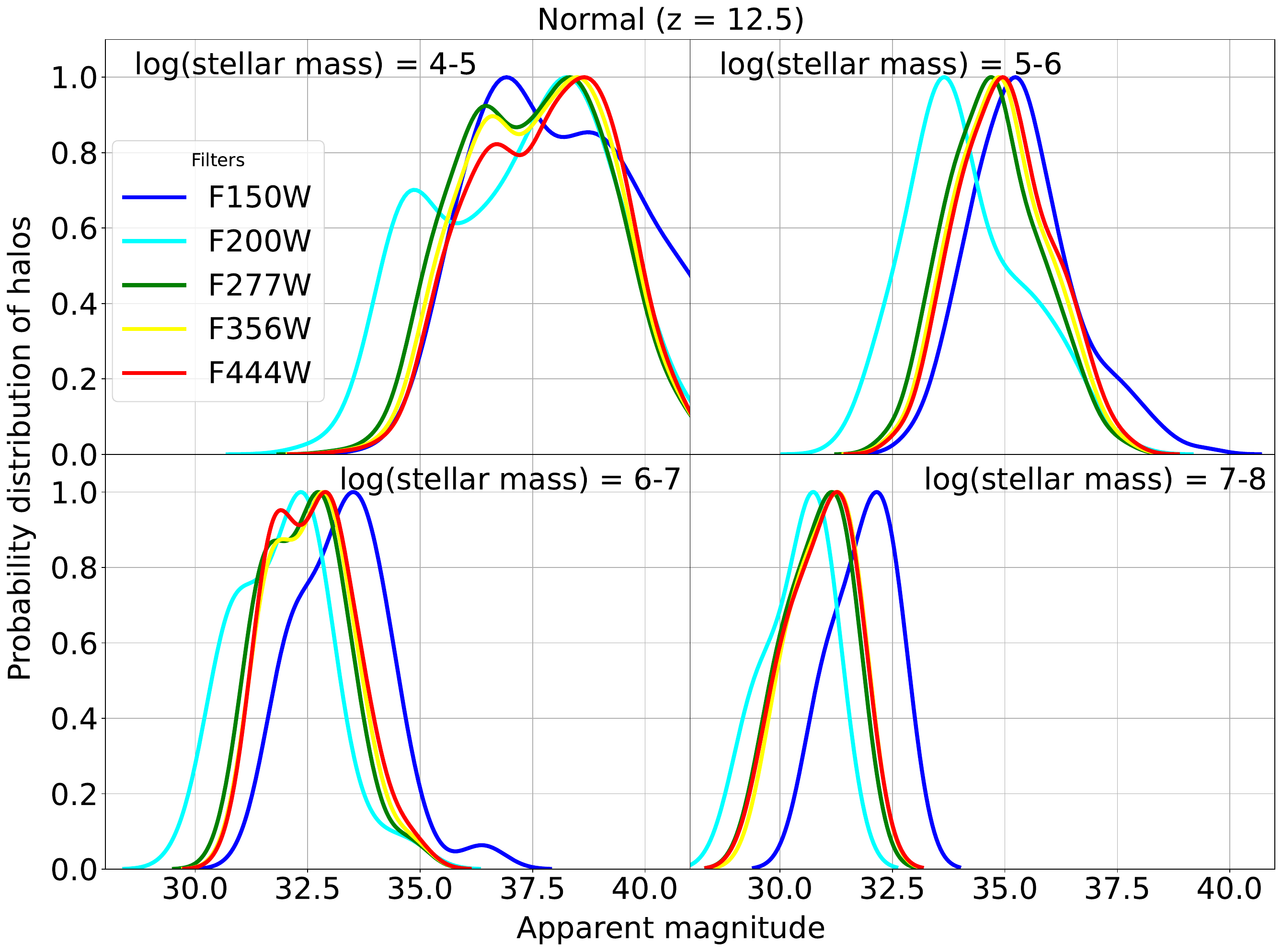}
\caption{Same as Figure \ref{fig:app_mag_prob_RP}, but for the Normal region. 
\label{fig:app_mag_prob_Normal}}
\end{figure}

\begin{figure}[t!]
\includegraphics[width = \columnwidth]{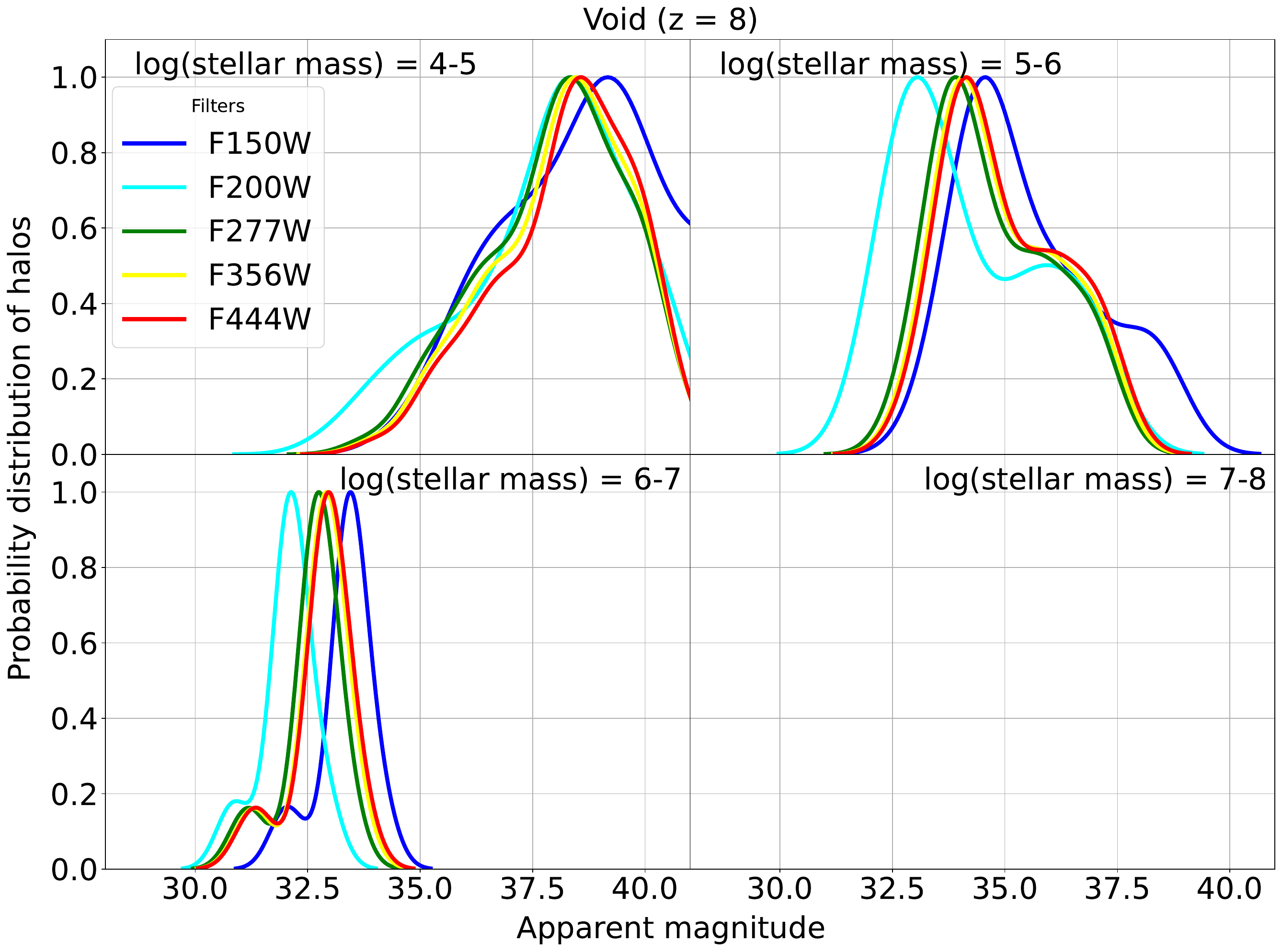}
\caption{Same as Figure \ref{fig:app_mag_prob_RP}, but for the Void region.
\label{fig:app_mag_prob_Void}}
\end{figure}

Figures \ref{fig:app_mag_prob_RP} -- \ref{fig:app_mag_prob_Void} show the probability distributions of apparent magnitude in various filters of galaxies in each region split into four stellar mass cuts, where the colors depict different NIRCam filters. Expectedly, the apparent magnitudes decrease with increasing stellar mass, ranging from $\sim 38$ in galaxies with $M_* = 10^{4-5} \ \textup{M}_{\odot}$ down to $\sim 31$ at $M_* = 10^{7-8} \ \textup{M}_{\odot}$ independent of the region. The majority of the filters have extremely similar trends with only F200W shifting towards being noticeably brighter than most of the other filters. With increasing stellar mass, the F150W magnitude increases relative to the other filters. We exemplify this through the mean and standard deviation in the F150W filter and its adjacent F200W filter. In the Rarepeak region, over the stellar mass decades from least ($10^{4-5} \ \textup{M}_{\odot}$) to greatest ($10^{7-8} \ \textup{M}_{\odot}$), the F150W magnitudes have a mean and standard deviation $37.80 \pm 1.6$, $35.03 \pm 1.3$, $33.27 \pm 0.9$, and $32.42 \pm 0.9$. In the Rarepeak region, over the stellar mass decades from least to greatest, F200W has a mean of $36.73 \pm 1.8$, $33.77 \pm 1.4$, $31.83 \pm 0.9$, and $30.89 \pm 0.8$. We see an increasing difference between the mean in the F150W and F200W filters with decreasing standard deviations with respect to stellar mass for each. This trend is also seen in the Normal and Void regions with similar differences between means over stellar mass ranges. The only exception is a decrease in the differences of the means in the Void region in the stellar mass range of $10^{5-6} \ \textup{M}_{\odot}$ with F150W having $35.53 \pm 1.5$ and F200W having $34.23 \pm 1.6$ and in the $10^{6 - 7} \ \textup{M}_{\odot}$ range with F150W having $33.33 \pm 0.5$ and F200W having $32.06 \pm 0.5$. Although these more massive galaxies have increasing SFRs, they now host older, redder stars and are less dominated by young, blue stars because it probes the UV redward of Ly$\alpha$ at these redshifts, causing this relative F150W change. We also see this trend toward older stellar populations exemplified in the increasing SFR and stellar masses of the \renaissance Simulations compared against the JWST observations in Figure \ref{fig:stellar_mass_SFR}. \par

The distribution in the three regions follow similar patterns, but there are some noticeable differences in the Void region (Figure \ref{fig:app_mag_prob_Void}) from the extended evolution of the galaxies relative to the other regions, allowing for some galaxies to have an older stellar population. Overall, these magnitude distributions give us an overview of trends of the low-mass progenitors of the JWST galaxies over a large stellar mass range, giving context to how the age and stellar mass of these galaxies affect their brightnesses in different filters.


\subsection{Absolute UV Magnitude versus stellar mass}

\begin{figure*}[ht!]
\includegraphics[width = \textwidth]{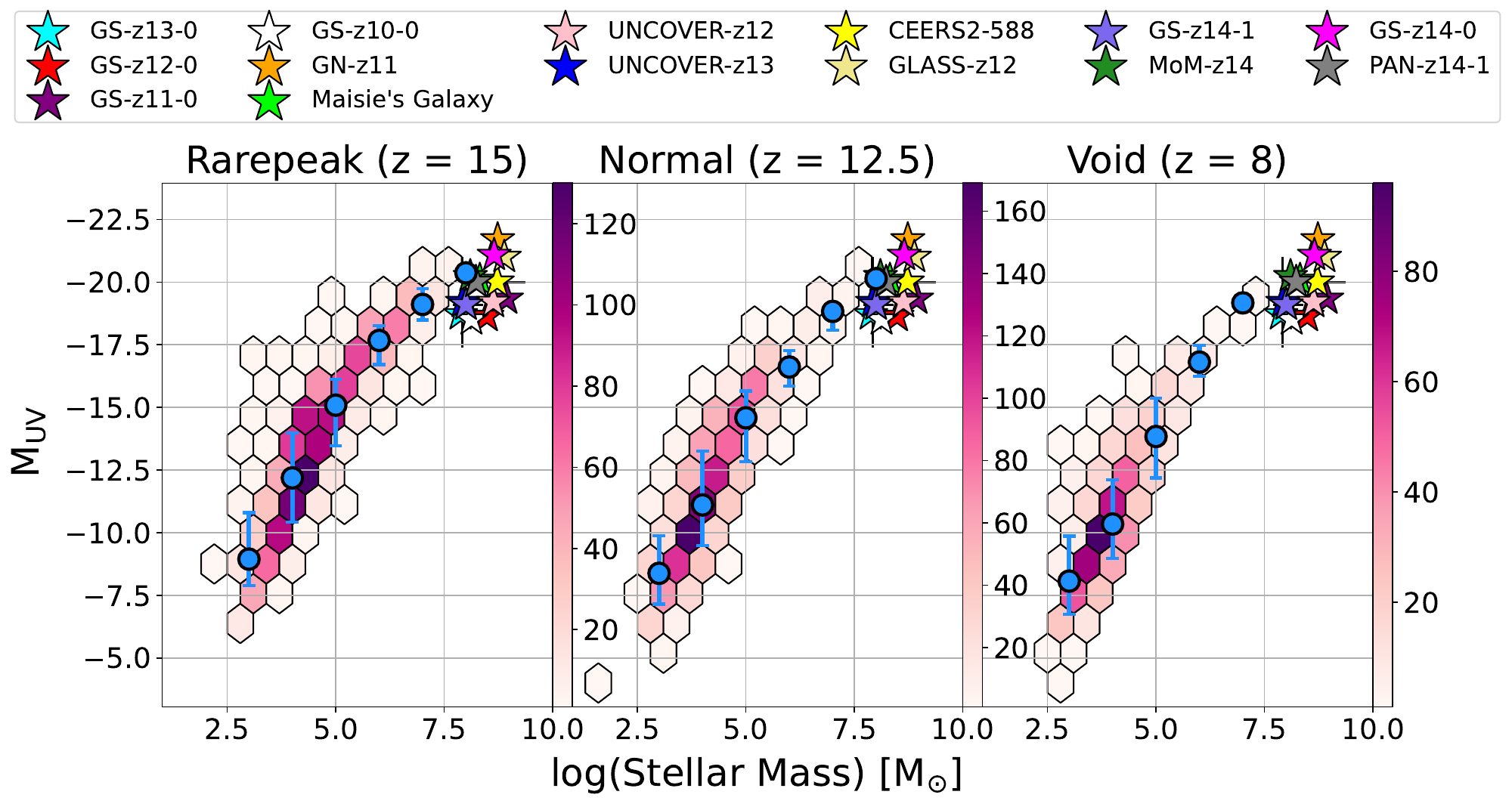}
\caption{The absolute UV magnitude as a function of stellar mass represented by red hexbins for the galaxies in the Rarepeak, Normal, and Void regions of \renaissance from left to right respectively. The absolute UV magnitude was averaged over 100 viewing angles of the \renaissance galaxies, as included in the catalog. The absolute UV magnitudes of the JWST galaxies are represented on the figure as stars of various colors with error bars, some of which are not visible in this scale. The blue points are represent the median of the absolute UV magnitude with error bars showing $1\sigma$.  
\label{fig:abs_mag_stellar_mass}}
\end{figure*}

Figure \ref{fig:abs_mag_stellar_mass} shows the absolute UV magnitude as a function of the stellar mass of the \renaissance galaxies in the Rarepeak, Normal, and Void regions from left to right, respectively. The hexbins are colored by the number of halos, and the stars represent the observed $z > 10$ galaxies. The blue points show the median of the absolute UV magnitudes with error bars representing the $1 \sigma$ confidence interval. We use $1\sigma$ because it has been found to strongly represent the scatter due to bursty star formation in $z \simeq 12 - 14$ galaxies \citep{shen_scatter, kravtsov2024stochasticstarformationabundance, gelli_mass_stochasticity_scatter, stark_scatter}. Overall, the \renaissance Simulations expectedly follow the trend of increasing intrinsic UV brightness with increasing stellar mass. Although the \renaissance Simulations only marginally overlap with the observed galaxies, this low-mass end trend and scatter will help interpret galaxies in lensed fields and at the sensitivity limit of JWST when spectroscopy is unavailable. We find that the standard deviation in $\textup{M}_{UV}$ can be up to 3 magnitudes in a stellar mass bin of $1$ dex. This indicates that the \renaissance galaxies have extremely bursty star formation that is overall consistent with prior results. It is generally expected for burstiness to decrease with decreasing redshift and increasing stellar mass as galaxies become less susceptible to feedback effects in deeper potential wells \citep{stellar_pop_dwarf_scatter, fulanetto_bursty_scatter, legrand_scatter, hopkins_scatter_bursty, stark_scatter}; we find this less obviously with decreasing redshift, but we do find that the $1\sigma$ spread decrease to $\sim 1$ mag as the stellar mass increases above $10^6 \ \textup{M}_{\odot}$ in each region. \par 

Figure \ref{fig:abs_mag_stellar_mass} depicts a smooth transition from the \renaissance galaxies to the JWST galaxies with a decrease in slope around a stellar mass of $10^{7} \ \textup{M}_{\odot}$ in each region, particularly for the Void. This slight decrease in slope may be caused by the increase in older stars as the Void galaxies at $z = 8$ have had more time to evolve. This trend suggests once again that the \renaissance galaxies represent the early stages of formation of these high redshift JWST galaxies.

\subsection{Color-magnitude diagrams}

\begin{figure*}[t!]
\includegraphics[width = \textwidth]{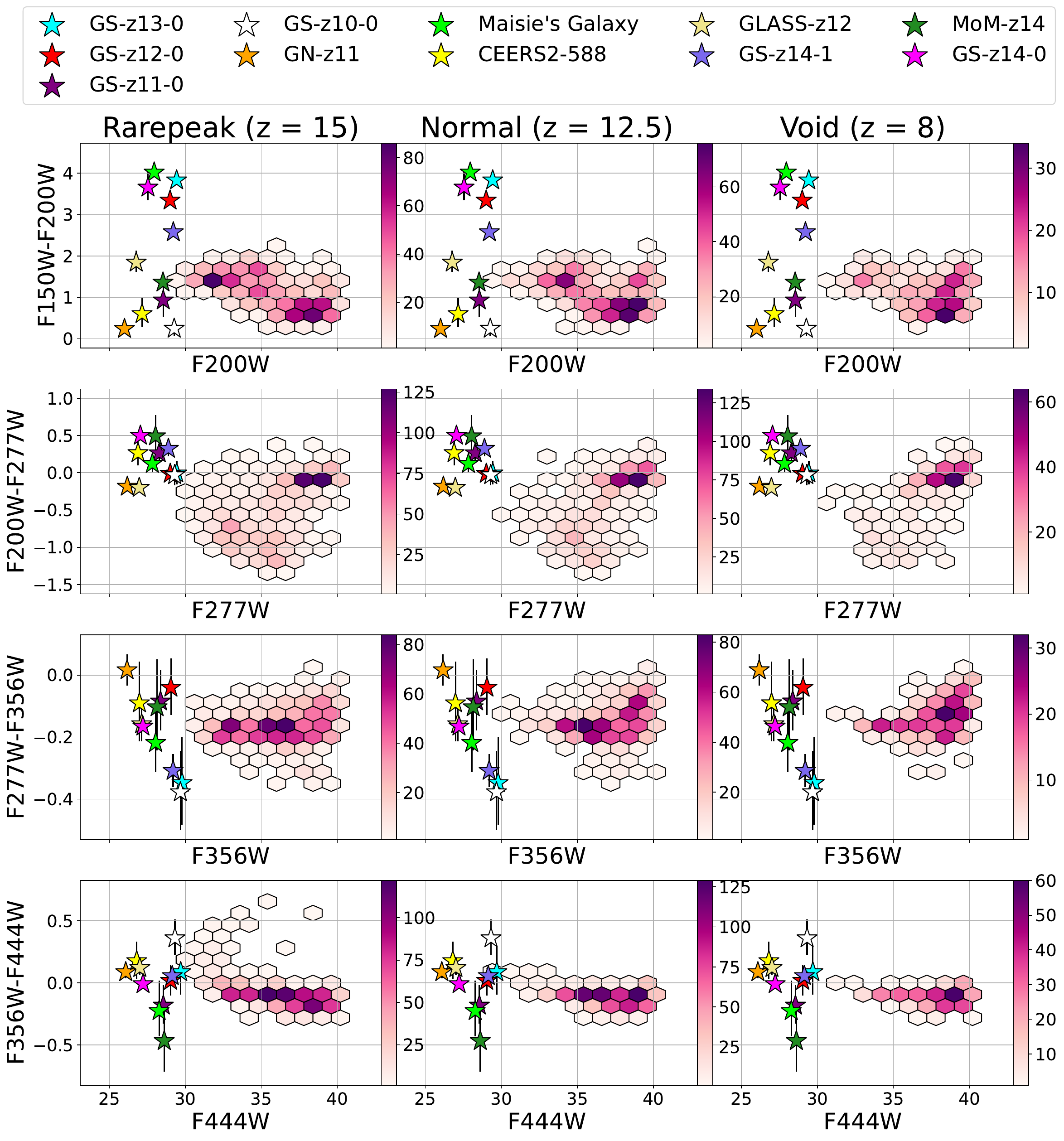}
\caption{The color-magnitude plots for each combination of JWST wideband filters with the hexbins colored by number of halos for the Rarepeak, Normal, and Void regions of \renaissance from left to right, respectively. The stars represent spectroscopically confirmed JWST galaxies at $z > 10$.
\label{fig:color_mag}}
\end{figure*}

Figure \ref{fig:color_mag} shows color-magnitude diagrams for each wide band JWST filter combination. The observed $z > 10$ galaxies are represented as stars of various colors. The colors of the hexbins represent the number of galaxies. Accordingly, these low-mass simulated galaxies are dimmer than the galaxies seen by JWST with only some small overlap in the F200W, F356W, and F444W filters in the Rarepeak region, and some small overlap in the F200W and F444W filters in the Normal region. There is no overlap in the Void region, which is caused by the large scale environment and lack of massive galaxies. The \renaissance galaxies are still in a range where they could reasonably transition in later evolutionary stages at higher masses into the brightness of the JWST galaxies. Across the \renaissance Simulations, there is a trend that while brightness increases, the colors do not change significantly. \par

The color of the JWST galaxies in the first row (F150W -- F200W) have a wide spread of 4 magnitudes with the \renaissance galaxies' colors on the blue side of this range. They are overall younger galaxies that still contain a majority of exclusively young stars in lower mass systems relative to the JWST galaxies. JWST galaxies that overlap into this blue range have intense star formation. This large scatter by the JWST galaxies in bluer filters is due to reddening by the neutral gas in their surrounding IGM below the Ly$\alpha$ break, whereas our mock observations do not include this effect.  \par 

The colors of the JWST galaxies in the second row (F200W -- F277W) have a smaller scatter than the \renaissance galaxies but still overlap on the red side. While there is a lot of scatter in these color-magnitude diagrams, the majority of the observed galaxies, have colors between $-0.5$ and $0.5$. The bluer simulated galaxies are dominated by very young stars, whereas JWST galaxies likely have significantly older stellar populations in higher mass systems. We similarly see in the third row (F277W -- F356W) and fourth row (F356W -- F444W) that the \renaissance Simulations align well the colors of the JWST galaxies with a similar spread in the galaxy samples. These filters are above the Ly$\alpha$ break, so they are not affected by any IGM reddening. If the \renaissance galaxies continue keeping a steady color as they grow, the \renaissance galaxies should smoothly transition into the JWST galaxies as they age and increase in stellar mass and SFR. \par

\subsection{Color-color diagrams}

\begin{figure*}[t!]
\includegraphics[width = \textwidth]{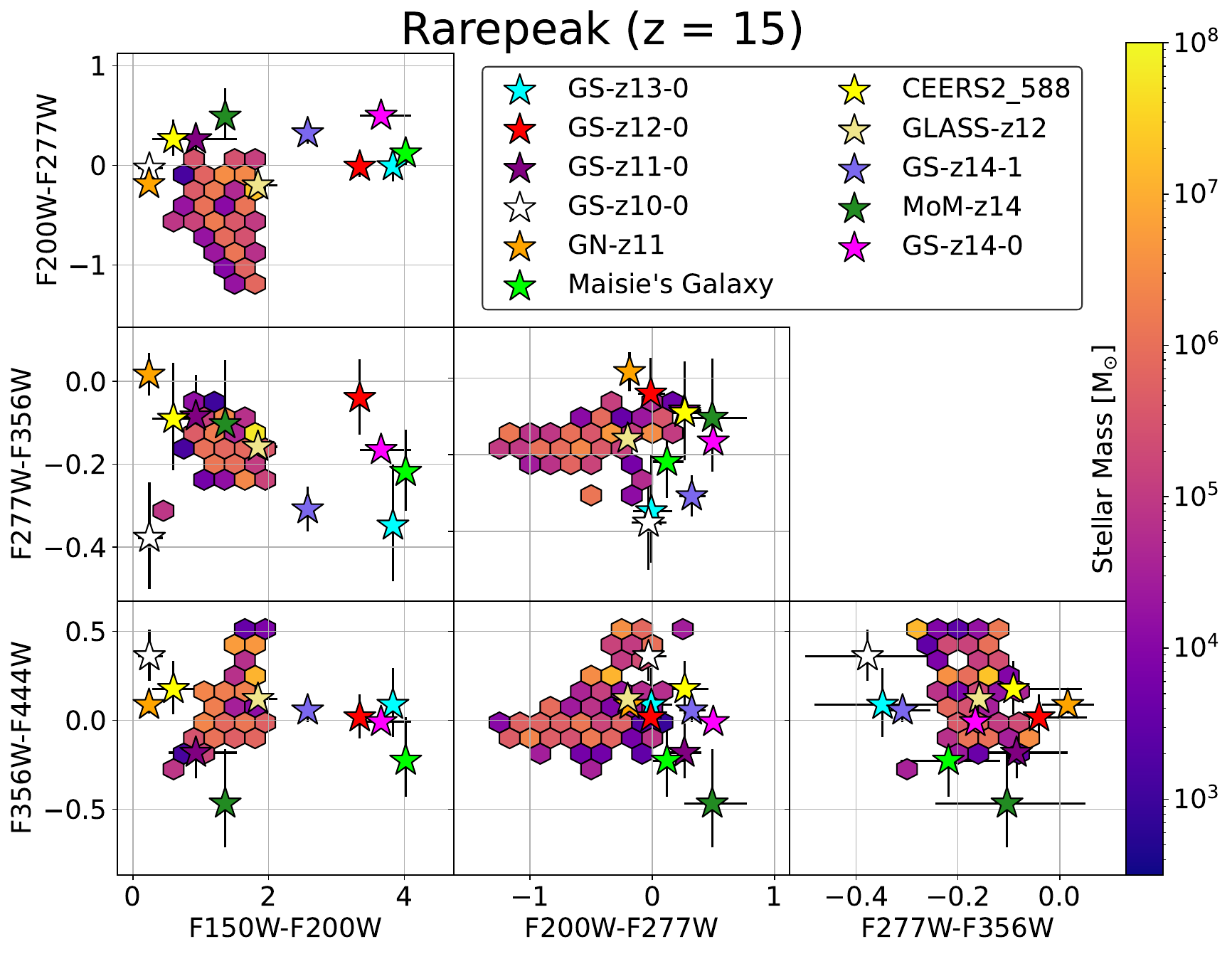}
\caption{Color-color diagrams of all combinations of the wideband NIRCam filters for the Rarepeak region of the \renaissance Simulation. The hexbins are colored by the mean stellar mass of the galaxies in each hexbin. The colors of the spectroscopically confirmed z $> 10$ JWST galaxies are included with error bars as stars of various colors.
\label{fig:color_color_stellar_mass_RP}}
\end{figure*}

\begin{figure*}[t!]
\includegraphics[width = \textwidth]{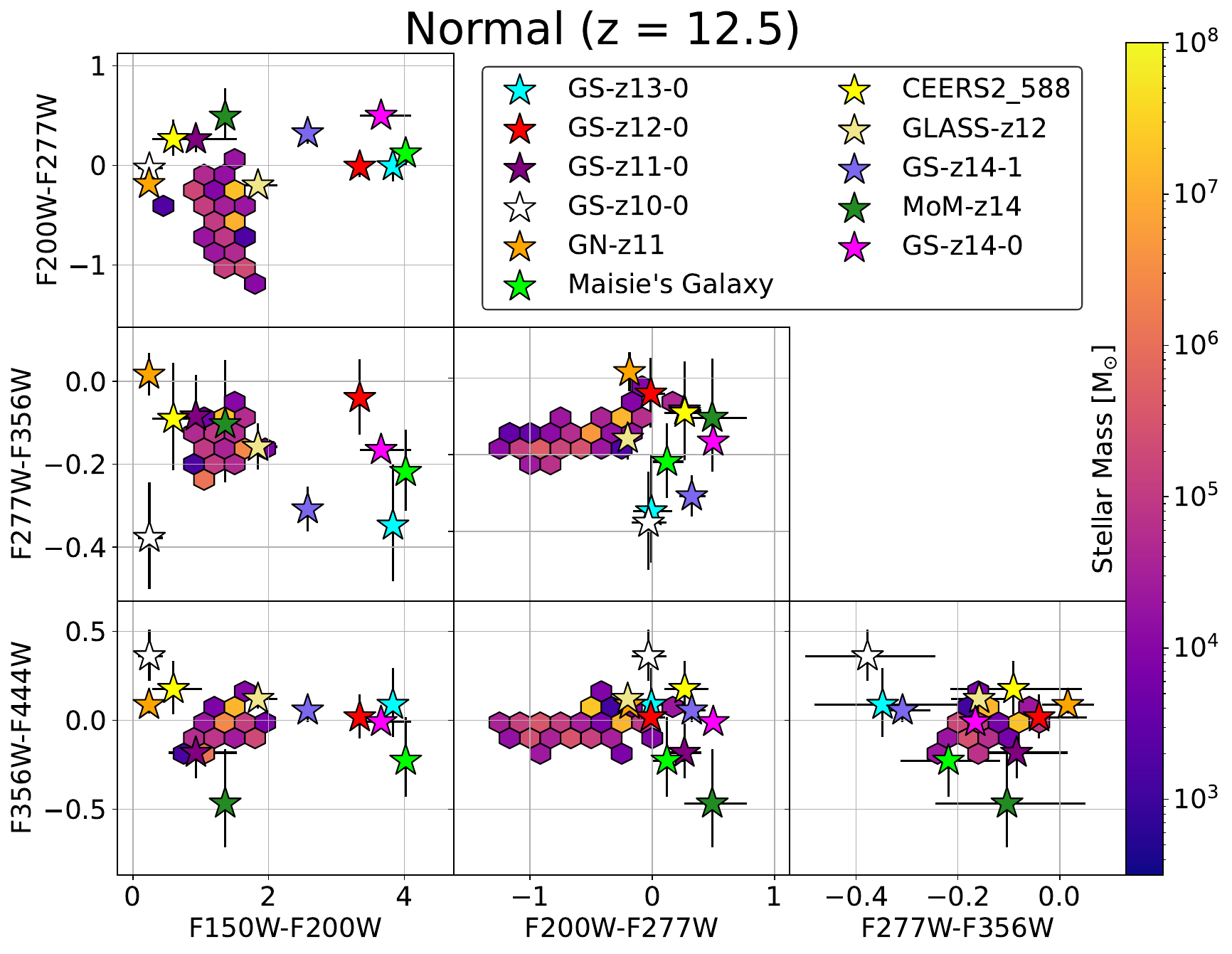}
\caption{Same as Figure \ref{fig:color_color_stellar_mass_RP} but for the Normal region. 
\label{fig:color_color_stellar_mass_Normal}}
\end{figure*}

\begin{figure*}[t!]
\includegraphics[width = \textwidth]{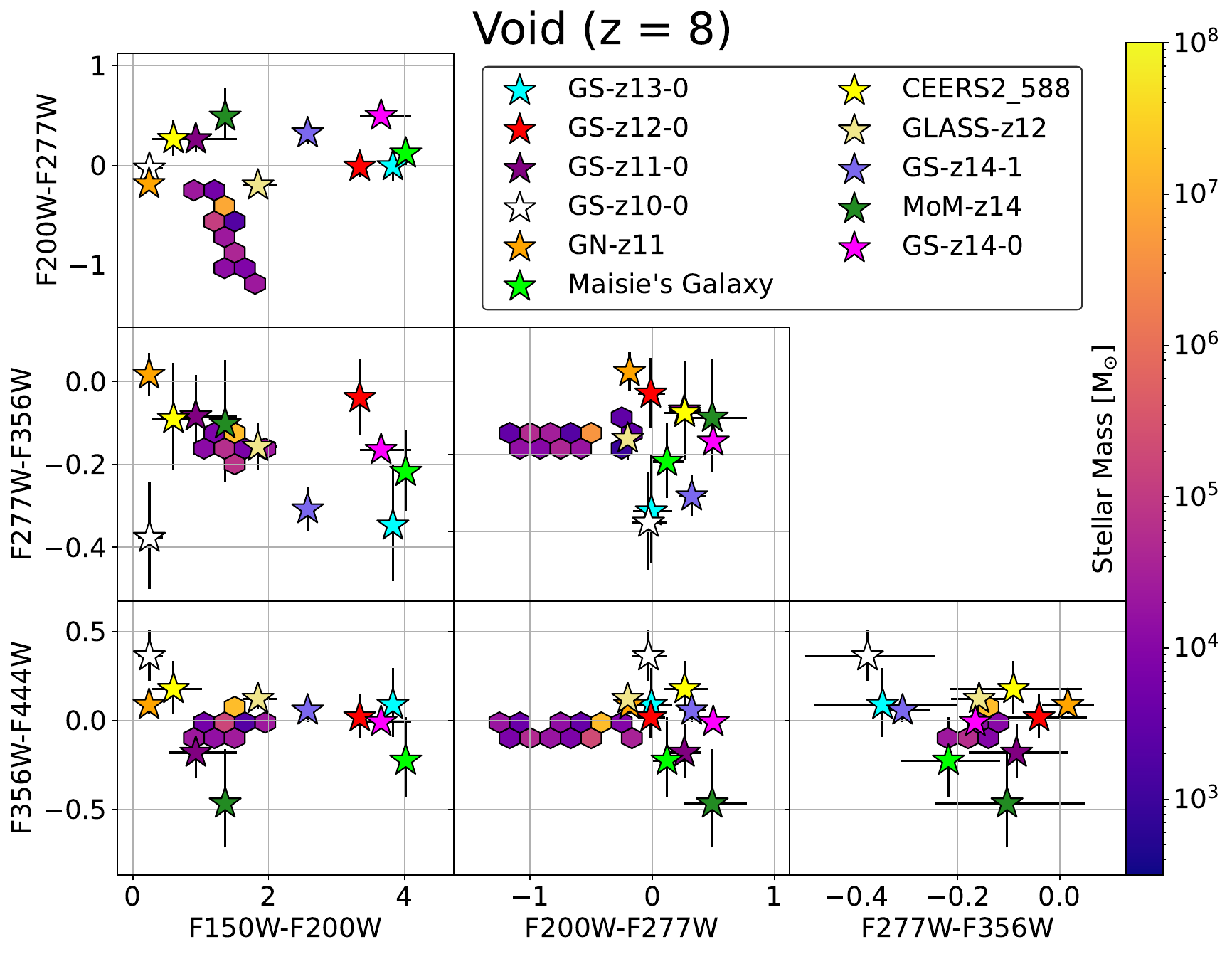}
\caption{Same as Figure \ref{fig:color_color_stellar_mass_RP} but for the Void region.
\label{fig:color_color_stellar_mass_Void}}
\end{figure*}

Figure \ref{fig:color_color_stellar_mass_RP} shows the color-color diagrams of the \renaissance galaxies in the Rarepeak region for all NIRCam wideband filter combinations, with each hexbin colored by the mean of the stellar masses of the galaxies in that hexbin. The stars represent $z > 10$ galaxies that are spectroscopically confirmed by JWST. Figure \ref{fig:color_color_stellar_mass_Normal} and Figure \ref{fig:color_color_stellar_mass_Void} show the same for the Normal and Void regions, respectively. For the Rarepeak region, there is no clear trend with stellar mass values in all color-color plots, indicating that in the early evolutionary stages galaxies do not preferentially cluster in color-color space according to their stellar masses. We see this similar behavior in the other regions, depicted in Figures \ref{fig:color_color_stellar_mass_Normal} and \ref{fig:color_color_stellar_mass_Void}. \par

For all of the color-color diagrams with respect to F150W -- F200W in Figure \ref{fig:color_color_stellar_mass_RP}, the \renaissance galaxies are bluer than five of the JWST galaxies in color with the exception of F200W -- F277W, again likely because of the Ly$\alpha$ causing a dropout in galaxies blueward of F200W. We see something similar to this in the Figure \ref{fig:color_mag} where the bluer filters F150W and F200W are outliers, with galaxies having much dimmer F150W magnitudes than F200W in each \renaissance region. The JWST galaxies are relatively on par with most of the \renaissance galaxies' colors in ``y-axis" filters. As stated before, the \renaissance Simulations capture a lower-mass population than the observed galaxies, showing them earlier in their formation sequence; thus, they have less older stars than the JWST galaxies. As the plots progress towards redder filters, the \renaissance galaxies tend to overlap better with the JWST galaxies. This trend is seen further in Figure \ref{fig:color_color_stellar_mass_RP} where, as the filters increase in wavelength, the \renaissance galaxies overlap with more JWST galaxies. \par

Figure \ref{fig:color_color_stellar_mass_Normal} shows a similar pattern, but with fewer massive galaxies in the Normal region, there is less overlap with the galaxies seen by JWST. Therefore, we see bluer galaxies in \renaissance in the Normal region than we see in the JWST galaxies, with minimal overlap in the plots with respect to F200W -- F277W and less overlap than the Rarepeak region. However, it still trends towards more overlap for redder filters. The Void region has a tighter trend than the galaxies in the Rarepeak or Normal region, as seen in Figure \ref{fig:color_color_stellar_mass_Void}, likely due to physical differences in their structure as compared to the Rarepeak and Normal galaxies. However, the trend still exists that redder filters trend towards greater overlap between the \renaissance and JWST galaxies.\par

\begin{figure*}[t!]
\includegraphics[width = \textwidth]{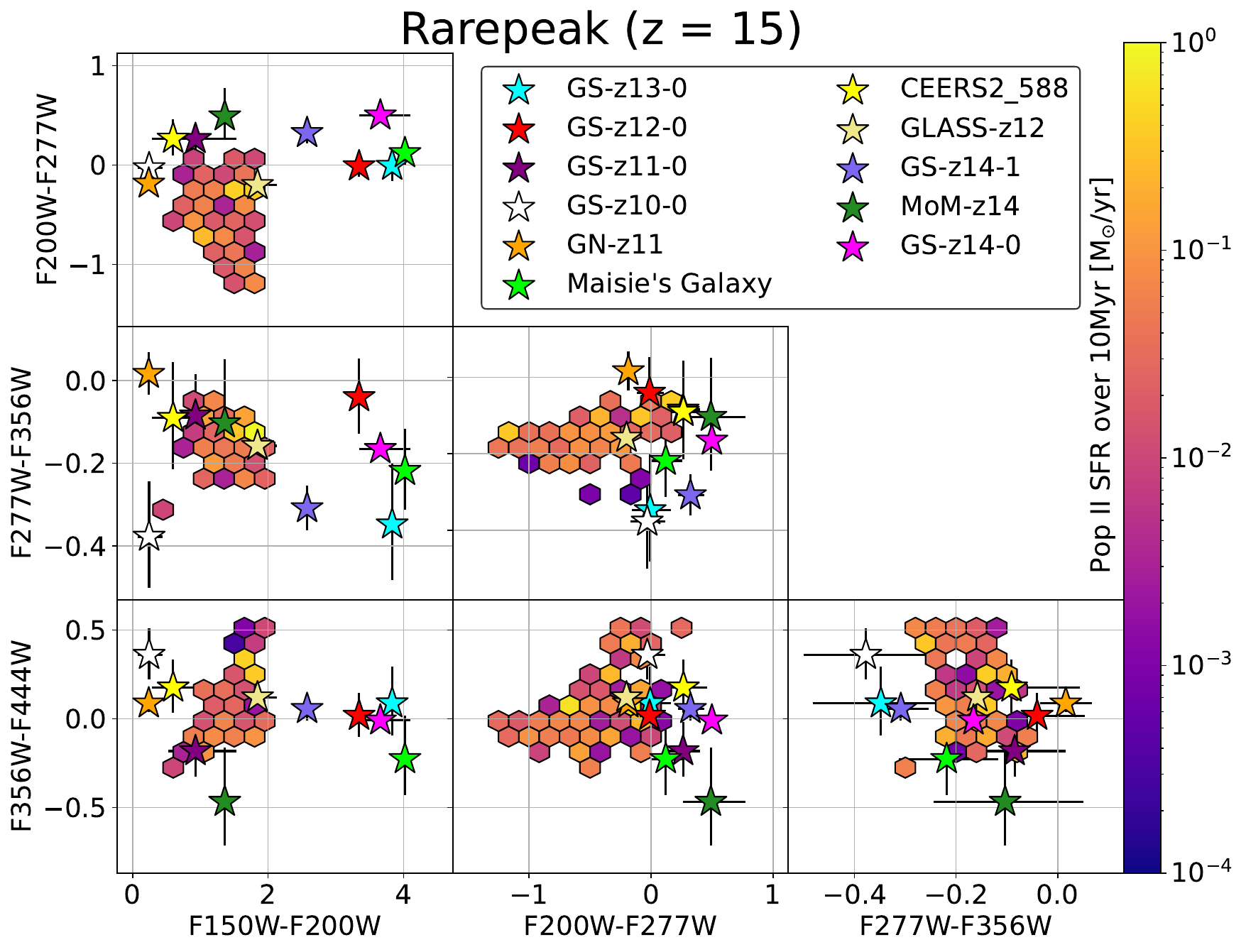}
\caption{Same as Figure \ref{fig:color_color_stellar_mass_RP} but colored by SFR. 
\label{fig:color_color_SFR_RP}}
\end{figure*}

\begin{figure*}[t!]
\includegraphics[width = \textwidth]{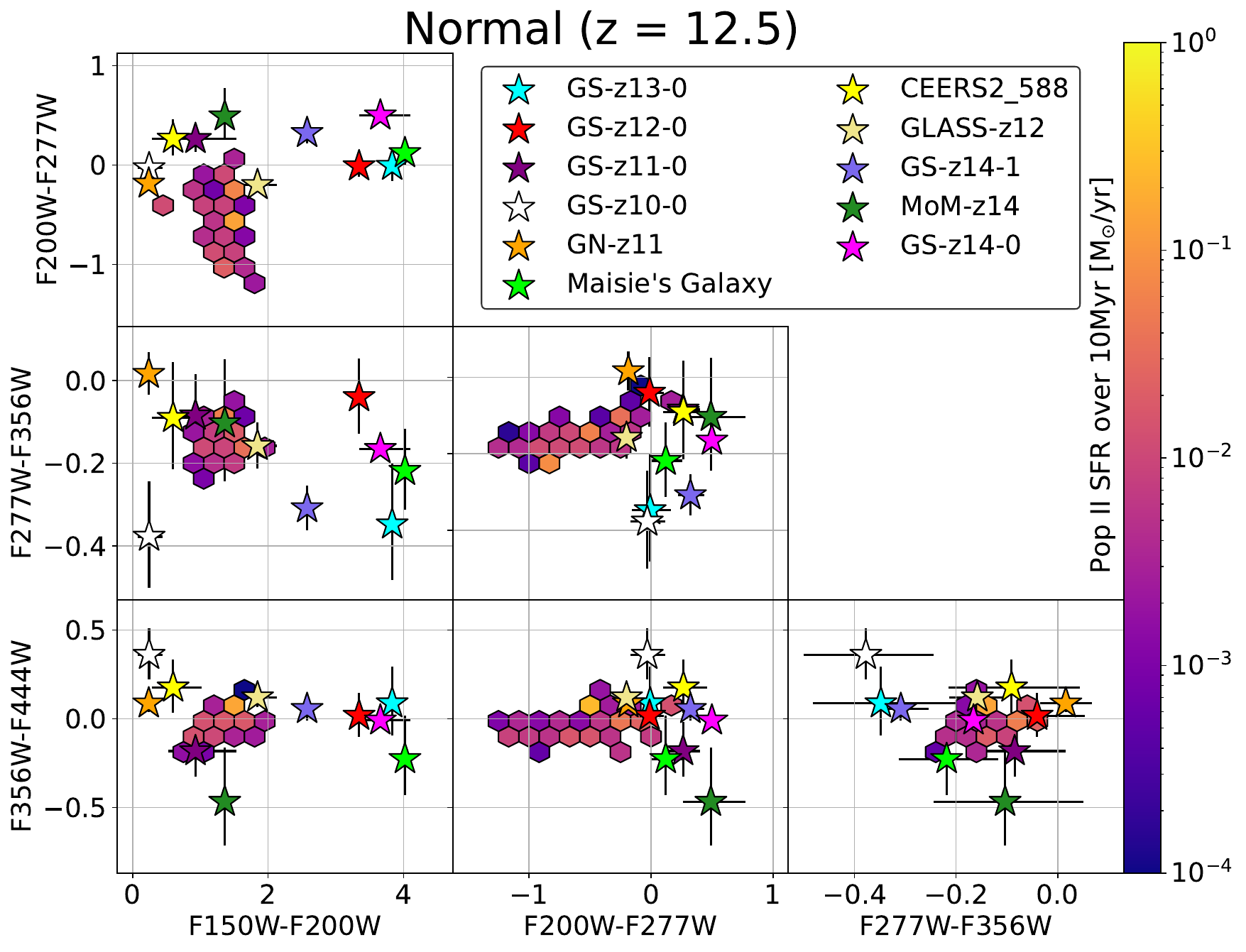}
\caption{Same as Figure \ref{fig:color_color_stellar_mass_Normal} but colored by SFR.
\label{fig:color_color_SFR_Normal}}
\end{figure*}

\begin{figure*}[t!]
\includegraphics[width = \textwidth]{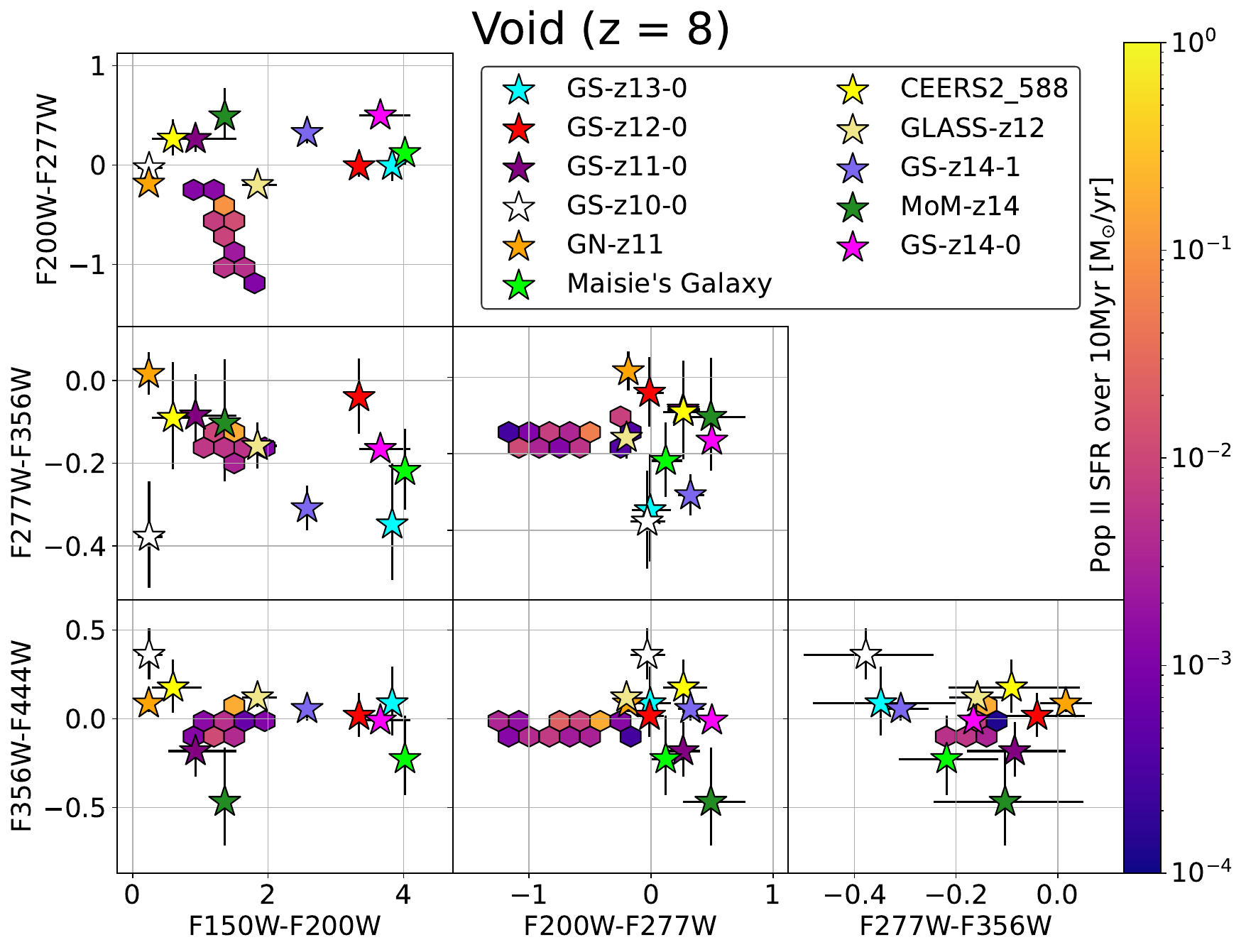}
\caption{Same as Figure \ref{fig:color_color_stellar_mass_Void} but colored by SFR.
\label{fig:color_color_SFRs_Void}}
\end{figure*}

Figures \ref{fig:color_color_SFR_RP}--\ref{fig:color_color_SFR_Normal} once again show the color-color diagrams of the \renaissance galaxies in the Rarepeak, Normal, and Void regions, respectively, but with the hexbins colored by the mean SFRs. For all of these figures, some hexbins existed where the SFR was zero within that hexbin, which we removed. The SFRs, much like the stellar masses, do not follow any specific trend, with higher SFRs spread sporadically throughout the bluer and redder areas. \par

\section{Discussion}
\label{discussion}

\subsection{Caveats}
In this paper, we compare the mock observables of the \renaissance Simulations with current JWST observations of $z > 10$ galaxies. We note that the \renaissance Simulations are zoom-in simulations focusing on high redshifts and low mass galaxies, allowing for pc-scale resolution. The \renaissance Simulations give us insight into  the formation history of the galaxies observed by JWST. However, there are some caveats worth noting. Firstly, we are not able to compare directly with the redshifts of any of the JWST galaxies; however, the evolutionary sequence of the simulated galaxies capture plausible physical and observable characteristics at these very high redshifts. The redshifts of $z > 10$ JWST galaxies range from the lowest being GS-z10-0 with $z = 10.38^{+0.07}_{-0.06}$ and the highest of those being MoM-z14 with $z = 14.44^{+0.02}_{-0.02}$. Additionally, the simulated galaxies have much lower masses than that of the $z > 10$ JWST galaxies, a sacrifice \renaissance makes for increased resolution and input physics. These caveats are the reason \renaissance acts as an insight into the formation history of the observed JWST galaxies rather a direct comparison. \par

Despite the pc-scale resolution of the \renaissance galaxies, it does have its limitations. Specifically very small, low mass galaxies are unresolvable by JWST. Thus, these galaxies are effectively point sources and their half light radii or Sersic indices cannot be calculated, so we do not include them in the morphology studies in this paper.

Finally, the \renaissance galaxies do not contain AGN that could be present in the observed JWST galaxies, leading to overall lower apparent and absolute UV magnitudes for the \renaissance Simulations. \par   

\subsection{Comparison with other high redshift simulations}

Here we compare the \renaissance Simulation results with other high redshift simulations that focus on the first galaxies. While these simulations do not always cover the same mass and redshift ranges as \renaissance, comparisons with them are important in validating what \renaissance has shown us about early galaxy formation. Overall, we find a strong agreement between \renaissance's findings and the results of other high redshift cosmological simulations presented next. 

\subsubsection{ASTRID}

The \astrid Simulations are high redshift simulations that end at $z = 3$ with a box size of $205 \ \textup{Mpc\ h}^{-1}$ \citep{astrid_galaxies, astrid_synthetic}, from which they created mock observations at $3 \leq z \leq 6$ and compared them directly with the first release of CEERS data. \par

\astrid has overall higher stellar masses of galaxies at $8 \leq z \leq 10$ than \renaissance with minimum overlap in the range of approximately $10^{7-10} \ \textup{M}_{\odot}$. The lowest stellar mass bin that \astrid presents at $10^{10} \ \textup{M}_{\odot}$ over the range $8 \leq z \leq 12$ has a SFR $\Psi = 10^{-3} - 10^{-1} \ \textup{M}_{\odot} \ \textup{yr}^{-1}$, while our highest mass galaxies in \renaissance have a $\Psi \approx 1 \ \textup{M}_{\odot} \ \textup{yr}^{-1}$, making their SFRs much lower than ours for higher mass galaxies at equivalent redshifts. \par

In \astrid's mock observations, they take a representative sample of stellar masses at $z = 4$ and find that they have an overabundance of galaxies compared to CEERS observations. In \astrid's comparison of their galaxies' Sersic indices and stellar masses, they find a trend towards lower Sersic indices for higher mass galaxies at higher redshifts compared to those same galaxies at lower redshifts. This indicates a more irregular shape for these galaxies at higher redshifts, aligning with what we see in the \renaissance Simulations. The \astrid galaxies also tend to maintain a consistent Sersic index until they reach a stellar mass between $10^{10.5 - 11} \ \textup{M}_{\odot}$, which aligns with our findings of a consistent Sersic index trend over increasing stellar masses below $\sim 10^{8} \ \textup{M}_{\odot}$. They find steady Sersic indices, despite changes in half light radii, that peak between $n = 0.5 - 1.0$, similar to the trend we find in the \renaissance Simulations. In their study of half light radii as a function of stellar mass, they find a similar trend with the half light radii increasing minimally until they reach a stellar mass between $10^{10.5 - 11} \ \textup{M}_{\odot}$ where the half light radius begins to spike, aligning with our findings of a consistent half light radii across stellar mass at low masses. Overall, we find very little morphology change at low masses over redshifts $z = 8, 12.5, $ and $15$, consistent with \astrid's findings of very little change of galaxies' Sersic indices at low masses across both mass and redshift. However, we find no trend in half light radii over the redshifts presented in our study, where \astrid finds an increase in half light radii as redshift increases from $z = 6$ to $z = 3$ with a peak at $z = 6$ around $1$ kpc. This is most likely due to intensely centralized star formation in the early Universe and the limited time between $z = 8 - 15$. \par

\astrid's study of the ratio between half light radii and half stellar mass radii find that the majority of the galaxies in their lower mass regime have a ratio $\approx 1$ over all redshifts. While they do have an outlying region with a ratio greater than 1, they state that this most likely comes from errors in fitting the Sersic fit in their pipeline. While they find a ratio $\approx 1$, we find that our ratio trends towards ratios less than unity. This may be due to a lack of resolution in our mock observations that are limited by the JWST PSF leading to a perceived larger half light radius or due to intense central star formation at \renaissance's high redshifts compared to \astrid's lower redshifts observed in this paper. \par


\subsubsection{THESAN}

\thesan and \thesanzoom are high redshift simulation suites focusing on $z \geq 5.5$ including various simulations with different levels of zooms \citep{thesan_intro, thesan, thesan_morphology, thesan_zoom, thesan_zoom_starbursts}. \thesanzoom contains 14 high redshift galaxies from \thesan at $z > 3$ with stellar masses of $\approx 10^{4 - 10} \ \textup{M}_{\odot}$. \thesan's 16 simulations have varying underlying physical models, volumes, and resolutions, with \thesan having a box volume $(95.5 \ \textup{cMpc)}^3$. \par 

At $z = 8$, \thesan has stellar masses in the range $\textup{M}_{*} \approx 10^{6.5 - 10} \ \textup{M}_{\odot}$, overlapping the tail end of the \renaissance galaxies. \thesan's SFR at $z = 8$ steadily increases from $\Psi \approx 10^{-2}$ to $1 \ \textup{M}_{\odot} \ \textup{yr}^{-1}$ as the stellar mass increases from $\textup{M}_{*}  \approx 10^{6.5}$ to $10^{8} \ \textup{M}_{\odot}$ while \renaissance exhibits nearly the same trend and normalization. The galaxies at $z \approx 15$ in the \thesanzoom Simulation have SFRs over $100 \ \textup{Myr}$ of $\Psi \approx 10^{-4} - 10^{-1} \ \textup{M}_{\odot} \ \textup{yr}^{-1}$ as the stellar masses increase from $\textup{M}_{*} \approx 10^{4}$ to $10^{7} \ \textup{M}_{\odot}$, agreeing very well with \renaissance. \par

\thesan studies the half light radii of its galaxies at a higher masses than \renaissance at $\textup{M}_{*} \geq 10^{7.5} \ \textup{M}_{\odot}$. Unsurprisingly, in this higher mass regime they find generally larger half light radii than we find at similar redshifts. \thesan's galaxies at $\textup{M}_{*} \approx 10^{7.5} \ \textup{M}_{\odot}$ and $z \simeq 10$ and $z \simeq 8$ have half light radii in a range of $r_{1/2} \approx 0.5$ to $5$ kpc. The \renaissance galaxies at $\textup{M}_{*} \approx 10^{7.5} \ \textup{M}_{\odot}$ and $z = 15$ have half light radii $r_{1/2} < 0.8$ kpc. In the Normal region in \renaissance, at $z = 12.5$, the galaxies at $\textup{M}_{*} \approx 10^{7.5} \ \textup{M}_{\odot}$ have half light radii $r_{1/2} < 0.5$ kpc. The Void region in \renaissance barely reaches $\textup{M}_{*} \approx 10^{7.5} \ \textup{M}_{\odot}$, but the most massive galaxies at $z = 8$ have half light radii around $r_{1/2} \approx 0.1$ kpc. \thesan overall finds its galaxies have larger half light radii than observed galaxies. \thesanzoom studies the stellar half mass radii of the galaxies from $3 < z < 13$ in the mass range $\textup{M}_{*} \approx 10^{6 - 10} \ \textup{M}_{\odot}$ finding that as the stellar mass increases and redshift decreases, the half stellar mass radius range is $\sim 0.1 - 10$ kpc at $\textup{M}_{*} \approx 10^6 \ \textup{M}_{\odot}$ to a range of $\sim 0.3 - 10$ kpc at $\textup{M}_{*} \approx 10^{10} \ \textup{M}_{\odot}$. We similarly see a loss of compact galaxies as stellar mass increases and redshift decreases in \renaissance going from a range of $10^{-3} - 1$ kpc at $\textup{M}_{*} \approx 10^4 \ \textup{M}_{\odot}$ to a range of $0.1 - 1$ kpc at $\textup{M}_{*} \approx 10^6 \ \textup{M}_{\odot}$ in the Rarepeak region. For the Void region in the same stellar mass ranges, the stellar half mass radius goes from a range of $0.03 - 1$ kpc to a range of $0.3 - 1$ kpc. \thesanzoom finds larger stellar half masses than \renaissance over similar stellar mass ranges. \par 


\subsubsection{MEGATRON}

\megatron is a high redshift cosmological simulation suite with seven different simulations, including a high redshift simulation suite and a lower-redshift simulation suite. Here, we focus on the high redshift simulation suite, whose galaxies have a range of $\textup{M}_{*} \approx 10^{2} - 10^{8.5} \ \textup{M}_{\odot}$ at $z = 8.5$ \citep{megatron_reproducingdiversityhighredshift, megatron_impactstarformationfeedback}. Within the high redshift suite, there are four simulations with varying star formation models. While \megatron has similar stellar masses to \renaissance, it has almost a factor of 100 more galaxies.  \par 

\megatron finds a similar increase in SFR with decreasing redshift for all regions, unlike \renaissance which maintains a similar star formation rate over similar redshifts for each of its regions, although \megatron's increase is minimal. The slight increase in \megatron's SFR may simply be because they have more high mass galaxies that better sample the burstiness native to galaxy formation at high redshift. At a redshifts of $z \approx 8.5$, they find that most of the simulations have a SFR $\Psi < 10 \ \textup{M}_{\odot} \ \textup{yr}^{-1}$, with the exception of the efficient star formation simulation that produces galaxies with $\Psi > 10 \ \textup{M}_{\odot}\textup{yr}^{-1}$. An SFR slightly below $\Psi = 10 \ \textup{M}_{\odot}\textup{yr}^{-1}$ is very similar to the most massive galaxies in \renaissance. \par


\subsubsection{SPHINX}

\sphinx is another high redshift hydrodynamics cosmological simulation \citep{sphinx_intro, sphinx}. Here we focus on \textsc{sphinx}$^{20}$, the largest volume simulation in the \sphinx simulation suite with a comoving volume of $(20 \ \textup{Mpc})^3$ presenting results in the range of $z = 4.64 - 10$. The stellar masses in \textsc{sphinx}$^{20}$ fall in the range of $\textup{M}_{*} \approx 10^{7 - 10} \ \textup{M}_{\odot}$ when accounting for all redshifts. At higher redshifts where $7 \leq z \leq 10$, the stellar masses do not exceed $\textup{M}_{*} \approx 10^{9} \ \textup{M}_{\odot}$. These fall in the range of the maximum stellar masses presented in the \renaissance Simulations. Furthermore, \sphinx finds SFRs over $10 \ \textup{Myr}$ of $\Psi \approx 10^{-0.5 - 0.5} \ \textup{M}_{\odot} \ \textup{yr}^{-1} $ in the stellar mass range $\textup{M}_{*} \approx 10^{7 - 10} \ \textup{M}_{\odot}$, aligning very well to \renaissance's trend in this same stellar mass range. \par


\subsection{Outlook}

In this paper, we focus on comparisons of the global properties, the morphologies, and the photometry of the \renaissance galaxies. This obviously leaves out an important capability of \powderday: nebular emission lines. In a follow-up paper, we will compare the mock emission lines of our galaxies, calculated by \cloudy on the fly \citep{Garg_cloudy_BPT} through \powderday, with the emission lines produced by JWST's $z > 10$ galaxies. We plan to study emission line indicators of galactic properties of high redshift galaxies and how they compare to commonly used relationships. \par

In a future study, we will also utilize the fact that \renaissance includes Pop III star formation and feedback. From this future study we will be able to predict observational indicators of Pop III stars in future JWST, 30-m class telescopes, and Roman Space Telescope observations. We note JWST may not be able to detect Pop III stars because there will likely be an abrupt change between Pop II and Pop III galaxies since Pop III stars, if mostly massive,  will explode in supernovae quickly and the galaxy will rapidly transition to Pop II after it is chemically enriched. Therefore, there will only be a short period of time where Pop III stars dominate the luminosity in a galaxy. JWST’s attempts to detect a Pop III galaxy may also be complicated by the fact that some metal rich galaxies will have an emission line ratio [O\textsc{iii}] 5007 $\mathring{\textup{A}}$/H$\beta < 10^{-2}$, testing JWST's sensitivity limits \citep{katz}. While this may be the case, the \renaissance galaxies will still give insight into the assembly history of high redshift galaxies possibly detected by JWST and these galaxies' properties when they are dominated by Pop III stars. \par

Overall, \renaissance is useful in providing us a glimpse into the progenitors and formation history of observed $z > 10$ high redshift galaxies. \renaissance's high resolution allows us to study these first galaxies in detail starting from their very beginnings. Our mock observations, calculated with \powderday, can guide future observational campaigns and establish an understanding of high redshift galaxies' formation and early lives. \par 

\section{Conclusion}
\label{conclusion}

We find that the \renaissance Simulation generally aligns with spectroscopically confirmed $z > 10$ galaxies observed by JWST. We also find that the physical properties, morphologies, and photometry we determine from the mock observations of \renaissance, calculated with \powderday, are in general agreement with other high redshift cosmological simulations, namely \astrid, \thesan, \megatron, and \sphinx. We present the \renaissance Simulations as a glimpse into the earliest galaxy formation phases, showing how the properties of the galaxies at the lowest masses transition into the galaxies that have been observed through JWST and that may be observed by future, more sensitive telescopes. In the following, we list the key conclusions we have reached in this paper. 

\begin{itemize}
    \item The SFRs of the \renaissance galaxies follow a similar trend of JWST's highest redshift observations. The SFRs from the low mass \renaissance galaxies will likely transition smoothly into these JWST galaxies at similar sSFRs. The \renaissance galaxies have a stellar mass range of $ \sim 10^{3 - 8} \ \textup{M}_{\odot}$ and SFRs of $\sim 10^{-4 - 1} \ \textup{M}_{\odot}\textup{yr}^{-1}$ in all regions that transition into the observed $z > 10$ JWST galaxies in the stellar mass range $\sim 10^{7 - 9} \ \textup{M}_{\odot}$ and SFRs of $1 - 20 \ \textup{M}_{\odot} \ \textup{yr}^{-1}$. Almost all of these galaxies fall in the sSFR range of $10^{-9}$ to $10^{-7} \ \textup{yr}^{-1}$. 
    \item The \renaissance galaxies are small and contain compact, central, bursty star formation as evident by the ratio of their half light radii to half stellar mass being majority less than unity. There is little change in their sizes with decreasing redshift. In general, our simulated galaxies are irregular with a vast spread in their Sersic index, spanning the range $0 - 5$ in the Rarepeak and Normal regions and the range $0 -4$ in the Void region, similar to the $z > 10$ JWST galaxies' Sersic indices spanning the range $0 - 4$. The \renaissance galaxies have minimal change in morphology as their redshifts decrease, suggesting the \renaissance galaxies could maintain their morphologies as they reach the masses of the JWST galaxies again showing lower mass galaxies smoothly transition into the JWST galaxies. There is likely minimal morphology change at these high redshifts due to their rapid assembly. 
    \item We find that the F150W magnitude distribution falls behind the other NIRCam photometry as stellar mass increases in every region, indicating again that the low-mass \renaissance galaxies grow in  bursty star formation events. In the Rarepeak region, the difference in the mean of the filters F150W and F200W magnitudes increases by $\sim 0.2$ per dex in stellar mass in galaxies with $\textup{M}_* = 10^{4-7} \ \textup{M}_{\odot}$ and $0.1$ per dex at higher masses. In the Normal region, the difference in the mean of the filters F150W and F200W increases by $0.1$ at all stellar masses explored by this work. In the Void region, the difference in the mean of the filters F150W and F200W increases by $\sim 0.1$ from the stellar mass range $10^{4-5} \ \textup{M}_{\odot}$ to $10^{5-6} \ \textup{M}_{\odot}$ and decreases $\sim 0.03$ from the stellar mass range $10^{5-6} \ \textup{M}_{\odot}$ to $10^{6-7} \ \textup{M}_{\odot}$. The standard deviation for both filters decreases in every increasing stellar mass range.
    \item The trend of absolute UV magnitude with respect to stellar mass of the \renaissance Simulations has a large scatter due to the galaxies burstiness, but transition into the JWST galaxies' trend well, overlapping at the high-mass end of the simulations. Some of the JWST galaxies fall below our predicted absolute UV magnitudes indicating \renaissance may be underpredicting the brightness of these galaxies, possibly due to an overabundance of dust in the mock observations or simply a result of limited sample of higher mass galaxies in \renaissance. 
    \item The \renaissance galaxies are unsurprisingly dimmer than the $z > 10$ JWST galaxies since the \renaissance galaxies have overall less stellar mass. In bluer filters, the \renaissance galaxies are on the blue end of the colors observed in the JWST galaxies, perhaps in part due to the limited number of more evolved stellar populations, but likely because of the dropout of the JWST galaxies blueward of the Ly$\alpha$ break. 
\end{itemize}

The \renaissance Simulations are an important tool that provides insight into the global properties, morphologies, and photometry of the progenitors $z > 10$ JWST galaxies. In the wake of JWST and other upcoming telescopes, our mock observations presents a timely database of high redshift, high resolution zoom-in simulation predictions to guide future observational campaigns and further validate our simulations against JWST. 

\section{Data Accessibility}
The data described in this paper and further information on the \renaissance Simulations is permanently available at \url{https://www.firstgalaxies.physics.gatech.edu/}. 
\begin{acknowledgments}
We acknowledge funding support from NSF grants AST-2108020 and AST-2510197 and NASA grant 80NSSC21K1053. The analysis for this work was performed on the Phoenix cluster within Georgia Tech's Partnership for an Advanced Computing Environment (PACE). The \textit{Renaissance} simulations were performed on Blue Waters operated by the National Center for Supercomputing Applications (NCSA) with PRAC allocation support by the NSF (awards ACI-0832662, ACI-1238993, and ACI-1514580). This research is part of the Blue Waters sustained-petascale computing project, which is supported by the NSF (awards OCI-0725070 and ACI-1238993) and the state of Illinois. Blue Waters was a joint effort of the University of Illinois at Urbana-Champaign and its NCSA. The freely available astrophysical analysis code \texttt{yt} \citep{yt} and plotting library \texttt{matplotlib} \citep{matplotlib} were used to construct numerous plots within this paper. Computations described in this work were performed using the publicly available \texttt{Enzo} code, which is the product of a collaborative effort of many independent scientists from numerous institutions around the world.
\end{acknowledgments}

\bibliography{sample701}{}

@article{Keller_2023,
	doi = {10.3847/2041-8213/acb148},
  
	url = {https://doi.org/10.3847%2F2041-8213%2Facb148},
  
	year = 2023,
	month = {feb},
  
	publisher = {American Astronomical Society},
  
	volume = {943},
  
	number = {2},
  
	pages = {L28},
  
	author = {B. W. Keller and F. Munshi and M. Trebitsch and M. Tremmel},
  
	title = {Can Cosmological Simulations Reproduce the Spectroscopically Confirmed Galaxies Seen at z $\geq$ 10?},
  
	journal = {The Astrophysical Journal Letters}
}

@ARTICLE{mccaffrey2023tension,
       author = {{McCaffrey}, Joe and {Hardin}, Samantha and {Wise}, John H. and {Regan}, John A.},
        title = "{No Tension: JWST Galaxies at z > 10 Consistent with Cosmological Simulations}",
      journal = {The Open Journal of Astrophysics},
         year = 2023,
        month = sep,
       volume = {6},
          eid = {47},
        pages = {47},
          doi = {10.21105/astro.2304.13755},
archivePrefix = {arXiv},
       eprint = {2304.13755},
 primaryClass = {astro-ph.GA},
       adsurl = {https://ui.adsabs.harvard.edu/abs/2023OJAp....6E..47M}
}

@article{Narayanan_2021,
	doi = {10.3847/1538-4365/abc487},
  
	url = {https://doi.org/10.3847%2F1538-4365%2Fabc487},
  
	year = 2021,
	month = {jan},
  
	publisher = {American Astronomical Society},
  
	volume = {252},
  
	number = {1},
  
	pages = {12},
  
	author = {Desika Narayanan and Matthew J. Turk and Thomas Robitaille and Ashley J. Kelly and B. Connor McClellan and Ray S Sharma and Prerak Garg and Matthew Abruzzo and Ena Choi and Charlie Conroy and Benjamin D. Johnson and Benjamin Kimock and Qi Li and Christopher C. Lovell and Sidney Lower and George C. Privon and Jonathan Roberts and Snigdaa Sethuram and Gregory F. Snyder and Robert Thompson and John H. Wise},
  
	title = {powderday: Dust Radiative Transfer for Galaxy Simulations},
  
	journal = {The Astrophysical Journal Supplement Series}
}

@ARTICLE{eisenstein2023overview,
       author = {{Eisenstein}, Daniel J. and {Willott}, Chris and {Alberts}, Stacey and {Arribas}, Santiago and {Bonaventura}, Nina and {Bunker}, Andrew J. and {Cameron}, Alex J. and {Carniani}, Stefano and {Charlot}, Stephane and {Curtis-Lake}, Emma and {D'Eugenio}, Francesco and {Ferruit}, Pierre and {Giardino}, Giovanna and {Hainline}, Kevin and {Hausen}, Ryan and {Jakobsen}, Peter and {Johnson}, Benjamin D. and {Maiolino}, Roberto and {Rauscher}, Bernard J. and {Rieke}, Marcia and {Rieke}, George and {Rix}, Hans-Walter and {Robertson}, Brant and {Stark}, Daniel P. and {Tacchella}, Sandro and {Williams}, Christina C. and {Willmer}, Christopher N.~A. and {Baker}, William M. and {Baum}, Stefi and {Bhatawdekar}, Rachana and {Boyett}, Kristan and {Chen}, Zuyi and {Chevallard}, Jacopo and {Circosta}, Chiara and {Curti}, Mirko and {Danhaive}, A. Lola and {DeCoursey}, Christa and {Endsley}, Ryan and {de Graaff}, Anna and {Dressler}, Alan and {Egami}, Eiichi and {Helton}, Jakob M. and {Hviding}, Raphael E. and {Ji}, Zhiyuan and {Jones}, Gareth C. and {Kumari}, Nimisha and {L{\"u}tzgendorf}, Nora and {Laseter}, Isaac and {Looser}, Tobias J. and {Lyu}, Jianwei and {Maseda}, Michael V. and {Nelson}, Erica and {Parlanti}, Eleonora and {Perna}, Michele and {Pusk{\'a}s}, D{\'a}vid and {Rawle}, Tim and {Rodr{\'\i}guez Del Pino}, Bruno and {Rujopakarn}, Wiphu and {Sandles}, Lester and {Saxena}, Aayush and {Scholtz}, Jan and {Sharpe}, Katherine and {Shivaei}, Irene and {Silcock}, Maddie S. and {Simmonds}, Charlotte and {Skarbinski}, Maya and {Smit}, Renske and {Stone}, Meredith and {Suess}, Katherine A. and {Sun}, Fengwu and {Tang}, Mengtao and {Topping}, Michael W. and {{\"U}bler}, Hannah and {Villanueva}, Natalia C. and {Wallace}, Imaan E.~B. and {Whitler}, Lily and {Witstok}, Joris and {Woodrum}, Charity},
        title = "{Overview of the JWST Advanced Deep Extragalactic Survey (JADES)}",
      journal = {\apjs},
         year = 2026,
        month = mar,
       volume = {283},
       number = {1},
          eid = {6},
        pages = {6},
          doi = {10.3847/1538-4365/ae3163},
archivePrefix = {arXiv},
       eprint = {2306.02465},
 primaryClass = {astro-ph.GA},
       adsurl = {https://ui.adsabs.harvard.edu/abs/2026ApJS..283....6E}
}

@ARTICLE{Xu_2016,
       author = {{Xu}, Hao and {Wise}, John H. and {Norman}, Michael L. and {Ahn}, Kyungjin and {O'Shea}, Brian W.},
        title = "{Galaxy Properties and UV Escape Fractions during the Epoch of Reionization: Results from the Renaissance Simulations}",
      journal = {\apj},
         year = 2016,
        month = dec,
       volume = {833},
       number = {1},
          eid = {84},
        pages = {84},
          doi = {10.3847/1538-4357/833/1/84},
archivePrefix = {arXiv},
       eprint = {1604.07842},
 primaryClass = {astro-ph.GA},
       adsurl = {https://ui.adsabs.harvard.edu/abs/2016ApJ...833...84X}
}

@ARTICLE{oshea2015probing,
       author = {{O'Shea}, Brian W. and {Wise}, John H. and {Xu}, Hao and {Norman}, Michael L.},
        title = "{Probing the Ultraviolet Luminosity Function of the Earliest Galaxies with the Renaissance Simulations}",
      journal = {\apjl},
         year = 2015,
        month = jul,
       volume = {807},
       number = {1},
          eid = {L12},
        pages = {L12},
          doi = {10.1088/2041-8205/807/1/L12},
archivePrefix = {arXiv},
       eprint = {1503.01110},
 primaryClass = {astro-ph.GA},
       adsurl = {https://ui.adsabs.harvard.edu/abs/2015ApJ...807L..12O}
}

@ARTICLE{Bunker_2023,
       author = {{Bunker}, Andrew J. and {Saxena}, Aayush and {Cameron}, Alex J. and {Willott}, Chris J. and {Curtis-Lake}, Emma and {Jakobsen}, Peter and {Carniani}, Stefano and {Smit}, Renske and {Maiolino}, Roberto and {Witstok}, Joris and {Curti}, Mirko and {D'Eugenio}, Francesco and {Jones}, Gareth C. and {Ferruit}, Pierre and {Arribas}, Santiago and {Charlot}, Stephane and {Chevallard}, Jacopo and {Giardino}, Giovanna and {de Graaff}, Anna and {Looser}, Tobias J. and {L{\"u}tzgendorf}, Nora and {Maseda}, Michael V. and {Rawle}, Tim and {Rix}, Hans-Walter and {Del Pino}, Bruno Rodr{\'\i}guez and {Alberts}, Stacey and {Egami}, Eiichi and {Eisenstein}, Daniel J. and {Endsley}, Ryan and {Hainline}, Kevin and {Hausen}, Ryan and {Johnson}, Benjamin D. and {Rieke}, George and {Rieke}, Marcia and {Robertson}, Brant E. and {Shivaei}, Irene and {Stark}, Daniel P. and {Sun}, Fengwu and {Tacchella}, Sandro and {Tang}, Mengtao and {Williams}, Christina C. and {Willmer}, Christopher N.~A. and {Baker}, William M. and {Baum}, Stefi and {Bhatawdekar}, Rachana and {Bowler}, Rebecca and {Boyett}, Kristan and {Chen}, Zuyi and {Circosta}, Chiara and {Helton}, Jakob M. and {Ji}, Zhiyuan and {Kumari}, Nimisha and {Lyu}, Jianwei and {Nelson}, Erica and {Parlanti}, Eleonora and {Perna}, Michele and {Sandles}, Lester and {Scholtz}, Jan and {Suess}, Katherine A. and {Topping}, Michael W. and {{\"U}bler}, Hannah and {Wallace}, Imaan E.~B. and {Whitler}, Lily},
        title = "{JADES NIRSpec Spectroscopy of GN-z11: Lyman-{\ensuremath{\alpha}} emission and possible enhanced nitrogen abundance in a z = 10.60 luminous galaxy}",
      journal = {\aap},
         year = 2023,
        month = sep,
       volume = {677},
          eid = {A88},
        pages = {A88},
          doi = {10.1051/0004-6361/202346159},
archivePrefix = {arXiv},
       eprint = {2302.07256},
 primaryClass = {astro-ph.GA},
       adsurl = {https://ui.adsabs.harvard.edu/abs/2023A&A...677A..88B}
}

@ARTICLE{kennicutt,
       author = {{Kennicutt}, Robert C., Jr.},
        title = "{Star Formation in Galaxies Along the Hubble Sequence}",
      journal = {\araa},
         year = 1998,
        month = jan,
       volume = {36},
        pages = {189-232},
          doi = {10.1146/annurev.astro.36.1.189},
archivePrefix = {arXiv},
       eprint = {astro-ph/9807187},
 primaryClass = {astro-ph},
       adsurl = {https://ui.adsabs.harvard.edu/abs/1998ARA&A..36..189K}
}

@ARTICLE{katz,
       author = {{Katz}, Harley and {Kimm}, Taysun and {Ellis}, Richard S. and {Devriendt}, Julien and {Slyz}, Adrianne},
        title = "{The challenges of identifying Population III stars in the early Universe}",
      journal = {\mnras},
         year = 2023,
        month = sep,
       volume = {524},
       number = {1},
        pages = {351-360},
          doi = {10.1093/mnras/stad1903},
archivePrefix = {arXiv},
       eprint = {2207.04751},
 primaryClass = {astro-ph.GA},
       adsurl = {https://ui.adsabs.harvard.edu/abs/2023MNRAS.524..351K}
}

@ARTICLE{adamo2024billionyearsaccordingjwst,
       author = {{Adamo}, Angela and {Atek}, Hakim and {Bagley}, Micaela B. and {Ba{\~n}ados}, Eduardo and {Barrow}, Kirk S.~S. and {Berg}, Danielle A. and {Bezanson}, Rachel and {Brada{\v{c}}}, Maru{\v{s}}a and {Brammer}, Gabriel and {Carnall}, Adam C. and {Chisholm}, John and {Coe}, Dan and {Dayal}, Pratika and {Eisenstein}, Daniel J. and {Eldridge}, Jan J. and {Ferrara}, Andrea and {Fujimoto}, Seiji and {Graaff}, Anna de and {Habouzit}, Melanie and {Hutchison}, Taylor A. and {Kartaltepe}, Jeyhan S. and {Kassin}, Susan A. and {Kriek}, Mariska and {Labb{\'e}}, Ivo and {Maiolino}, Roberto and {Marques-Chaves}, Rui and {Maseda}, Michael V. and {Mason}, Charlotte and {Matthee}, Jorryt and {McQuinn}, Kristen B.~W. and {Meynet}, Georges and {Naidu}, Rohan P. and {Oesch}, Pascal A. and {Pentericci}, Laura and {P{\'e}rez-Gonz{\'a}lez}, Pablo G. and {Rigby}, Jane R. and {Roberts-Borsani}, Guido and {Schaerer}, Daniel and {Shapley}, Alice E. and {Stark}, Daniel P. and {Stiavelli}, Massimo and {Strom}, Allison L. and {Vanzella}, Eros and {Wang}, Feige and {Wilkins}, Stephen M. and {Williams}, Christina C. and {Willott}, Chris J. and {Wylezalek}, Dominika and {Nota}, Antonella},
        title = "{The first billion years according to JWST}",
      journal = {Nature Astronomy},
         year = 2025,
        month = aug,
       volume = {9},
        pages = {1134-1147},
          doi = {10.1038/s41550-025-02624-5},
archivePrefix = {arXiv},
       eprint = {2405.21054},
 primaryClass = {astro-ph.GA},
       adsurl = {https://ui.adsabs.harvard.edu/abs/2025NatAs...9.1134A}
}

@ARTICLE{sphinx,
       author = {{Katz}, Harley and {Rosdahl}, Joki and {Kimm}, Taysun and {Blaizot}, Jeremy and {Choustikov}, Nicholas and {Farcy}, Marion and {Garel}, Thibault and {Haehnelt}, Martin G. and {Michel-Dansac}, Leo and {Ocvirk}, Pierre},
        title = "{The SPHINX Public Data Release: Forward Modelling High-Redshift JWST Observations with Cosmological Radiation Hydrodynamics Simulations}",
      journal = {The Open Journal of Astrophysics},
         year = 2023,
        month = dec,
       volume = {6},
          eid = {44},
        pages = {44},
          doi = {10.21105/astro.2309.03269},
archivePrefix = {arXiv},
       eprint = {2309.03269},
 primaryClass = {astro-ph.GA},
       adsurl = {https://ui.adsabs.harvard.edu/abs/2023OJAp....6E..44K}
}

@ARTICLE{thesan,
       author = {{Garaldi}, Enrico and {Kannan}, Rahul and {Smith}, Aaron and {Borrow}, Josh and {Vogelsberger}, Mark and {Pakmor}, R{\"u}diger and {Springel}, Volker and {Hernquist}, Lars and {Gal{\'a}rraga-Espinosa}, Daniela and {Yeh}, Jessica Y.-C. and {Shen}, Xuejian and {Xu}, Clara and {Neyer}, Meredith and {Spina}, Benedetta and {Almualla}, Mouza and {Zhao}, Yu},
        title = "{The THESAN project: public data release of radiation-hydrodynamic simulations matching reionization-era JWST observations}",
      journal = {\mnras},
         year = 2024,
        month = jun,
       volume = {530},
       number = {4},
        pages = {3765-3786},
          doi = {10.1093/mnras/stae839},
archivePrefix = {arXiv},
       eprint = {2309.06475},
 primaryClass = {astro-ph.CO},
       adsurl = {https://ui.adsabs.harvard.edu/abs/2024MNRAS.530.3765G}
}

@ARTICLE{hyperion,
       author = {{Robitaille}, T.~P.},
        title = "{HYPERION: an open-source parallelized three-dimensional dust continuum radiative transfer code}",
      journal = {\aap},
         year = 2011,
        month = dec,
       volume = {536},
          eid = {A79},
        pages = {A79},
          doi = {10.1051/0004-6361/201117150},
archivePrefix = {arXiv},
       eprint = {1112.1071},
 primaryClass = {astro-ph.IM},
       adsurl = {https://ui.adsabs.harvard.edu/abs/2011A&A...536A..79R}
}

@ARTICLE{conroy_gunn_2010,
       author = {{Conroy}, Charlie and {Gunn}, James E.},
        title = "{The Propagation of Uncertainties in Stellar Population Synthesis Modeling. III. Model Calibration, Comparison, and Evaluation}",
      journal = {\apj},
         year = 2010,
        month = apr,
       volume = {712},
       number = {2},
        pages = {833-857},
          doi = {10.1088/0004-637X/712/2/833},
archivePrefix = {arXiv},
       eprint = {0911.3151},
 primaryClass = {astro-ph.CO},
       adsurl = {https://ui.adsabs.harvard.edu/abs/2010ApJ...712..833C}
}

@ARTICLE{conroy_gunn_2009,
       author = {{Conroy}, Charlie and {Gunn}, James E. and {White}, Martin},
        title = "{The Propagation of Uncertainties in Stellar Population Synthesis Modeling. I. The Relevance of Uncertain Aspects of Stellar Evolution and the Initial Mass Function to the Derived Physical Properties of Galaxies}",
      journal = {\apj},
         year = 2009,
        month = jul,
       volume = {699},
       number = {1},
        pages = {486-506},
          doi = {10.1088/0004-637X/699/1/486},
archivePrefix = {arXiv},
       eprint = {0809.4261},
 primaryClass = {astro-ph},
       adsurl = {https://ui.adsabs.harvard.edu/abs/2009ApJ...699..486C}
}

@ARTICLE{cloudy,
       author = {{Byler}, Nell and {Dalcanton}, Julianne J. and {Conroy}, Charlie and {Johnson}, Benjamin D.},
        title = "{Nebular Continuum and Line Emission in Stellar Population Synthesis Models}",
      journal = {\apj},
         year = 2017,
        month = may,
       volume = {840},
       number = {1},
          eid = {44},
        pages = {44},
          doi = {10.3847/1538-4357/aa6c66},
archivePrefix = {arXiv},
       eprint = {1611.08305},
 primaryClass = {astro-ph.GA},
       adsurl = {https://ui.adsabs.harvard.edu/abs/2017ApJ...840...44B}
}

@ARTICLE{wiseandabel,
       author = {{Wise}, John H. and {Abel}, Tom},
        title = "{ENZO+MORAY: radiation hydrodynamics adaptive mesh refinement simulations with adaptive ray tracing}",
      journal = {\mnras},
         year = 2011,
        month = jul,
       volume = {414},
       number = {4},
        pages = {3458-3491},
          doi = {10.1111/j.1365-2966.2011.18646.x},
archivePrefix = {arXiv},
       eprint = {1012.2865},
 primaryClass = {astro-ph.IM},
       adsurl = {https://ui.adsabs.harvard.edu/abs/2011MNRAS.414.3458W}
}

@ARTICLE{schaerer_2002,
       author = {{Schaerer}, D.},
        title = "{On the properties of massive Population III stars and metal-free stellar populations}",
      journal = {\aap},
         year = 2002,
        month = jan,
       volume = {382},
        pages = {28-42},
          doi = {10.1051/0004-6361:20011619},
archivePrefix = {arXiv},
       eprint = {astro-ph/0110697},
 primaryClass = {astro-ph},
       adsurl = {https://ui.adsabs.harvard.edu/abs/2002A&A...382...28S}
}

@ARTICLE{conroy_2009,
       author = {{Conroy}, Charlie and {Gunn}, James E. and {White}, Martin},
        title = "{The Propagation of Uncertainties in Stellar Population Synthesis Modeling. I. The Relevance of Uncertain Aspects of Stellar Evolution and the Initial Mass Function to the Derived Physical Properties of Galaxies}",
      journal = {\apj},
         year = 2009,
        month = jul,
       volume = {699},
       number = {1},
        pages = {486-506},
          doi = {10.1088/0004-637X/699/1/486},
archivePrefix = {arXiv},
       eprint = {0809.4261},
 primaryClass = {astro-ph},
       adsurl = {https://ui.adsabs.harvard.edu/abs/2009ApJ...699..486C}
}

@ARTICLE{vogelsberger,
       author = {{Vogelsberger}, Mark and {Nelson}, Dylan and {Pillepich}, Annalisa and {Shen}, Xuejian and {Marinacci}, Federico and {Springel}, Volker and {Pakmor}, R{\"u}diger and {Tacchella}, Sandro and {Weinberger}, Rainer and {Torrey}, Paul and {Hernquist}, Lars},
        title = "{High-redshift JWST predictions from IllustrisTNG: dust modelling and galaxy luminosity functions}",
      journal = {\mnras},
         year = 2020,
        month = mar,
       volume = {492},
       number = {4},
        pages = {5167-5201},
          doi = {10.1093/mnras/staa137},
archivePrefix = {arXiv},
       eprint = {1904.07238},
 primaryClass = {astro-ph.GA},
       adsurl = {https://ui.adsabs.harvard.edu/abs/2020MNRAS.492.5167V}
}

@ARTICLE{heger,
       author = {{Heger}, A. and {Fryer}, C.~L. and {Woosley}, S.~E. and {Langer}, N. and {Hartmann}, D.~H.},
        title = "{How Massive Single Stars End Their Life}",
      journal = {\apj},
         year = 2003,
        month = jul,
       volume = {591},
       number = {1},
        pages = {288-300},
          doi = {10.1086/375341},
archivePrefix = {arXiv},
       eprint = {astro-ph/0212469},
 primaryClass = {astro-ph},
       adsurl = {https://ui.adsabs.harvard.edu/abs/2003ApJ...591..288H}
}

@ARTICLE{fake_news,
       author = {{Haslbauer}, Moritz and {Kroupa}, Pavel and {Zonoozi}, Akram Hasani and {Haghi}, Hosein},
        title = "{Has JWST Already Falsified Dark-matter-driven Galaxy Formation?}",
      journal = {\apjl},
         year = 2022,
        month = nov,
       volume = {939},
       number = {2},
          eid = {L31},
        pages = {L31},
          doi = {10.3847/2041-8213/ac9a50},
archivePrefix = {arXiv},
       eprint = {2210.14915},
 primaryClass = {astro-ph.GA},
       adsurl = {https://ui.adsabs.harvard.edu/abs/2022ApJ...939L..31H}
}

@ARTICLE{boylan,
       author = {{Boylan-Kolchin}, Michael},
        title = "{Stress testing {\ensuremath{\Lambda}}CDM with high-redshift galaxy candidates}",
      journal = {Nature Astronomy},
         year = 2023,
        month = jun,
       volume = {7},
        pages = {731-735},
          doi = {10.1038/s41550-023-01937-7},
archivePrefix = {arXiv},
       eprint = {2208.01611},
 primaryClass = {astro-ph.CO},
       adsurl = {https://ui.adsabs.harvard.edu/abs/2023NatAs...7..731B}
}

@ARTICLE{astrid,
       author = {{Ni}, Yueying and {Di Matteo}, Tiziana and {Bird}, Simeon and {Croft}, Rupert and {Feng}, Yu and {Chen}, Nianyi and {Tremmel}, Michael and {DeGraf}, Colin and {Li}, Yin},
        title = "{The ASTRID simulation: the evolution of supermassive black holes}",
      journal = {\mnras},
         year = 2022,
        month = jun,
       volume = {513},
       number = {1},
        pages = {670-692},
          doi = {10.1093/mnras/stac351},
archivePrefix = {arXiv},
       eprint = {2110.14154},
 primaryClass = {astro-ph.GA},
       adsurl = {https://ui.adsabs.harvard.edu/abs/2022MNRAS.513..670N}
}

@article{Carniani_2024,
   title={Spectroscopic confirmation of two luminous galaxies at a redshift of 14},
   volume={633},
   ISSN={1476-4687},
   url={http://dx.doi.org/10.1038/s41586-024-07860-9},
   DOI={10.1038/s41586-024-07860-9},
   number={8029},
   journal={Nature},
   publisher={Springer Science and Business Media LLC},
   author={Carniani, Stefano and Hainline, Kevin and D’Eugenio, Francesco and Eisenstein, Daniel J. and Jakobsen, Peter and Witstok, Joris and Johnson, Benjamin D. and Chevallard, Jacopo and Maiolino, Roberto and Helton, Jakob M. and Willott, Chris and Robertson, Brant and Alberts, Stacey and Arribas, Santiago and Baker, William M. and Bhatawdekar, Rachana and Boyett, Kristan and Bunker, Andrew J. and Cameron, Alex J. and Cargile, Phillip A. and Charlot, Stéphane and Curti, Mirko and Curtis-Lake, Emma and Egami, Eiichi and Giardino, Giovanna and Isaak, Kate and Ji, Zhiyuan and Jones, Gareth C. and Kumari, Nimisha and Maseda, Michael V. and Parlanti, Eleonora and Pérez-González, Pablo G. and Rawle, Tim and Rieke, George and Rieke, Marcia and Del Pino, Bruno Rodríguez and Saxena, Aayush and Scholtz, Jan and Smit, Renske and Sun, Fengwu and Tacchella, Sandro and Übler, Hannah and Venturi, Giacomo and Williams, Christina C. and Willmer, Christopher N. A.},
   year={2024},
   month=jul, pages={318–322} }

@article{Harvey_2024,
   title={EPOCHS. IV. SED Modeling Assumptions and Their Impact on the Stellar Mass Function at 6.5 ≤ z ≤ 13.5 Using PEARLS and Public JWST Observations},
   volume={978},
   ISSN={1538-4357},
   url={http://dx.doi.org/10.3847/1538-4357/ad8c29},
   DOI={10.3847/1538-4357/ad8c29},
   number={1},
   journal={The Astrophysical Journal},
   publisher={American Astronomical Society},
   author={Harvey, Thomas and Conselice, Christopher J. and Adams, Nathan J. and Austin, Duncan and Juodžbalis, Ignas and Trussler, James and Li, Qiong and Ormerod, Katherine and Ferreira, Leonardo and Lovell, Christopher C. and Duan, Qiao and Westcott, Lewi and Harris, Honor and Bhatawdekar, Rachana and Coe, Dan and Cohen, Seth H. and Caruana, Joseph and Cheng, Cheng and Driver, Simon P. and Frye, Brenda and Furtak, Lukas J. and Grogin, Norman A. and Hathi, Nimish P. and Holwerda, Benne W. and Jansen, Rolf A. and Koekemoer, Anton M. and Marshall, Madeline A. and Nonino, Mario and Vijayan, Aswin P. and Wilkins, Stephen M. and Windhorst, Rogier and Willmer, Christopher N. A. and Yan, Haojing and Zitrin, Adi},
   year={2024},
   month=dec, pages={89} }

@article{Weibel_2024,
   title={Galaxy build-up in the first 1.5 Gyr of cosmic history: insights from the stellar mass function at z ~ 4–9 from JWST NIRCam observations},
   volume={533},
   ISSN={1365-2966},
   url={http://dx.doi.org/10.1093/mnras/stae1891},
   DOI={10.1093/mnras/stae1891},
   number={2},
   journal={Monthly Notices of the Royal Astronomical Society},
   publisher={Oxford University Press (OUP)},
   author={Weibel, Andrea and Oesch, Pascal A and Barrufet, Laia and Gottumukkala, Rashmi and Ellis, Richard S and Santini, Paola and Weaver, John R and Allen, Natalie and Bouwens, Rychard and Bowler, Rebecca A A and Brammer, Gabe and Carnall, Adam C and Cullen, Fergus and Dayal, Pratika and Dickinson, Mark and Donnan, Callum T and Dunlop, James S and Giavalisco, Mauro and Grogin, Norman A and Illingworth, Garth D and Koekemoer, Anton M and Labbe, Ivo and Marchesini, Danilo and McLeod, Derek J and McLure, Ross J and Naidu, Rohan P and Pérez-González, Pablo G and Shuntov, Marko and Stefanon, Mauro and Toft, Sune and Xiao, Mengyuan},
   year={2024},
   month=aug, pages={1808–1838} }

@ARTICLE{stanway,
       author = {{Stanway}, Elizabeth R. and {Eldridge}, J.~J.},
        title = "{Exploring the impact of IMF and binary parameter stochasticity with a binary population synthesis code}",
      journal = {\mnras},
         year = 2023,
        month = jul,
       volume = {522},
       number = {3},
        pages = {4430-4443},
          doi = {10.1093/mnras/stad1185},
archivePrefix = {arXiv},
       eprint = {2304.09549},
 primaryClass = {astro-ph.GA},
       adsurl = {https://ui.adsabs.harvard.edu/abs/2023MNRAS.522.4430S}
}

@inproceedings{Pickering_2016,
   title={Pandeia: a multi-mission exposure time calculator for JWST and WFIRST},
   url={http://dx.doi.org/10.1117/12.2231768},
   DOI={10.1117/12.2231768},
   booktitle={Observatory Operations: Strategies, Processes, and Systems VI},
   publisher={SPIE},
   author={Pickering, Timothy E. and Pontoppidan, Klaus M. and Laidler, Victoria G. and Sontag, Christopher D. and Robberto, Massimo and Karakla, Diane M. and Hanley, Christopher and Gilbert, Karoline and Slocum, Christine and Earl, Nicholas M. and Pueyo, Laurent and Ravindranath, Swara and Noriega-Crespo, Alberto and Barker, Elizabeth and Sienkiewicz, Mark J.},
   editor={Peck, Alison B. and Benn, Chris R. and Seaman, Robert L.},
   year={2016},
   month=jul, pages={44} }

@ARTICLE{bryan_enzo_2014,
       author = {{Bryan}, Greg L. and {Norman}, Michael L. and {O'Shea}, Brian W. and {Abel}, Tom and {Wise}, John H. and {Turk}, Matthew J. and {Reynolds}, Daniel R. and {Collins}, David C. and {Wang}, Peng and {Skillman}, Samuel W. and {Smith}, Britton and {Harkness}, Robert P. and {Bordner}, James and {Kim}, Ji-hoon and {Kuhlen}, Michael and {Xu}, Hao and {Goldbaum}, Nathan and {Hummels}, Cameron and {Kritsuk}, Alexei G. and {Tasker}, Elizabeth and {Skory}, Stephen and {Simpson}, Christine M. and {Hahn}, Oliver and {Oishi}, Jeffrey S. and {So}, Geoffrey C. and {Zhao}, Fen and {Cen}, Renyue and {Li}, Yuan and {Enzo Collaboration}},
        title = "{ENZO: An Adaptive Mesh Refinement Code for Astrophysics}",
      journal = {\apjs},
         year = 2014,
        month = apr,
       volume = {211},
       number = {2},
          eid = {19},
        pages = {19},
          doi = {10.1088/0067-0049/211/2/19},
archivePrefix = {arXiv},
       eprint = {1307.2265},
 primaryClass = {astro-ph.IM},
       adsurl = {https://ui.adsabs.harvard.edu/abs/2014ApJS..211...19B}
}

@article{Brummel-Smith2019, doi = {10.21105/joss.01636}, url = {https://doi.org/10.21105/joss.01636}, year = {2019}, publisher = {The Open Journal}, volume = {4}, number = {42}, pages = {1636}, author = {Corey Brummel-Smith and Greg Bryan and Iryna Butsky and Lauren Corlies and Andrew Emerick and John Forbes and Yusuke Fujimoto and Nathan J. Goldbaum and Philipp Grete and Cameron B. Hummels and Ji-hoon Kim and Daegene Koh and Miao Li and Yuan Li and Xinyu Li and Brian OShea and Molly S. Peeples and John A. Regan and Munier Salem and Wolfram Schmidt and Christine M. Simpson and Britton D. Smith and Jason Tumlinson and Matthew J. Turk and John H. Wise and Tom Abel and James Bordner and Renyue Cen and David C. Collins and Brian Crosby and Philipp Edelmann and Oliver Hahn and Robert Harkness and Elizabeth Harper-Clark and Shuo Kong and Alexei G. Kritsuk and Michael Kuhlen and James Larrue and Eve Lee and Greg Meece and Michael L. Norman and Jeffrey S. Oishi and Pascal Paschos and Carolyn Peruta and Alex Razoumov and Daniel R. Reynolds and Devin Silvia and Samuel W. Skillman and Stephen Skory and Geoffrey C. So and Elizabeth Tasker and Rick Wagner and Peng Wang and Hao Xu and Fen Zhao}, title = {ENZO: An Adaptive Mesh Refinement Code for Astrophysics (Version 2.6)}, journal = {Journal of Open Source Software} }

@ARTICLE{naidu2025cosmicmiracleremarkablyluminous,
       author = {{Naidu}, Rohan P. and {Oesch}, Pascal A. and {Brammer}, Gabriel and {Weibel}, Andrea and {Li}, Yijia and {Matthee}, Jorryt and {Chisolm}, John and {Pollock}, Clara L. and {Heintz}, Kasper E. and {Johnson}, Benjamin D. and {Shen}, Xuejian and {Hviding}, Raphael E. and {Leja}, Joel and {Tacchella}, Sandro and {Ganguly}, Arpita and {Witten}, Callum and {Atek}, Hakim and {Belli}, Siro and {Bose}, Sownak and {Bouwens}, Rychard and {Dayal}, Pratika and {Decarli}, Roberto and {de Graaff}, Anna and {Fudamoto}, Yoshinobu and {Giovinazzo}, Emma and {Greene}, Jenny E. and {Illingworth}, Garth and {Inoue}, Akio K. and {Kane}, Sarah G. and {Labbe}, Ivo and {Leonova}, Ecaterina and {Marques-Chaves}, Rui and {Meyer}, Roman A. and {Nelson}, Erica J. and {Roberts-Borsani}, Guido and {Schaerer}, Daniel and {Simcoe}, Robert A. and {Stefanon}, Mauro and {Sugahara}, Yuma and {Toft}, Sune and {van der Wel}, Arjen and {van Dokkum}, Pieter and {Walter}, Fabian and {Watson}, Darrach and {Weaver}, John R. and {Whitaker}, Katherine E.},
        title = "{A Cosmic Miracle: A Remarkably Luminous Galaxy at zspec = 14.44 Confirmed with JWST}",
      journal = {The Open Journal of Astrophysics},
         year = 2026,
        month = jan,
       volume = {9},
        pages = {56033},
          doi = {10.33232/001c.156033},
archivePrefix = {arXiv},
       eprint = {2505.11263},
 primaryClass = {astro-ph.GA},
       adsurl = {https://ui.adsabs.harvard.edu/abs/2026OJAp....956033N}
}

@ARTICLE{oesch_gnz11_hubble,
       author = {{Oesch}, P.~A. and {Brammer}, G. and {van Dokkum}, P.~G. and {Illingworth}, G.~D. and {Bouwens}, R.~J. and {Labb{\'e}}, I. and {Franx}, M. and {Momcheva}, I. and {Ashby}, M.~L.~N. and {Fazio}, G.~G. and {Gonzalez}, V. and {Holden}, B. and {Magee}, D. and {Skelton}, R.~E. and {Smit}, R. and {Spitler}, L.~R. and {Trenti}, M. and {Willner}, S.~P.},
        title = "{A Remarkably Luminous Galaxy at z=11.1 Measured with Hubble Space Telescope Grism Spectroscopy}",
      journal = {\apj},
         year = 2016,
        month = mar,
       volume = {819},
       number = {2},
          eid = {129},
        pages = {129},
          doi = {10.3847/0004-637X/819/2/129},
archivePrefix = {arXiv},
       eprint = {1603.00461},
 primaryClass = {astro-ph.GA},
       adsurl = {https://ui.adsabs.harvard.edu/abs/2016ApJ...819..129O}
}

@INPROCEEDINGS{JADES_overview,
       author = {{Bunker}, Andrew J. and {NIRSPEC Instrument Science Team} and {JAESs Collaboration}},
        title = "{Spectroscopy with the JWST Advanced Deep Extragalactic Survey (JADES) - the NIRSpec/NIRCAM GTO galaxy evolution project}",
    booktitle = {Uncovering Early Galaxy Evolution in the ALMA and JWST Era},
         year = 2020,
       editor = {{da Cunha}, Elisabete and {Hodge}, Jacqueline and {Afonso}, Jos{\'e} and {Pentericci}, Laura and {Sobral}, David},
       series = {IAU Symposium},
       volume = {352},
        month = jan,
        pages = {342-346},
          doi = {10.1017/S1743921319009463},
archivePrefix = {arXiv},
       eprint = {2112.15207},
 primaryClass = {astro-ph.GA},
       adsurl = {https://ui.adsabs.harvard.edu/abs/2020IAUS..352..342B}
}

@ARTICLE{CEERS_overview,
       author = {{Finkelstein}, Steven L. and {Bagley}, Micaela B. and {Ferguson}, Henry C. and {Wilkins}, Stephen M. and {Kartaltepe}, Jeyhan S. and {Papovich}, Casey and {Yung}, L.~Y. Aaron and {Arrabal Haro}, Pablo and {Behroozi}, Peter and {Dickinson}, Mark and {Kocevski}, Dale D. and {Koekemoer}, Anton M. and {Larson}, Rebecca L. and {Le Bail}, Aur{\'e}lien and {Morales}, Alexa M. and {P{\'e}rez-Gonz{\'a}lez}, Pablo G. and {Burgarella}, Denis and {Dav{\'e}}, Romeel and {Hirschmann}, Michaela and {Somerville}, Rachel S. and {Wuyts}, Stijn and {Bromm}, Volker and {Casey}, Caitlin M. and {Fontana}, Adriano and {Fujimoto}, Seiji and {Gardner}, Jonathan P. and {Giavalisco}, Mauro and {Grazian}, Andrea and {Grogin}, Norman A. and {Hathi}, Nimish P. and {Hutchison}, Taylor A. and {Jha}, Saurabh W. and {Jogee}, Shardha and {Kewley}, Lisa J. and {Kirkpatrick}, Allison and {Long}, Arianna S. and {Lotz}, Jennifer M. and {Pentericci}, Laura and {Pierel}, Justin D.~R. and {Pirzkal}, Nor and {Ravindranath}, Swara and {Ryan}, Russell E. and {Trump}, Jonathan R. and {Yang}, Guang and {Bhatawdekar}, Rachana and {Bisigello}, Laura and {Buat}, V{\'e}ronique and {Calabr{\`o}}, Antonello and {Castellano}, Marco and {Cleri}, Nikko J. and {Cooper}, M.~C. and {Croton}, Darren and {Daddi}, Emanuele and {Dekel}, Avishai and {Elbaz}, David and {Franco}, Maximilien and {Gawiser}, Eric and {Holwerda}, Benne W. and {Huertas-Company}, Marc and {Jaskot}, Anne E. and {Leung}, Gene C.~K. and {Lucas}, Ray A. and {Mobasher}, Bahram and {Pandya}, Viraj and {Tacchella}, Sandro and {Weiner}, Benjamin J. and {Zavala}, Jorge A.},
        title = "{CEERS Key Paper. I. An Early Look into the First 500 Myr of Galaxy Formation with JWST}",
      journal = {\apjl},
         year = 2023,
        month = mar,
       volume = {946},
       number = {1},
          eid = {L13},
        pages = {L13},
          doi = {10.3847/2041-8213/acade4},
archivePrefix = {arXiv},
       eprint = {2211.05792},
 primaryClass = {astro-ph.GA},
       adsurl = {https://ui.adsabs.harvard.edu/abs/2023ApJ...946L..13F}
}

@ARTICLE{robertson_2022,
       author = {{Robertson}, B.~E. and {Tacchella}, S. and {Johnson}, B.~D. and {Hainline}, K. and {Whitler}, L. and {Eisenstein}, D.~J. and {Endsley}, R. and {Rieke}, M. and {Stark}, D.~P. and {Alberts}, S. and {Dressler}, A. and {Egami}, E. and {Hausen}, R. and {Rieke}, G. and {Shivaei}, I. and {Williams}, C.~C. and {Willmer}, C.~N.~A. and {Arribas}, S. and {Bonaventura}, N. and {Bunker}, A. and {Cameron}, A.~J. and {Carniani}, S. and {Charlot}, S. and {Chevallard}, J. and {Curti}, M. and {Curtis-Lake}, E. and {D'Eugenio}, F. and {Jakobsen}, P. and {Looser}, T.~J. and {L{\"u}tzgendorf}, N. and {Maiolino}, R. and {Maseda}, M.~V. and {Rawle}, T. and {Rix}, H. -W. and {Smit}, R. and {{\"U}bler}, H. and {Willott}, C. and {Witstok}, J. and {Baum}, S. and {Bhatawdekar}, R. and {Boyett}, K. and {Chen}, Z. and {de Graaff}, A. and {Florian}, M. and {Helton}, J.~M. and {Hviding}, R.~E. and {Ji}, Z. and {Kumari}, N. and {Lyu}, J. and {Nelson}, E. and {Sandles}, L. and {Saxena}, A. and {Suess}, K.~A. and {Sun}, F. and {Topping}, M. and {Wallace}, I.~E.~B.},
        title = "{Identification and properties of intense star-forming galaxies at redshifts z > 10}",
      journal = {Nature Astronomy},
         year = 2023,
        month = may,
       volume = {7},
        pages = {611-621},
          doi = {10.1038/s41550-023-01921-1},
archivePrefix = {arXiv},
       eprint = {2212.04480},
 primaryClass = {astro-ph.GA},
       adsurl = {https://ui.adsabs.harvard.edu/abs/2023NatAs...7..611R}
}

@ARTICLE{curtislake2023spectroscopicconfirmationmetalpoorgalaxies,
       author = {{Curtis-Lake}, Emma and {Carniani}, Stefano and {Cameron}, Alex and {Charlot}, Stephane and {Jakobsen}, Peter and {Maiolino}, Roberto and {Bunker}, Andrew and {Witstok}, Joris and {Smit}, Renske and {Chevallard}, Jacopo and {Willott}, Chris and {Ferruit}, Pierre and {Arribas}, Santiago and {Bonaventura}, Nina and {Curti}, Mirko and {D'Eugenio}, Francesco and {Franx}, Marijn and {Giardino}, Giovanna and {Looser}, Tobias J. and {L{\"u}tzgendorf}, Nora and {Maseda}, Michael V. and {Rawle}, Tim and {Rix}, Hans-Walter and {Rodr{\'\i}guez del Pino}, Bruno and {{\"U}bler}, Hannah and {Sirianni}, Marco and {Dressler}, Alan and {Egami}, Eiichi and {Eisenstein}, Daniel J. and {Endsley}, Ryan and {Hainline}, Kevin and {Hausen}, Ryan and {Johnson}, Benjamin D. and {Rieke}, Marcia and {Robertson}, Brant and {Shivaei}, Irene and {Stark}, Daniel P. and {Tacchella}, Sandro and {Williams}, Christina C. and {Willmer}, Christopher N.~A. and {Bhatawdekar}, Rachana and {Bowler}, Rebecca and {Boyett}, Kristan and {Chen}, Zuyi and {de Graaff}, Anna and {Helton}, Jakob M. and {Hviding}, Raphael E. and {Jones}, Gareth C. and {Kumari}, Nimisha and {Lyu}, Jianwei and {Nelson}, Erica and {Perna}, Michele and {Sandles}, Lester and {Saxena}, Aayush and {Suess}, Katherine A. and {Sun}, Fengwu and {Topping}, Michael W. and {Wallace}, Imaan E.~B. and {Whitler}, Lily},
        title = "{Spectroscopic confirmation of four metal-poor galaxies at z = 10.3-13.2}",
      journal = {Nature Astronomy},
         year = 2023,
        month = may,
       volume = {7},
        pages = {622-632},
          doi = {10.1038/s41550-023-01918-w},
archivePrefix = {arXiv},
       eprint = {2212.04568},
 primaryClass = {astro-ph.GA},
       adsurl = {https://ui.adsabs.harvard.edu/abs/2023NatAs...7..622C}
}

@ARTICLE{arrabel_maisies_galaxy,
       author = {{Arrabal Haro}, Pablo and {Dickinson}, Mark and {Finkelstein}, Steven L. and {Kartaltepe}, Jeyhan S. and {Donnan}, Callum T. and {Burgarella}, Denis and {Carnall}, Adam C. and {Cullen}, Fergus and {Dunlop}, James S. and {Fern{\'a}ndez}, Vital and {Fujimoto}, Seiji and {Jung}, Intae and {Krips}, Melanie and {Larson}, Rebecca L. and {Papovich}, Casey and {P{\'e}rez-Gonz{\'a}lez}, Pablo G. and {Amor{\'\i}n}, Ricardo O. and {Bagley}, Micaela B. and {Buat}, V{\'e}ronique and {Casey}, Caitlin M. and {Chworowsky}, Katherine and {Cohen}, Seth H. and {Ferguson}, Henry C. and {Giavalisco}, Mauro and {Huertas-Company}, Marc and {Hutchison}, Taylor A. and {Kocevski}, Dale D. and {Koekemoer}, Anton M. and {Lucas}, Ray A. and {McLeod}, Derek J. and {McLure}, Ross J. and {Pirzkal}, Norbert and {Seill{\'e}}, Lise-Marie and {Trump}, Jonathan R. and {Weiner}, Benjamin J. and {Wilkins}, Stephen M. and {Zavala}, Jorge A.},
        title = "{Confirmation and refutation of very luminous galaxies in the early Universe}",
      journal = {\nat},
         year = 2023,
        month = oct,
       volume = {622},
       number = {7984},
        pages = {707-711},
          doi = {10.1038/s41586-023-06521-7},
archivePrefix = {arXiv},
       eprint = {2303.15431},
 primaryClass = {astro-ph.GA},
       adsurl = {https://ui.adsabs.harvard.edu/abs/2023Natur.622..707A}
}

@ARTICLE{grogin_candels,
       author = {{Grogin}, Norman A. and {Kocevski}, Dale D. and {Faber}, S.~M. and {Ferguson}, Henry C. and {Koekemoer}, Anton M. and {Riess}, Adam G. and {Acquaviva}, Viviana and {Alexander}, David M. and {Almaini}, Omar and {Ashby}, Matthew L.~N. and {Barden}, Marco and {Bell}, Eric F. and {Bournaud}, Fr{\'e}d{\'e}ric and {Brown}, Thomas M. and {Caputi}, Karina I. and {Casertano}, Stefano and {Cassata}, Paolo and {Castellano}, Marco and {Challis}, Peter and {Chary}, Ranga-Ram and {Cheung}, Edmond and {Cirasuolo}, Michele and {Conselice}, Christopher J. and {Roshan Cooray}, Asantha and {Croton}, Darren J. and {Daddi}, Emanuele and {Dahlen}, Tomas and {Dav{\'e}}, Romeel and {de Mello}, Du{\'\i}lia F. and {Dekel}, Avishai and {Dickinson}, Mark and {Dolch}, Timothy and {Donley}, Jennifer L. and {Dunlop}, James S. and {Dutton}, Aaron A. and {Elbaz}, David and {Fazio}, Giovanni G. and {Filippenko}, Alexei V. and {Finkelstein}, Steven L. and {Fontana}, Adriano and {Gardner}, Jonathan P. and {Garnavich}, Peter M. and {Gawiser}, Eric and {Giavalisco}, Mauro and {Grazian}, Andrea and {Guo}, Yicheng and {Hathi}, Nimish P. and {H{\"a}ussler}, Boris and {Hopkins}, Philip F. and {Huang}, Jia-Sheng and {Huang}, Kuang-Han and {Jha}, Saurabh W. and {Kartaltepe}, Jeyhan S. and {Kirshner}, Robert P. and {Koo}, David C. and {Lai}, Kamson and {Lee}, Kyoung-Soo and {Li}, Weidong and {Lotz}, Jennifer M. and {Lucas}, Ray A. and {Madau}, Piero and {McCarthy}, Patrick J. and {McGrath}, Elizabeth J. and {McIntosh}, Daniel H. and {McLure}, Ross J. and {Mobasher}, Bahram and {Moustakas}, Leonidas A. and {Mozena}, Mark and {Nandra}, Kirpal and {Newman}, Jeffrey A. and {Niemi}, Sami-Matias and {Noeske}, Kai G. and {Papovich}, Casey J. and {Pentericci}, Laura and {Pope}, Alexandra and {Primack}, Joel R. and {Rajan}, Abhijith and {Ravindranath}, Swara and {Reddy}, Naveen A. and {Renzini}, Alvio and {Rix}, Hans-Walter and {Robaina}, Aday R. and {Rodney}, Steven A. and {Rosario}, David J. and {Rosati}, Piero and {Salimbeni}, Sara and {Scarlata}, Claudia and {Siana}, Brian and {Simard}, Luc and {Smidt}, Joseph and {Somerville}, Rachel S. and {Spinrad}, Hyron and {Straughn}, Amber N. and {Strolger}, Louis-Gregory and {Telford}, Olivia and {Teplitz}, Harry I. and {Trump}, Jonathan R. and {van der Wel}, Arjen and {Villforth}, Carolin and {Wechsler}, Risa H. and {Weiner}, Benjamin J. and {Wiklind}, Tommy and {Wild}, Vivienne and {Wilson}, Grant and {Wuyts}, Stijn and {Yan}, Hao-Jing and {Yun}, Min S.},
        title = "{CANDELS: The Cosmic Assembly Near-infrared Deep Extragalactic Legacy Survey}",
      journal = {\apjs},
         year = 2011,
        month = dec,
       volume = {197},
       number = {2},
          eid = {35},
        pages = {35},
          doi = {10.1088/0067-0049/197/2/35},
archivePrefix = {arXiv},
       eprint = {1105.3753},
 primaryClass = {astro-ph.CO},
       adsurl = {https://ui.adsabs.harvard.edu/abs/2011ApJS..197...35G}
}

@ARTICLE{koekemoer_candels,
       author = {{Koekemoer}, Anton M. and {Faber}, S.~M. and {Ferguson}, Henry C. and {Grogin}, Norman A. and {Kocevski}, Dale D. and {Koo}, David C. and {Lai}, Kamson and {Lotz}, Jennifer M. and {Lucas}, Ray A. and {McGrath}, Elizabeth J. and {Ogaz}, Sara and {Rajan}, Abhijith and {Riess}, Adam G. and {Rodney}, Steve A. and {Strolger}, Louis and {Casertano}, Stefano and {Castellano}, Marco and {Dahlen}, Tomas and {Dickinson}, Mark and {Dolch}, Timothy and {Fontana}, Adriano and {Giavalisco}, Mauro and {Grazian}, Andrea and {Guo}, Yicheng and {Hathi}, Nimish P. and {Huang}, Kuang-Han and {van der Wel}, Arjen and {Yan}, Hao-Jing and {Acquaviva}, Viviana and {Alexander}, David M. and {Almaini}, Omar and {Ashby}, Matthew L.~N. and {Barden}, Marco and {Bell}, Eric F. and {Bournaud}, Fr{\'e}d{\'e}ric and {Brown}, Thomas M. and {Caputi}, Karina I. and {Cassata}, Paolo and {Challis}, Peter J. and {Chary}, Ranga-Ram and {Cheung}, Edmond and {Cirasuolo}, Michele and {Conselice}, Christopher J. and {Roshan Cooray}, Asantha and {Croton}, Darren J. and {Daddi}, Emanuele and {Dav{\'e}}, Romeel and {de Mello}, Duilia F. and {de Ravel}, Loic and {Dekel}, Avishai and {Donley}, Jennifer L. and {Dunlop}, James S. and {Dutton}, Aaron A. and {Elbaz}, David and {Fazio}, Giovanni G. and {Filippenko}, Alexei V. and {Finkelstein}, Steven L. and {Frazer}, Chris and {Gardner}, Jonathan P. and {Garnavich}, Peter M. and {Gawiser}, Eric and {Gruetzbauch}, Ruth and {Hartley}, Will G. and {H{\"a}ussler}, Boris and {Herrington}, Jessica and {Hopkins}, Philip F. and {Huang}, Jia-Sheng and {Jha}, Saurabh W. and {Johnson}, Andrew and {Kartaltepe}, Jeyhan S. and {Khostovan}, Ali A. and {Kirshner}, Robert P. and {Lani}, Caterina and {Lee}, Kyoung-Soo and {Li}, Weidong and {Madau}, Piero and {McCarthy}, Patrick J. and {McIntosh}, Daniel H. and {McLure}, Ross J. and {McPartland}, Conor and {Mobasher}, Bahram and {Moreira}, Heidi and {Mortlock}, Alice and {Moustakas}, Leonidas A. and {Mozena}, Mark and {Nandra}, Kirpal and {Newman}, Jeffrey A. and {Nielsen}, Jennifer L. and {Niemi}, Sami and {Noeske}, Kai G. and {Papovich}, Casey J. and {Pentericci}, Laura and {Pope}, Alexandra and {Primack}, Joel R. and {Ravindranath}, Swara and {Reddy}, Naveen A. and {Renzini}, Alvio and {Rix}, Hans-Walter and {Robaina}, Aday R. and {Rosario}, David J. and {Rosati}, Piero and {Salimbeni}, Sara and {Scarlata}, Claudia and {Siana}, Brian and {Simard}, Luc and {Smidt}, Joseph and {Snyder}, Diana and {Somerville}, Rachel S. and {Spinrad}, Hyron and {Straughn}, Amber N. and {Telford}, Olivia and {Teplitz}, Harry I. and {Trump}, Jonathan R. and {Vargas}, Carlos and {Villforth}, Carolin and {Wagner}, Cory R. and {Wandro}, Pat and {Wechsler}, Risa H. and {Weiner}, Benjamin J. and {Wiklind}, Tommy and {Wild}, Vivienne and {Wilson}, Grant and {Wuyts}, Stijn and {Yun}, Min S.},
        title = "{CANDELS: The Cosmic Assembly Near-infrared Deep Extragalactic Legacy Survey{\textemdash}The Hubble Space Telescope Observations, Imaging Data Products, and Mosaics}",
      journal = {\apjs},
         year = 2011,
        month = dec,
       volume = {197},
       number = {2},
          eid = {36},
        pages = {36},
          doi = {10.1088/0067-0049/197/2/36},
archivePrefix = {arXiv},
       eprint = {1105.3754},
 primaryClass = {astro-ph.CO},
       adsurl = {https://ui.adsabs.harvard.edu/abs/2011ApJS..197...36K}
}

@ARTICLE{illustris_synthetic_img_torrey,
       author = {{Torrey}, Paul and {Snyder}, Gregory F. and {Vogelsberger}, Mark and {Hayward}, Christopher C. and {Genel}, Shy and {Sijacki}, Debora and {Springel}, Volker and {Hernquist}, Lars and {Nelson}, Dylan and {Kriek}, Mariska and {Pillepich}, Annalisa and {Sales}, Laura V. and {McBride}, Cameron K.},
        title = "{Synthetic galaxy images and spectra from the Illustris simulation}",
      journal = {\mnras},
         year = 2015,
        month = mar,
       volume = {447},
       number = {3},
        pages = {2753-2771},
          doi = {10.1093/mnras/stu2592},
archivePrefix = {arXiv},
       eprint = {1411.3717},
 primaryClass = {astro-ph.GA},
       adsurl = {https://ui.adsabs.harvard.edu/abs/2015MNRAS.447.2753T}
}

@ARTICLE{sphinx_intro,
       author = {{Rosdahl}, Joakim and {Katz}, Harley and {Blaizot}, J{\'e}r{\'e}my and {Kimm}, Taysun and {Michel-Dansac}, L{\'e}o and {Garel}, Thibault and {Haehnelt}, Martin and {Ocvirk}, Pierre and {Teyssier}, Romain},
        title = "{The SPHINX cosmological simulations of the first billion years: the impact of binary stars on reionization}",
      journal = {\mnras},
         year = 2018,
        month = sep,
       volume = {479},
       number = {1},
        pages = {994-1016},
          doi = {10.1093/mnras/sty1655},
archivePrefix = {arXiv},
       eprint = {1801.07259},
 primaryClass = {astro-ph.GA},
       adsurl = {https://ui.adsabs.harvard.edu/abs/2018MNRAS.479..994R}
}

@ARTICLE{astrid_galaxies,
       author = {{Bird}, Simeon and {Ni}, Yueying and {Di Matteo}, Tiziana and {Croft}, Rupert and {Feng}, Yu and {Chen}, Nianyi},
        title = "{The ASTRID simulation: galaxy formation and reionization}",
      journal = {\mnras},
         year = 2022,
        month = may,
       volume = {512},
       number = {3},
        pages = {3703-3716},
          doi = {10.1093/mnras/stac648},
archivePrefix = {arXiv},
       eprint = {2111.01160},
 primaryClass = {astro-ph.GA},
       adsurl = {https://ui.adsabs.harvard.edu/abs/2022MNRAS.512.3703B}
}

@ARTICLE{illustris_tng,
       author = {{Nelson}, Dylan and {Pillepich}, Annalisa and {Springel}, Volker and {Weinberger}, Rainer and {Hernquist}, Lars and {Pakmor}, R{\"u}diger and {Genel}, Shy and {Torrey}, Paul and {Vogelsberger}, Mark and {Kauffmann}, Guinevere and {Marinacci}, Federico and {Naiman}, Jill},
        title = "{First results from the IllustrisTNG simulations: the galaxy colour bimodality}",
      journal = {\mnras},
         year = 2018,
        month = mar,
       volume = {475},
       number = {1},
        pages = {624-647},
          doi = {10.1093/mnras/stx3040},
archivePrefix = {arXiv},
       eprint = {1707.03395},
 primaryClass = {astro-ph.GA},
       adsurl = {https://ui.adsabs.harvard.edu/abs/2018MNRAS.475..624N}
}

@ARTICLE{astrid_lrd,
       author = {{LaChance}, Patrick and {Croft}, Rupert A.~C. and {Di Matteo}, Tiziana and {Zhou}, Yihao and {Pacucci}, Fabio and {Ni}, Yueying and {Chen}, Nianyi and {Bird}, Simeon},
        title = "{The Properties of Little Red Dot Galaxies in the ASTRID Simulation}",
      journal = {arXiv e-prints},
         year = 2025,
        month = may,
          eid = {arXiv:2505.20439},
        pages = {arXiv:2505.20439},
          doi = {10.48550/arXiv.2505.20439},
archivePrefix = {arXiv},
       eprint = {2505.20439},
 primaryClass = {astro-ph.CO},
       adsurl = {https://ui.adsabs.harvard.edu/abs/2025arXiv250520439L}
}

@ARTICLE{astrid_synthetic,
       author = {{LaChance}, Patrick and {Croft}, Rupert and {Ni}, Yueying and {Chen}, Nianyi and {Matteo}, Tiziana Di and {Bird}, Simeon},
        title = "{The evolution of galaxy morphology from redshift z=6 to 3: Mock JWST observations of galaxies in the ASTRID simulation}",
      journal = {The Open Journal of Astrophysics},
         year = 2025,
        month = feb,
       volume = {8},
          eid = {20},
        pages = {20},
          doi = {10.33232/001c.129991},
archivePrefix = {arXiv},
       eprint = {2401.16608},
 primaryClass = {astro-ph.GA},
       adsurl = {https://ui.adsabs.harvard.edu/abs/2025OJAp....8E..20L}
}

@ARTICLE{thesan_intro,
       author = {{Kannan}, R. and {Garaldi}, E. and {Smith}, A. and {Pakmor}, R. and {Springel}, V. and {Vogelsberger}, M. and {Hernquist}, L.},
        title = "{Introducing the THESAN project: radiation-magnetohydrodynamic simulations of the epoch of reionization}",
      journal = {\mnras},
         year = 2022,
        month = apr,
       volume = {511},
       number = {3},
        pages = {4005-4030},
          doi = {10.1093/mnras/stab3710},
archivePrefix = {arXiv},
       eprint = {2110.00584},
 primaryClass = {astro-ph.GA},
       adsurl = {https://ui.adsabs.harvard.edu/abs/2022MNRAS.511.4005K}
}

@INPROCEEDINGS{STPSF_2014,
       author = {{Perrin}, Marshall D. and {Sivaramakrishnan}, Anand and {Lajoie}, Charles-Philippe and {Elliott}, Erin and {Pueyo}, Laurent and {Ravindranath}, Swara and {Albert}, Lo{\"\i}c.},
        title = "{Updated point spread function simulations for JWST with WebbPSF}",
    booktitle = {Space Telescopes and Instrumentation 2014: Optical, Infrared, and Millimeter Wave},
         year = 2014,
       editor = {{Oschmann}, Jr., Jacobus M. and {Clampin}, Mark and {Fazio}, Giovanni G. and {MacEwen}, Howard A.},
       series = {Society of Photo-Optical Instrumentation Engineers (SPIE) Conference Series},
       volume = {9143},
        month = aug,
          eid = {91433X},
        pages = {91433X},
          doi = {10.1117/12.2056689},
       adsurl = {https://ui.adsabs.harvard.edu/abs/2014SPIE.9143E..3XP}
}

@INPROCEEDINGS{STPSF_2012,
       author = {{Perrin}, Marshall D. and {Soummer}, R{\'e}mi and {Elliott}, Erin M. and {Lallo}, Matthew D. and {Sivaramakrishnan}, Anand},
        title = "{Simulating point spread functions for the James Webb Space Telescope with WebbPSF}",
    booktitle = {Space Telescopes and Instrumentation 2012: Optical, Infrared, and Millimeter Wave},
         year = 2012,
       editor = {{Clampin}, Mark C. and {Fazio}, Giovanni G. and {MacEwen}, Howard A. and {Oschmann}, Jr., Jacobus M.},
       series = {Society of Photo-Optical Instrumentation Engineers (SPIE) Conference Series},
       volume = {8442},
        month = sep,
          eid = {84423D},
        pages = {84423D},
          doi = {10.1117/12.925230},
       adsurl = {https://ui.adsabs.harvard.edu/abs/2012SPIE.8442E..3DP}
}

@ARTICLE{astropy_1,
       author = {{Astropy Collaboration} and {Robitaille}, Thomas P. and {Tollerud}, Erik J. and {Greenfield}, Perry and {Droettboom}, Michael and {Bray}, Erik and {Aldcroft}, Tom and {Davis}, Matt and {Ginsburg}, Adam and {Price-Whelan}, Adrian M. and {Kerzendorf}, Wolfgang E. and {Conley}, Alexander and {Crighton}, Neil and {Barbary}, Kyle and {Muna}, Demitri and {Ferguson}, Henry and {Grollier}, Fr{\'e}d{\'e}ric and {Parikh}, Madhura M. and {Nair}, Prasanth H. and {Unther}, Hans M. and {Deil}, Christoph and {Woillez}, Julien and {Conseil}, Simon and {Kramer}, Roban and {Turner}, James E.~H. and {Singer}, Leo and {Fox}, Ryan and {Weaver}, Benjamin A. and {Zabalza}, Victor and {Edwards}, Zachary I. and {Azalee Bostroem}, K. and {Burke}, D.~J. and {Casey}, Andrew R. and {Crawford}, Steven M. and {Dencheva}, Nadia and {Ely}, Justin and {Jenness}, Tim and {Labrie}, Kathleen and {Lim}, Pey Lian and {Pierfederici}, Francesco and {Pontzen}, Andrew and {Ptak}, Andy and {Refsdal}, Brian and {Servillat}, Mathieu and {Streicher}, Ole},
        title = "{Astropy: A community Python package for astronomy}",
      journal = {\aap},
         year = 2013,
        month = oct,
       volume = {558},
          eid = {A33},
        pages = {A33},
          doi = {10.1051/0004-6361/201322068},
archivePrefix = {arXiv},
       eprint = {1307.6212},
 primaryClass = {astro-ph.IM},
       adsurl = {https://ui.adsabs.harvard.edu/abs/2013A&A...558A..33A}
}

@ARTICLE{astropy_2,
       author = {{Astropy Collaboration} and {Price-Whelan}, A.~M. and {Sip{\H{o}}cz}, B.~M. and {G{\"u}nther}, H.~M. and {Lim}, P.~L. and {Crawford}, S.~M. and {Conseil}, S. and {Shupe}, D.~L. and {Craig}, M.~W. and {Dencheva}, N. and {Ginsburg}, A. and {VanderPlas}, J.~T. and {Bradley}, L.~D. and {P{\'e}rez-Su{\'a}rez}, D. and {de Val-Borro}, M. and {Aldcroft}, T.~L. and {Cruz}, K.~L. and {Robitaille}, T.~P. and {Tollerud}, E.~J. and {Ardelean}, C. and {Babej}, T. and {Bach}, Y.~P. and {Bachetti}, M. and {Bakanov}, A.~V. and {Bamford}, S.~P. and {Barentsen}, G. and {Barmby}, P. and {Baumbach}, A. and {Berry}, K.~L. and {Biscani}, F. and {Boquien}, M. and {Bostroem}, K.~A. and {Bouma}, L.~G. and {Brammer}, G.~B. and {Bray}, E.~M. and {Breytenbach}, H. and {Buddelmeijer}, H. and {Burke}, D.~J. and {Calderone}, G. and {Cano Rodr{\'\i}guez}, J.~L. and {Cara}, M. and {Cardoso}, J.~V.~M. and {Cheedella}, S. and {Copin}, Y. and {Corrales}, L. and {Crichton}, D. and {D'Avella}, D. and {Deil}, C. and {Depagne}, {\'E}. and {Dietrich}, J.~P. and {Donath}, A. and {Droettboom}, M. and {Earl}, N. and {Erben}, T. and {Fabbro}, S. and {Ferreira}, L.~A. and {Finethy}, T. and {Fox}, R.~T. and {Garrison}, L.~H. and {Gibbons}, S.~L.~J. and {Goldstein}, D.~A. and {Gommers}, R. and {Greco}, J.~P. and {Greenfield}, P. and {Groener}, A.~M. and {Grollier}, F. and {Hagen}, A. and {Hirst}, P. and {Homeier}, D. and {Horton}, A.~J. and {Hosseinzadeh}, G. and {Hu}, L. and {Hunkeler}, J.~S. and {Ivezi{\'c}}, {\v{Z}}. and {Jain}, A. and {Jenness}, T. and {Kanarek}, G. and {Kendrew}, S. and {Kern}, N.~S. and {Kerzendorf}, W.~E. and {Khvalko}, A. and {King}, J. and {Kirkby}, D. and {Kulkarni}, A.~M. and {Kumar}, A. and {Lee}, A. and {Lenz}, D. and {Littlefair}, S.~P. and {Ma}, Z. and {Macleod}, D.~M. and {Mastropietro}, M. and {McCully}, C. and {Montagnac}, S. and {Morris}, B.~M. and {Mueller}, M. and {Mumford}, S.~J. and {Muna}, D. and {Murphy}, N.~A. and {Nelson}, S. and {Nguyen}, G.~H. and {Ninan}, J.~P. and {N{\"o}the}, M. and {Ogaz}, S. and {Oh}, S. and {Parejko}, J.~K. and {Parley}, N. and {Pascual}, S. and {Patil}, R. and {Patil}, A.~A. and {Plunkett}, A.~L. and {Prochaska}, J.~X. and {Rastogi}, T. and {Reddy Janga}, V. and {Sabater}, J. and {Sakurikar}, P. and {Seifert}, M. and {Sherbert}, L.~E. and {Sherwood-Taylor}, H. and {Shih}, A.~Y. and {Sick}, J. and {Silbiger}, M.~T. and {Singanamalla}, S. and {Singer}, L.~P. and {Sladen}, P.~H. and {Sooley}, K.~A. and {Sornarajah}, S. and {Streicher}, O. and {Teuben}, P. and {Thomas}, S.~W. and {Tremblay}, G.~R. and {Turner}, J.~E.~H. and {Terr{\'o}n}, V. and {van Kerkwijk}, M.~H. and {de la Vega}, A. and {Watkins}, L.~L. and {Weaver}, B.~A. and {Whitmore}, J.~B. and {Woillez}, J. and {Zabalza}, V. and {Astropy Contributors}},
        title = "{The Astropy Project: Building an Open-science Project and Status of the v2.0 Core Package}",
      journal = {\aj},
         year = 2018,
        month = sep,
       volume = {156},
       number = {3},
          eid = {123},
        pages = {123},
          doi = {10.3847/1538-3881/aabc4f},
archivePrefix = {arXiv},
       eprint = {1801.02634},
 primaryClass = {astro-ph.IM},
       adsurl = {https://ui.adsabs.harvard.edu/abs/2018AJ....156..123A}
}

@ARTICLE{astropy_3, 
       author = {{Astropy Collaboration} and {Price-Whelan}, Adrian M. and {Lim}, Pey Lian and {Earl}, Nicholas and {Starkman}, Nathaniel and {Bradley}, Larry and {Shupe}, David L. and {Patil}, Aarya A. and {Corrales}, Lia and {Brasseur}, C.~E. and {N{\"o}the}, Maximilian and {Donath}, Axel and {Tollerud}, Erik and {Morris}, Brett M. and {Ginsburg}, Adam and {Vaher}, Eero and {Weaver}, Benjamin A. and {Tocknell}, James and {Jamieson}, William and {van Kerkwijk}, Marten H. and {Robitaille}, Thomas P. and {Merry}, Bruce and {Bachetti}, Matteo and {G{\"u}nther}, H. Moritz and {Aldcroft}, Thomas L. and {Alvarado-Montes}, Jaime A. and {Archibald}, Anne M. and {B{\'o}di}, Attila and {Bapat}, Shreyas and {Barentsen}, Geert and {Baz{\'a}n}, Juanjo and {Biswas}, Manish and {Boquien}, M{\'e}d{\'e}ric and {Burke}, D.~J. and {Cara}, Daria and {Cara}, Mihai and {Conroy}, Kyle E. and {Conseil}, Simon and {Craig}, Matthew W. and {Cross}, Robert M. and {Cruz}, Kelle L. and {D'Eugenio}, Francesco and {Dencheva}, Nadia and {Devillepoix}, Hadrien A.~R. and {Dietrich}, J{\"o}rg P. and {Eigenbrot}, Arthur Davis and {Erben}, Thomas and {Ferreira}, Leonardo and {Foreman-Mackey}, Daniel and {Fox}, Ryan and {Freij}, Nabil and {Garg}, Suyog and {Geda}, Robel and {Glattly}, Lauren and {Gondhalekar}, Yash and {Gordon}, Karl D. and {Grant}, David and {Greenfield}, Perry and {Groener}, Austen M. and {Guest}, Steve and {Gurovich}, Sebastian and {Handberg}, Rasmus and {Hart}, Akeem and {Hatfield-Dodds}, Zac and {Homeier}, Derek and {Hosseinzadeh}, Griffin and {Jenness}, Tim and {Jones}, Craig K. and {Joseph}, Prajwel and {Kalmbach}, J. Bryce and {Karamehmetoglu}, Emir and {Ka{\l}uszy{\'n}ski}, Miko{\l}aj and {Kelley}, Michael S.~P. and {Kern}, Nicholas and {Kerzendorf}, Wolfgang E. and {Koch}, Eric W. and {Kulumani}, Shankar and {Lee}, Antony and {Ly}, Chun and {Ma}, Zhiyuan and {MacBride}, Conor and {Maljaars}, Jakob M. and {Muna}, Demitri and {Murphy}, N.~A. and {Norman}, Henrik and {O'Steen}, Richard and {Oman}, Kyle A. and {Pacifici}, Camilla and {Pascual}, Sergio and {Pascual-Granado}, J. and {Patil}, Rohit R. and {Perren}, Gabriel I. and {Pickering}, Timothy E. and {Rastogi}, Tanuj and {Roulston}, Benjamin R. and {Ryan}, Daniel F. and {Rykoff}, Eli S. and {Sabater}, Jose and {Sakurikar}, Parikshit and {Salgado}, Jes{\'u}s and {Sanghi}, Aniket and {Saunders}, Nicholas and {Savchenko}, Volodymyr and {Schwardt}, Ludwig and {Seifert-Eckert}, Michael and {Shih}, Albert Y. and {Jain}, Anany Shrey and {Shukla}, Gyanendra and {Sick}, Jonathan and {Simpson}, Chris and {Singanamalla}, Sudheesh and {Singer}, Leo P. and {Singhal}, Jaladh and {Sinha}, Manodeep and {Sip{\H{o}}cz}, Brigitta M. and {Spitler}, Lee R. and {Stansby}, David and {Streicher}, Ole and {{\v{S}}umak}, Jani and {Swinbank}, John D. and {Taranu}, Dan S. and {Tewary}, Nikita and {Tremblay}, Grant R. and {de Val-Borro}, Miguel and {Van Kooten}, Samuel J. and {Vasovi{\'c}}, Zlatan and {Verma}, Shresth and {de Miranda Cardoso}, Jos{\'e} Vin{\'\i}cius and {Williams}, Peter K.~G. and {Wilson}, Tom J. and {Winkel}, Benjamin and {Wood-Vasey}, W.~M. and {Xue}, Rui and {Yoachim}, Peter and {Zhang}, Chen and {Zonca}, Andrea and {Astropy Project Contributors}},
        title = "{The Astropy Project: Sustaining and Growing a Community-oriented Open-source Project and the Latest Major Release (v5.0) of the Core Package}",
      journal = {\apj},
         year = 2022,
        month = aug,
       volume = {935},
       number = {2},
          eid = {167},
        pages = {167},
          doi = {10.3847/1538-4357/ac7c74},
archivePrefix = {arXiv},
       eprint = {2206.14220},
 primaryClass = {astro-ph.IM},
       adsurl = {https://ui.adsabs.harvard.edu/abs/2022ApJ...935..167A}
}

@ARTICLE{mock_galaxy_hst_jwst_illustris_tng,
       author = {{Snyder}, Gregory F. and {Pe{\~n}a}, Theodore and {Yung}, L.~Y. Aaron and {Rose}, Caitlin and {Kartaltepe}, Jeyhan and {Ferguson}, Harry},
        title = "{Mock galaxy surveys for HST and JWST from the IllustrisTNG simulations}",
      journal = {\mnras},
         year = 2023,
        month = feb,
       volume = {518},
       number = {4},
        pages = {6318-6324},
          doi = {10.1093/mnras/stac3397},
archivePrefix = {arXiv},
       eprint = {2211.09677},
 primaryClass = {astro-ph.GA},
       adsurl = {https://ui.adsabs.harvard.edu/abs/2023MNRAS.518.6318S}
}

@article{ytree,
  doi = {10.21105/joss.01881},
  url = {https://doi.org/10.21105/joss.01881},
  year  = {2019},
  month = {dec},
  publisher = {The Open Journal},
  volume = {4},
  number = {44},
  pages = {1881},
  author = {Britton D. Smith and Meagan Lang},
  title = {ytree: A Python package for analyzing merger trees},
  journal = {Journal of Open Source Software}
}

@ARTICLE{yt,
   author = {{Turk}, M.~J. and {Smith}, B.~D. and {Oishi}, J.~S. and {Skory}, S. and
     {Skillman}, S.~W. and {Abel}, T. and {Norman}, M.~L.},
    title = "{yt: A Multi-code Analysis Toolkit for Astrophysical Simulation Data}",
  journal = {The Astrophysical Journal Supplement Series},
archivePrefix = "arXiv",
   eprint = {1011.3514},
 primaryClass = "astro-ph.IM",
     year = 2011,
    month = jan,
   volume = 192,
      eid = {9},
    pages = {9},
      doi = {10.1088/0067-0049/192/1/9},
   adsurl = {https://ui.adsabs.harvard.edu/abs/2011ApJS..192....9T}
}

@ARTICLE{2020SciPy-NMeth,
  author  = {Virtanen, Pauli and Gommers, Ralf and Oliphant, Travis E. and
            Haberland, Matt and Reddy, Tyler and Cournapeau, David and
            Burovski, Evgeni and Peterson, Pearu and Weckesser, Warren and
            Bright, Jonathan and {van der Walt}, St{\'e}fan J. and
            Brett, Matthew and Wilson, Joshua and Millman, K. Jarrod and
            Mayorov, Nikolay and Nelson, Andrew R. J. and Jones, Eric and
            Kern, Robert and Larson, Eric and Carey, C J and
            Polat, {\.I}lhan and Feng, Yu and Moore, Eric W. and
            {VanderPlas}, Jake and Laxalde, Denis and Perktold, Josef and
            Cimrman, Robert and Henriksen, Ian and Quintero, E. A. and
            Harris, Charles R. and Archibald, Anne M. and
            Ribeiro, Ant{\^o}nio H. and Pedregosa, Fabian and
            {van Mulbregt}, Paul and {SciPy 1.0 Contributors}},
  title   = {{{SciPy} 1.0: Fundamental Algorithms for Scientific
            Computing in Python}},
  journal = {Nature Methods},
  year    = {2020},
  volume  = {17},
  pages   = {261--272},
  adsurl  = {https://rdcu.be/b08Wh},
  doi     = {10.1038/s41592-019-0686-2},
}

@ARTICLE{donnan_corey,
       author = {{Donnan}, C.~T. and {McLeod}, D.~J. and {Dunlop}, J.~S. and {McLure}, R.~J. and {Carnall}, A.~C. and {Begley}, R. and {Cullen}, F. and {Hamadouche}, M.~L. and {Bowler}, R.~A.~A. and {Magee}, D. and {McCracken}, H.~J. and {Milvang-Jensen}, B. and {Moneti}, A. and {Targett}, T.},
        title = "{The evolution of the galaxy UV luminosity function at redshifts z ≃ 8 - 15 from deep JWST and ground-based near-infrared imaging}",
      journal = {\mnras},
         year = 2023,
        month = feb,
       volume = {518},
       number = {4},
        pages = {6011-6040},
          doi = {10.1093/mnras/stac3472},
archivePrefix = {arXiv},
       eprint = {2207.12356},
 primaryClass = {astro-ph.GA},
       adsurl = {https://ui.adsabs.harvard.edu/abs/2023MNRAS.518.6011D}
}

@ARTICLE{UV_continuum_slope,
       author = {{Dottorini}, D. and {Calabr{\`o}}, A. and {Pentericci}, L. and {Mascia}, S. and {Llerena}, M. and {Napolitano}, L. and {Santini}, P. and {Roberts-Borsani}, G. and {Castellano}, M. and {Amorin}, R. and {Dickinson}, M. and {Fontana}, A. and {Hathi}, N. and {Hirschmann}, M. and {Koekemoer}, A.~M. and {Lucas}, R.~A. and {Merlin}, E. and {Morales}, A. and {Pacucci}, F. and {Wilkins}, S. and {Arrabal Haro}, P. and {Bagley}, M. and {Finkelstein}, S.~L. and {Kartaltepe}, J. and {Papovich}, C. and {Pirzkal}, N.},
        title = "{Evolution of the UV slope of galaxies at cosmic morning (z > 4): The properties of extremely blue galaxies}",
      journal = {\aap},
         year = 2025,
        month = jun,
       volume = {698},
          eid = {A234},
        pages = {A234},
          doi = {10.1051/0004-6361/202453267},
archivePrefix = {arXiv},
       eprint = {2412.01623},
 primaryClass = {astro-ph.GA},
       adsurl = {https://ui.adsabs.harvard.edu/abs/2025A&A...698A.234D}
}

@ARTICLE{UV_continuum_slope_windows,
       author = {{Calzetti}, Daniela and {Kinney}, Anne L. and {Storchi-Bergmann}, Thaisa},
        title = "{Dust Extinction of the Stellar Continua in Starburst Galaxies: The Ultraviolet and Optical Extinction Law}",
      journal = {\apj},
         year = 1994,
        month = jul,
       volume = {429},
        pages = {582},
          doi = {10.1086/174346},
       adsurl = {https://ui.adsabs.harvard.edu/abs/1994ApJ...429..582C}
}

@ARTICLE{donnan2026spectroscopicconfirmationlargeluminous,
       author = {{Donnan}, Callum T. and {McLeod}, Derek J. and {McLure}, Ross J. and {Dunlop}, James S. and {Cullen}, Fergus and {Dickinson}, Mark and {Arrabal Haro}, Pablo and {Taylor}, Anthony J. and {Bondestam}, Cecilia and {Liu}, Feng-Yuan and {Arellano-C{\'o}rdova}, Karla Z. and {Barrufet}, Laia and {Begley}, Ryan and {Carnall}, Adam C. and {Golawska}, Hanna and {Leung}, Ho-Hin and {Scholte}, Dirk and {Stanton}, Thomas M.},
        title = "{Spectroscopic Confirmation of a Large and Luminous Galaxy with Weak Emission Lines at z = 13.53}",
      journal = {\apj},
         year = 2026,
        month = may,
       volume = {1002},
       number = {2},
          eid = {134},
        pages = {134},
          doi = {10.3847/1538-4357/ae5c05},
archivePrefix = {arXiv},
       eprint = {2601.11515},
 primaryClass = {astro-ph.GA},
       adsurl = {https://ui.adsabs.harvard.edu/abs/2026ApJ..1002..134D}
}

@ARTICLE{thesan_zoom,
       author = {{Kannan}, Rahul and {Puchwein}, Ewald and {Smith}, Aaron and {Borrow}, Josh and {Garaldi}, Enrico and {Keating}, Laura and {Vogelsberger}, Mark and {Zier}, Oliver and {McClymont}, William and {Shen}, Xuejian and {Popovic}, Filip and {Tacchella}, Sandro and {Hernquist}, Lars and {Springel}, Volker},
        title = "{Introducing the THESAN-ZOOM project: radiation-hydrodynamic simulations of high-redshift galaxies with a multi-phase interstellar medium}",
      journal = {The Open Journal of Astrophysics},
         year = 2025,
        month = oct,
       volume = {8},
          eid = {153},
        pages = {153},
          doi = {10.33232/001c.145804},
archivePrefix = {arXiv},
       eprint = {2502.20437},
 primaryClass = {astro-ph.GA},
       adsurl = {https://ui.adsabs.harvard.edu/abs/2025OJAp....8E.153K}
}

@ARTICLE{thesan_morphology,
       author = {{Shen}, Xuejian and {Vogelsberger}, Mark and {Borrow}, Josh and {Hu}, Yongao and {Erickson}, Evan and {Kannan}, Rahul and {Smith}, Aaron and {Garaldi}, Enrico and {Hernquist}, Lars and {Morishita}, Takahiro and {Tacchella}, Sandro and {Zier}, Oliver and {Sun}, Guochao and {Eilers}, Anna-Christina and {Wang}, Hui},
        title = "{The THESAN project: galaxy sizes during the epoch of reionization}",
      journal = {\mnras},
         year = 2024,
        month = oct,
       volume = {534},
       number = {2},
        pages = {1433-1458},
          doi = {10.1093/mnras/stae2156},
archivePrefix = {arXiv},
       eprint = {2402.08717},
 primaryClass = {astro-ph.GA},
       adsurl = {https://ui.adsabs.harvard.edu/abs/2024MNRAS.534.1433S}
}

@ARTICLE{thesan_zoom_starbursts,
       author = {{McClymont}, William and {Tacchella}, Sandro and {Smith}, Aaron and {Kannan}, Rahul and {Puchwein}, Ewald and {Borrow}, Josh and {Garaldi}, Enrico and {Keating}, Laura and {Vogelsberger}, Mark and {Zier}, Oliver and {Shen}, Xuejian and {Popovic}, Filip},
        title = "{The THESAN-ZOOM project: central starbursts and inside-out quenching govern galaxy sizes in the early Universe}",
      journal = {\mnras},
         year = 2025,
        month = dec,
       volume = {544},
       number = {2},
        pages = {1732-1747},
          doi = {10.1093/mnras/staf1861},
archivePrefix = {arXiv},
       eprint = {2503.04894},
 primaryClass = {astro-ph.GA},
       adsurl = {https://ui.adsabs.harvard.edu/abs/2025MNRAS.544.1732M}
}

@ARTICLE{megatron_impactstarformationfeedback,
       author = {{Katz}, Harley and {Rey}, Martin P. and {Cadiou}, Corentin and {Kimm}, Taysun and {Agertz}, Oscar},
        title = "{The Impact of Star Formation and Feedback Recipes on the Stellar Mass and Interstellar Medium of High-Redshift Galaxies}",
      journal = {The Open Journal of Astrophysics},
         year = 2026,
        month = feb,
       volume = {9},
        pages = {56097},
          doi = {10.33232/001c.156097},
archivePrefix = {arXiv},
       eprint = {2411.07282},
 primaryClass = {astro-ph.GA},
       adsurl = {https://ui.adsabs.harvard.edu/abs/2026OJAp....956097K}
}

@ARTICLE{megatron_reproducingdiversityhighredshift,
       author = {{Katz}, Harley and {Rey}, Martin P. and {Cadiou}, Corentin and {Agertz}, Oscar and {Blaizot}, Jeremy and {Cameron}, Alex J. and {Choustikov}, Nicholas and {Devriendt}, Julien and {Hauk}, Uliana and {Jones}, Gareth C. and {Kimm}, Taysun and {Laseter}, Isaac and {Martin-Alvarez}, Sergio and {Matsumoto}, Kosei and {Pearce}, Autumn and {Rodr{\'\i}guez Montero}, Francisco and {Rosdahl}, Joki and {Sanati}, Mahsa and {Saxena}, Aayush and {Slyz}, Adrianne and {Stiskalek}, Richard and {Storck}, Anatole and {Veenema}, Oscar and {Yee}, Wonjae},
        title = "{MEGATRON: Reproducing the Diversity of High-Redshift Galaxy Spectra with Cosmological Radiation Hydrodynamics Simulations}",
      journal = {arXiv e-prints},
         year = 2025,
        month = oct,
          eid = {arXiv:2510.05201},
        pages = {arXiv:2510.05201},
          doi = {10.48550/arXiv.2510.05201},
archivePrefix = {arXiv},
       eprint = {2510.05201},
 primaryClass = {astro-ph.GA},
       adsurl = {https://ui.adsabs.harvard.edu/abs/2025arXiv251005201K}
}

@ARTICLE{naidu2025_BHstar,
       author = {{Naidu}, Rohan P. and {Matthee}, Jorryt and {Katz}, Harley and {de Graaff}, Anna and {Oesch}, Pascal and {Smith}, Aaron and {Greene}, Jenny E. and {Brammer}, Gabriel and {Weibel}, Andrea and {Hviding}, Raphael and {Chisholm}, John and {Labb\textbackslash'e}, Ivo and {Simcoe}, Robert A. and {Witten}, Callum and {Atek}, Hakim and {Baggen}, Josephine F.~W. and {Belli}, Sirio and {Bezanson}, Rachel and {Boogaard}, Leindert A. and {Bose}, Sownak and {Covelo-Paz}, Alba and {Dayal}, Pratika and {Fudamoto}, Yoshinobu and {Furtak}, Lukas J. and {Giovinazzo}, Emma and {Goulding}, Andy and {Gronke}, Max and {Heintz}, Kasper E. and {Hirschmann}, Michaela and {Illingworth}, Garth and {Inoue}, Akio K. and {Johnson}, Benjamin D. and {Leja}, Joel and {Leonova}, Ecaterina and {McConachie}, Ian and {Maseda}, Michael V. and {Natarajan}, Priyamvada and {Nelson}, Erica and {Setton}, David J. and {Shivaei}, Irene and {Sobral}, David and {Stefanon}, Mauro and {Tacchella}, Sandro and {Toft}, Sune and {Torralba}, Alberto and {van Dokkum}, Pieter and {van der Wel}, Arjen and {Volonteri}, Marta and {Walter}, Fabian and {Wang}, Bingjie and {Watson}, Darach},
        title = "{A ``Black Hole Star'' Reveals the Remarkable Gas-Enshrouded Hearts of the Little Red Dots}",
      journal = {arXiv e-prints},
         year = 2025,
        month = mar,
          eid = {arXiv:2503.16596},
        pages = {arXiv:2503.16596},
          doi = {10.48550/arXiv.2503.16596},
archivePrefix = {arXiv},
       eprint = {2503.16596},
 primaryClass = {astro-ph.GA},
       adsurl = {https://ui.adsabs.harvard.edu/abs/2025arXiv250316596N}
}

@ARTICLE{behroozi_2012_rockstar,
       author = {{Behroozi}, Peter S. and {Wechsler}, Risa H. and {Wu}, Hao-Yi},
        title = "{The ROCKSTAR Phase-space Temporal Halo Finder and the Velocity Offsets of Cluster Cores}",
      journal = {\apj},
         year = 2013,
        month = jan,
       volume = {762},
       number = {2},
          eid = {109},
        pages = {109},
          doi = {10.1088/0004-637X/762/2/109},
archivePrefix = {arXiv},
       eprint = {1110.4372},
 primaryClass = {astro-ph.CO},
       adsurl = {https://ui.adsabs.harvard.edu/abs/2013ApJ...762..109B}
}

@ARTICLE{harikane_spectroscopic_constraints,
       author = {{Harikane}, Yuichi and {Nakajima}, Kimihiko and {Ouchi}, Masami and {Umeda}, Hiroya and {Isobe}, Yuki and {Ono}, Yoshiaki and {Xu}, Yi and {Zhang}, Yechi},
        title = "{Pure Spectroscopic Constraints on UV Luminosity Functions and Cosmic Star Formation History from 25 Galaxies at z $_{spec}$ = 8.61-13.20 Confirmed with JWST/NIRSpec}",
      journal = {\apj},
         year = 2024,
        month = jan,
       volume = {960},
       number = {1},
          eid = {56},
        pages = {56},
          doi = {10.3847/1538-4357/ad0b7e},
archivePrefix = {arXiv},
       eprint = {2304.06658},
 primaryClass = {astro-ph.GA},
       adsurl = {https://ui.adsabs.harvard.edu/abs/2024ApJ...960...56H}
}

@ARTICLE{hainline_cosmos_infancy,
       author = {{Hainline}, Kevin N. and {Johnson}, Benjamin D. and {Robertson}, Brant and {Tacchella}, Sandro and {Helton}, Jakob M. and {Sun}, Fengwu and {Eisenstein}, Daniel J. and {Simmonds}, Charlotte and {Topping}, Michael W. and {Whitler}, Lily and {Willmer}, Christopher N.~A. and {Rieke}, Marcia and {Suess}, Katherine A. and {Hviding}, Raphael E. and {Cameron}, Alex J. and {Alberts}, Stacey and {Baker}, William M. and {Baum}, Stefi and {Bhatawdekar}, Rachana and {Bonaventura}, Nina and {Boyett}, Kristan and {Bunker}, Andrew J. and {Carniani}, Stefano and {Charlot}, Stephane and {Chevallard}, Jacopo and {Chen}, Zuyi and {Curti}, Mirko and {Curtis-Lake}, Emma and {D'Eugenio}, Francesco and {Egami}, Eiichi and {Endsley}, Ryan and {Hausen}, Ryan and {Ji}, Zhiyuan and {Looser}, Tobias J. and {Lyu}, Jianwei and {Maiolino}, Roberto and {Nelson}, Erica and {Pusk{\'a}s}, D{\'a}vid and {Rawle}, Tim and {Sandles}, Lester and {Saxena}, Aayush and {Smit}, Renske and {Stark}, Daniel P. and {Williams}, Christina C. and {Willott}, Chris and {Witstok}, Joris},
        title = "{The Cosmos in Its Infancy: JADES Galaxy Candidates at z > 8 in GOODS-S and GOODS-N}",
      journal = {\apj},
         year = 2024,
        month = mar,
       volume = {964},
       number = {1},
          eid = {71},
        pages = {71},
          doi = {10.3847/1538-4357/ad1ee4},
archivePrefix = {arXiv},
       eprint = {2306.02468},
 primaryClass = {astro-ph.GA},
       adsurl = {https://ui.adsabs.harvard.edu/abs/2024ApJ...964...71H}
}

@ARTICLE{tacchella_JADES_imaging,
       author = {{Tacchella}, Sandro and {Eisenstein}, Daniel J. and {Hainline}, Kevin and {Johnson}, Benjamin D. and {Baker}, William M. and {Helton}, Jakob M. and {Robertson}, Brant and {Suess}, Katherine A. and {Chen}, Zuyi and {Nelson}, Erica and {Pusk{\'a}s}, D{\'a}vid and {Sun}, Fengwu and {Alberts}, Stacey and {Egami}, Eiichi and {Hausen}, Ryan and {Rieke}, George and {Rieke}, Marcia and {Shivaei}, Irene and {Williams}, Christina C. and {Willmer}, Christopher N.~A. and {Bunker}, Andrew and {Cameron}, Alex J. and {Carniani}, Stefano and {Charlot}, Stephane and {Curti}, Mirko and {Curtis-Lake}, Emma and {Looser}, Tobias J. and {Maiolino}, Roberto and {Maseda}, Michael V. and {Rawle}, Tim and {Rix}, Hans-Walter and {Smit}, Renske and {{\"U}bler}, Hannah and {Willott}, Chris and {Witstok}, Joris and {Baum}, Stefi and {Bhatawdekar}, Rachana and {Boyett}, Kristan and {Danhaive}, A. Lola and {de Graaff}, Anna and {Endsley}, Ryan and {Ji}, Zhiyuan and {Lyu}, Jianwei and {Sandles}, Lester and {Saxena}, Aayush and {Scholtz}, Jan and {Topping}, Michael W. and {Whitler}, Lily},
        title = "{JADES Imaging of GN-z11: Revealing the Morphology and Environment of a Luminous Galaxy 430 Myr after the Big Bang}",
      journal = {\apj},
         year = 2023,
        month = jul,
       volume = {952},
       number = {1},
          eid = {74},
        pages = {74},
          doi = {10.3847/1538-4357/acdbc6},
archivePrefix = {arXiv},
       eprint = {2302.07234},
 primaryClass = {astro-ph.GA},
       adsurl = {https://ui.adsabs.harvard.edu/abs/2023ApJ...952...74T}
}

@ARTICLE{Baldwin_GNz11_size_estimate,
       author = {{Baldwin}, James O. and {Nelson}, Erica and {Johnson}, Benjamin D. and {Oesch}, Pascal A. and {Tacchella}, Sandro and {Illingworth}, Garth D. and {Gibson}, Justus and {Hartley}, Abby},
        title = "{A Size Estimate for Galaxy GN-z11}",
      journal = {Research Notes of the American Astronomical Society},
         year = 2024,
        month = jan,
       volume = {8},
       number = {1},
          eid = {29},
        pages = {29},
          doi = {10.3847/2515-5172/ad220a},
archivePrefix = {arXiv},
       eprint = {2401.04186},
 primaryClass = {astro-ph.GA},
       adsurl = {https://ui.adsabs.harvard.edu/abs/2024RNAAS...8...29B}
}

@ARTICLE{Wang_UNCOVER,
       author = {{Wang}, Bingjie and {Fujimoto}, Seiji and {Labb{\'e}}, Ivo and {Furtak}, Lukas J. and {Miller}, Tim B. and {Setton}, David J. and {Zitrin}, Adi and {Atek}, Hakim and {Bezanson}, Rachel and {Brammer}, Gabriel and {Leja}, Joel and {Oesch}, Pascal A. and {Price}, Sedona H. and {Chemerynska}, Iryna and {Cutler}, Sam E. and {Dayal}, Pratika and {van Dokkum}, Pieter and {Goulding}, Andy D. and {Greene}, Jenny E. and {Fudamoto}, Y. and {Khullar}, Gourav and {Kokorev}, Vasily and {Marchesini}, Danilo and {Pan}, Richard and {Weaver}, John R. and {Whitaker}, Katherine E. and {Williams}, Christina C.},
        title = "{UNCOVER: Illuminating the Early Universe-JWST/NIRSpec Confirmation of z > 12 Galaxies}",
      journal = {\apjl},
         year = 2023,
        month = nov,
       volume = {957},
       number = {2},
          eid = {L34},
        pages = {L34},
          doi = {10.3847/2041-8213/acfe07},
archivePrefix = {arXiv},
       eprint = {2308.03745},
 primaryClass = {astro-ph.GA},
       adsurl = {https://ui.adsabs.harvard.edu/abs/2023ApJ...957L..34W}
}

@ARTICLE{finkelstein_maisies,
       author = {{Finkelstein}, Steven L. and {Bagley}, Micaela B. and {Arrabal Haro}, Pablo and {Dickinson}, Mark and {Ferguson}, Henry C. and {Kartaltepe}, Jeyhan S. and {Papovich}, Casey and {Burgarella}, Denis and {Kocevski}, Dale D. and {Huertas-Company}, Marc and {Iyer}, Kartheik G. and {Koekemoer}, Anton M. and {Larson}, Rebecca L. and {P{\'e}rez-Gonz{\'a}lez}, Pablo G. and {Rose}, Caitlin and {Tacchella}, Sandro and {Wilkins}, Stephen M. and {Chworowsky}, Katherine and {Medrano}, Aubrey and {Morales}, Alexa M. and {Somerville}, Rachel S. and {Yung}, L.~Y. Aaron and {Fontana}, Adriano and {Giavalisco}, Mauro and {Grazian}, Andrea and {Grogin}, Norman A. and {Kewley}, Lisa J. and {Kirkpatrick}, Allison and {Kurczynski}, Peter and {Lotz}, Jennifer M. and {Pentericci}, Laura and {Pirzkal}, Nor and {Ravindranath}, Swara and {Ryan}, Russell E. and {Trump}, Jonathan R. and {Yang}, Guang and {Almaini}, Omar and {Amor{\'\i}n}, Ricardo O. and {Annunziatella}, Marianna and {Backhaus}, Bren E. and {Barro}, Guillermo and {Behroozi}, Peter and {Bell}, Eric F. and {Bhatawdekar}, Rachana and {Bisigello}, Laura and {Bromm}, Volker and {Buat}, V{\'e}ronique and {Buitrago}, Fernando and {Calabr{\`o}}, Antonello and {Casey}, Caitlin M. and {Castellano}, Marco and {Ch{\'a}vez Ortiz}, {\'O}scar A. and {Ciesla}, Laure and {Cleri}, Nikko J. and {Cohen}, Seth H. and {Cole}, Justin W. and {Cooke}, Kevin C. and {Cooper}, M.~C. and {Cooray}, Asantha R. and {Costantin}, Luca and {Cox}, Isabella G. and {Croton}, Darren and {Daddi}, Emanuele and {Dav{\'e}}, Romeel and {de La Vega}, Alexander and {Dekel}, Avishai and {Elbaz}, David and {Estrada-Carpenter}, Vicente and {Faber}, Sandra M. and {Fern{\'a}ndez}, Vital and {Finkelstein}, Keely D. and {Freundlich}, Jonathan and {Fujimoto}, Seiji and {Garc{\'\i}a-Argum{\'a}nez}, {\'A}ngela and {Gardner}, Jonathan P. and {Gawiser}, Eric and {G{\'o}mez-Guijarro}, Carlos and {Guo}, Yuchen and {Hamblin}, Kurt and {Hamilton}, Timothy S. and {Hathi}, Nimish P. and {Holwerda}, Benne W. and {Hirschmann}, Michaela and {Hutchison}, Taylor A. and {Jaskot}, Anne E. and {Jha}, Saurabh W. and {Jogee}, Shardha and {Juneau}, St{\'e}phanie and {Jung}, Intae and {Kassin}, Susan A. and {Le Bail}, Aur{\'e}lien and {Leung}, Gene C.~K. and {Lucas}, Ray A. and {Magnelli}, Benjamin and {Mantha}, Kameswara Bharadwaj and {Matharu}, Jasleen and {McGrath}, Elizabeth J. and {McIntosh}, Daniel H. and {Merlin}, Emiliano and {Mobasher}, Bahram and {Newman}, Jeffrey A. and {Nicholls}, David C. and {Pandya}, Viraj and {Rafelski}, Marc and {Ronayne}, Kaila and {Santini}, Paola and {Seill{\'e}}, Lise-Marie and {Shah}, Ekta A. and {Shen}, Lu and {Simons}, Raymond C. and {Snyder}, Gregory F. and {Stanway}, Elizabeth R. and {Straughn}, Amber N. and {Teplitz}, Harry I. and {Vanderhoof}, Brittany N. and {Vega-Ferrero}, Jes{\'u}s and {Wang}, Weichen and {Weiner}, Benjamin J. and {Willmer}, Christopher N.~A. and {Wuyts}, Stijn and {Zavala}, Jorge A. and {Ceers Team}},
        title = "{A Long Time Ago in a Galaxy Far, Far Away: A Candidate z {\ensuremath{\sim}} 12 Galaxy in Early JWST CEERS Imaging}",
      journal = {\apjl},
         year = 2022,
        month = dec,
       volume = {940},
       number = {2},
          eid = {L55},
        pages = {L55},
          doi = {10.3847/2041-8213/ac966e},
archivePrefix = {arXiv},
       eprint = {2207.12474},
 primaryClass = {astro-ph.GA},
       adsurl = {https://ui.adsabs.harvard.edu/abs/2022ApJ...940L..55F}
}

@ARTICLE{harikane_maisies,
       author = {{Harikane}, Yuichi and {Ouchi}, Masami and {Oguri}, Masamune and {Ono}, Yoshiaki and {Nakajima}, Kimihiko and {Isobe}, Yuki and {Umeda}, Hiroya and {Mawatari}, Ken and {Zhang}, Yechi},
        title = "{A Comprehensive Study of Galaxies at z   9-16 Found in the Early JWST Data: Ultraviolet Luminosity Functions and Cosmic Star Formation History at the Pre-reionization Epoch}",
      journal = {\apjs},
         year = 2023,
        month = mar,
       volume = {265},
       number = {1},
          eid = {5},
        pages = {5},
          doi = {10.3847/1538-4365/acaaa9},
archivePrefix = {arXiv},
       eprint = {2208.01612},
 primaryClass = {astro-ph.GA},
       adsurl = {https://ui.adsabs.harvard.edu/abs/2023ApJS..265....5H}
}

@ARTICLE{Ono_maisies_morphology,
       author = {{Ono}, Yoshiaki and {Harikane}, Yuichi and {Ouchi}, Masami and {Yajima}, Hidenobu and {Abe}, Makito and {Isobe}, Yuki and {Shibuya}, Takatoshi and {Wise}, John H. and {Zhang}, Yechi and {Nakajima}, Kimihiko and {Umeda}, Hiroya},
        title = "{Morphologies of Galaxies at z {\ensuremath{\gtrsim}} 9 Uncovered by JWST/NIRCam Imaging: Cosmic Size Evolution and an Identification of an Extremely Compact Bright Galaxy at z 12}",
      journal = {\apj},
         year = 2023,
        month = jul,
       volume = {951},
       number = {1},
          eid = {72},
        pages = {72},
          doi = {10.3847/1538-4357/acd44a},
archivePrefix = {arXiv},
       eprint = {2208.13582},
 primaryClass = {astro-ph.GA},
       adsurl = {https://ui.adsabs.harvard.edu/abs/2023ApJ...951...72O}
}

@ARTICLE{bouwens_UV_luminosity_CEERS,
       author = {{Bouwens}, Rychard and {Illingworth}, Garth and {Oesch}, Pascal and {Stefanon}, Mauro and {Naidu}, Rohan and {van Leeuwen}, Ivana and {Magee}, Dan},
        title = "{UV luminosity density results at z > 8 from the first JWST/NIRCam fields: limitations of early data sets and the need for spectroscopy}",
      journal = {\mnras},
         year = 2023,
        month = jul,
       volume = {523},
       number = {1},
        pages = {1009-1035},
          doi = {10.1093/mnras/stad1014},
archivePrefix = {arXiv},
       eprint = {2212.06683},
 primaryClass = {astro-ph.CO},
       adsurl = {https://ui.adsabs.harvard.edu/abs/2023MNRAS.523.1009B}
}

@ARTICLE{bakx_glass,
       author = {{Bakx}, Tom J.~L.~C. and {Zavala}, Jorge A. and {Mitsuhashi}, Ikki and {Treu}, Tommaso and {Fontana}, Adriano and {Tadaki}, Ken-ichi and {Casey}, Caitlin M. and {Castellano}, Marco and {Glazebrook}, Karl and {Hagimoto}, Masato and {Ikeda}, Ryota and {Jones}, Tucker and {Leethochawalit}, Nicha and {Mason}, Charlotte and {Morishita}, Takahiro and {Nanayakkara}, Themiya and {Pentericci}, Laura and {Roberts-Borsani}, Guido and {Santini}, Paola and {Serjeant}, Stephen and {Tamura}, Yoichi and {Trenti}, Michele and {Vanzella}, Eros},
        title = "{Deep ALMA redshift search of a z {\ensuremath{\sim}} 12 GLASS-JWST galaxy candidate}",
      journal = {\mnras},
         year = 2023,
        month = mar,
       volume = {519},
       number = {4},
        pages = {5076-5085},
          doi = {10.1093/mnras/stac3723},
archivePrefix = {arXiv},
       eprint = {2208.13642},
 primaryClass = {astro-ph.GA},
       adsurl = {https://ui.adsabs.harvard.edu/abs/2023MNRAS.519.5076B}
}

@ARTICLE{naidu_glass,
       author = {{Naidu}, Rohan P. and {Oesch}, Pascal A. and {van Dokkum}, Pieter and {Nelson}, Erica J. and {Suess}, Katherine A. and {Brammer}, Gabriel and {Whitaker}, Katherine E. and {Illingworth}, Garth and {Bouwens}, Rychard and {Tacchella}, Sandro and {Matthee}, Jorryt and {Allen}, Natalie and {Bezanson}, Rachel and {Conroy}, Charlie and {Labbe}, Ivo and {Leja}, Joel and {Leonova}, Ecaterina and {Magee}, Dan and {Price}, Sedona H. and {Setton}, David J. and {Strait}, Victoria and {Stefanon}, Mauro and {Toft}, Sune and {Weaver}, John R. and {Weibel}, Andrea},
        title = "{Two Remarkably Luminous Galaxy Candidates at z {\ensuremath{\approx}} 10-12 Revealed by JWST}",
      journal = {\apjl},
         year = 2022,
        month = nov,
       volume = {940},
       number = {1},
          eid = {L14},
        pages = {L14},
          doi = {10.3847/2041-8213/ac9b22},
archivePrefix = {arXiv},
       eprint = {2207.09434},
 primaryClass = {astro-ph.GA},
       adsurl = {https://ui.adsabs.harvard.edu/abs/2022ApJ...940L..14N}
}

@ARTICLE{Ono_morphology_GSz14,
       author = {{Ono}, Yoshiaki and {Ouchi}, Masami and {Harikane}, Yuichi and {Yajima}, Hidenobu and {Nakajima}, Kimihiko and {Fujimoto}, Seiji and {Nakane}, Minami and {Xu}, Yi},
        title = "{Morphological Demographics of Galaxies at z {\ensuremath{\sim}} 10{\textendash}16: Log-normal Size Distribution and Exponential Profiles Consistent with the Disk Formation Scenario}",
      journal = {\apj},
         year = 2025,
        month = oct,
       volume = {991},
       number = {2},
          eid = {222},
        pages = {222},
          doi = {10.3847/1538-4357/adfc4d},
archivePrefix = {arXiv},
       eprint = {2502.08885},
 primaryClass = {astro-ph.GA},
       adsurl = {https://ui.adsabs.harvard.edu/abs/2025ApJ...991..222O}
}

@ARTICLE{helton_photodetection_GSz14_1,
       author = {{Helton}, Jakob M. and {Rieke}, George H. and {Alberts}, Stacey and {Wu}, Zihao and {Eisenstein}, Daniel J. and {Hainline}, Kevin N. and {Carniani}, Stefano and {Ji}, Zhiyuan and {Baker}, William M. and {Bhatawdekar}, Rachana and {Bunker}, Andrew J. and {Cargile}, Phillip A. and {Charlot}, St{\'e}phane and {Chevallard}, Jacopo and {D'Eugenio}, Francesco and {Egami}, Eiichi and {Johnson}, Benjamin D. and {Jones}, Gareth C. and {Lyu}, Jianwei and {Maiolino}, Roberto and {P{\'e}rez-Gonz{\'a}lez}, Pablo G. and {Rieke}, Marcia J. and {Robertson}, Brant and {Saxena}, Aayush and {Scholtz}, Jan and {Shivaei}, Irene and {Sun}, Fengwu and {Tacchella}, Sandro and {Whitler}, Lily and {Williams}, Christina C. and {Willmer}, Christopher N.~A. and {Willott}, Chris and {Witstok}, Joris and {Zhu}, Yongda},
        title = "{Photometric detection at 7.7 {\ensuremath{\mu}}m of a galaxy beyond redshift 14 with JWST/MIRI}",
      journal = {Nature Astronomy},
         year = 2025,
        month = may,
       volume = {9},
        pages = {729-740},
          doi = {10.1038/s41550-025-02503-z},
archivePrefix = {arXiv},
       eprint = {2405.18462},
 primaryClass = {astro-ph.GA},
       adsurl = {https://ui.adsabs.harvard.edu/abs/2025NatAs...9..729H}
}

@ARTICLE{Wu_GSz14_1,
       author = {{Wu}, Zihao and {Eisenstein}, Daniel J. and {Johnson}, Benjamin D. and {Jakobsen}, Peter and {Alberts}, Stacey and {Arribas}, Santiago and {Baker}, William M. and {Bunker}, Andrew J. and {Carniani}, Stefano and {Charlot}, St{\'e}phane and {Chevallard}, Jacopo and {Curti}, Mirko and {Curtis-Lake}, Emma and {D'Eugenio}, Francesco and {Hainline}, Kevin and {Helton}, Jakob M. and {Hsiao}, Tiger Yu-Yang and {Ji}, Xihan and {Ji}, Zhiyuan and {Looser}, Tobias J. and {Rieke}, George and {Rinaldi}, Pierluigi and {Robertson}, Brant and {Scholtz}, Jan and {Sun}, Fengwu and {Tacchella}, Sandro and {Trussler}, James A.~A. and {Williams}, Christina C. and {Willmer}, Christopher N.~A. and {Willott}, Chris and {Witstok}, Joris and {Zhu}, Yongda},
        title = "{JADES-GS-z14-1: A Compact, Faint Galaxy at z {\ensuremath{\approx}} 14 with Weak Metal Lines from Extremely Deep JWST MIRI, NIRCam, and NIRSpec Observations}",
      journal = {\apj},
         year = 2025,
        month = oct,
       volume = {992},
       number = {2},
          eid = {212},
        pages = {212},
          doi = {10.3847/1538-4357/ae01a1},
archivePrefix = {arXiv},
       eprint = {2507.22858},
 primaryClass = {astro-ph.GA},
       adsurl = {https://ui.adsabs.harvard.edu/abs/2025ApJ...992..212W}
}
\bibliographystyle{aasjournalv7}



\end{document}